\documentclass[11pt,english,1p,preprint]{elsarticle}
\usepackage[T1]{fontenc}
\usepackage[latin9]{inputenc}
\usepackage{babel}
\usepackage{array}
\usepackage{varioref}
\usepackage{float}
\usepackage{multirow}
\usepackage{amsmath}
\usepackage{amssymb}
\usepackage{graphicx}
\PassOptionsToPackage{normalem}{ulem}
\usepackage{ulem}

\makeatletter

\providecommand{\tabularnewline}{\\}
\floatstyle{ruled}
\newfloat{algorithm}{tbp}{loa}
\providecommand{\algorithmname}{Algorithm}
\floatname{algorithm}{\protect\algorithmname}

\newenvironment{lyxlist}[1]
	{\begin{list}{}
		{\settowidth{\labelwidth}{#1}
		 \setlength{\leftmargin}{\labelwidth}
		 \addtolength{\leftmargin}{\labelsep}
		 }}
	{\end{list}}

\usepackage{babel}
\usepackage{amsfonts}
\usepackage{amsthm}
\usepackage[percent]{overpic}
\usepackage{mathrsfs}
\usepackage{eqparbox}
\usepackage{stfloats}
\usepackage{pdfsync}
\usepackage{algorithm}
\usepackage{algorithmicx}
\usepackage{algpseudocode}
\usepackage{url}
\usepackage{color,soul}%
\usepackage[title]{appendix}
\usepackage[normalem]{ulem}
\usepackage{tabulary}
\usepackage{hyperref}
\usepackage{subfig}\usepackage{caption}
\usepackage{verbatim}
\usepackage[author={Loreto}]{pdfcomment}
\usepackage{environ}
\RenewEnviron{comment}{\pdfcomment{\BODY}}

\providecommand{\tabularnewline}{\\}
\newtheorem{thm}{Theorem}
\newtheorem{lem}[thm]{Lemma}
\newproof{pf}{Proof}

\newcommand{\proofoflemma}{}

\newproof{pflemma}{Proof of Lemma \proofoflemma}

\newcommand{\proofoftheorem}{}
\newproof{pfthm}{Proof of Theorem \proofoftheorem}
\newenvironment{proofofT}[1]
 {\renewcommand{\proofoftheorem}{#1}\pfthm}
 {\endthm}

\journal{Computer Communications}

\@ifundefined{showcaptionsetup}{}{%
 \PassOptionsToPackage{caption=false}{subfig}}
\usepackage{subfig}
\makeatother

\begin{document}
\begin{frontmatter}{}

\title{D2D Data Offloading in Vehicular Environments\\ with Optimal Delivery
Time Selection}

\author{Loreto Pescosolido, Marco Conti, Andrea Passarella}

\address{Italian National Research Council,}

\address{Institute for Informatics and Telematics (CNR-IIT), Pisa, Italy\vspace{-10mm}
}

\ead{loreto.pescosolido@iit.cnr.it, marco.conti@iit.cnr.it, andrea.passarella@iit.cnr.it}
\begin{abstract}
Within the framework of a Device-to-Device (D2D) data offloading system
for cellular networks, we propose a Content Delivery Management System
(CDMS) in which the instant for transmitting a content to a requesting
node, through a D2D communication, is selected to minimize the energy
consumption required for transmission. The proposed system is particularly
fit to highly dynamic scenarios, such as vehicular networks, where
the network topology changes at a rate which is comparable with the
order of magnitude of the delay tolerance. We present an analytical
framework able to predict the system performance, in terms of energy
consumption, using tools from the theory of point processes, validating
it through simulations, and provide a thorough performance evaluation
of the proposed CDMS, in terms of energy consumption and spectrum
use. Our performance analysis compares the energy consumption and
spectrum use obtained with the proposed scheme with the performance
of two benchmark systems. The first one is a plain classic cellular
scheme, the second is a D2D data offloading scheme (that we proposed
in previous works) in which the D2D transmissions are performed as
soon as there is a device with the required content within the maximum
D2D transmission range. The results show that, in specific scenarios,
the proposed scheme achieves an overall, i.e., including both cellular
and D2D communications, reduction of the energy consumption of up
to 70\% with respect to the plain cellular scheme and of up to 18\%
with respect to the benchmark D2D offloading scheme. Furthermore,
compared to the benchmark D2D offloading scheme in which the transmission
instant is not optimized, the reduction of the energy consumed for
the D2D transmissions only, is almost always above 90\%, peaking at
a 97\% percent reduction. Regarding spectrum use, the proposed scheme
allows to achieve an average fraction of the available radio resources
used per control interval which ranges between 40\% and 55\% less
than those used by the cellular scheme.
\end{abstract}
\end{frontmatter}{} \let\today\relax ~

\section{Introduction}

Device-to-Device (D2D) data offloading in cellular networks \citep{Rebecchi2015}
is a powerful means to decrease congestion at the base stations, reduce
the energy consumption of the overall system, and increase spectral
efficiency. The idea is that, whenever a content is requested by a
node, if the content is available at any of its neighbors, it should
be obtained from it, rather than through a network infrastructure
node. We define the nodes that can potentially hand the desired content
to the requesting node as Potential Content Providers (PCPs). The
set of PCPs depends on scenario parameters like the node density and
the content popularity, and on the specific protocol design. For
delay-tolerant applications, an interesting protocol design option
is that, in case a node issuing a content request has no PCP in its
neighborhood at the time of request, it waits for a predefined interval,
known as\emph{ content timeout}, within which it is still possible
to obtain the content from a new neighbor, encountered in the meantime
\citep{Whitbeck2012,Bruno2014}. Only at the expiration of the content
timeout, if the content has not yet been obtained, it is transmitted
by the infrastructure nodes, which retrieve it from a remote source,
through an Infrastructure-to-Device (I2D) transmission. This approach
is particularly effective in highly dynamic scenarios, such as vehicular
networks, where the network topology changes at a fast rate. The
use of a content timeout allows to increase the population of PCPs
beyond the set of the requesting node's neighbors at the request time,
extending such population to the nodes that will become its neighbors
in the future. In this way, the system may obtain an increase of the
offloading efficiency, defined as the percentage of contents delivered
by using D2D communications between peer nodes (vehicles), rather
than using I2D transmissions from the infrastructure nodes. 

In our prior works \citep{Pescosolido2018WoWMoM,Pescosolido2018AdHocNetworks}
we have shown that the considered type of D2D data offloading protocols
are also very effective in reducing the overall energy consumption
by exploiting the short-range D2D transmissions among nodes (provided
that the popular contents are kept in their caches by the nodes that
receive them), which require less transmit power (on average) than
the conventional I2D ones performed by the eNBs. While this is true
for most D2D data offloading protocols, especially when power control
is in use, there is still room for a significant performance improvement,
by taking \emph{full} advantage of the delay tolerance of requests,
with respect solution proposed in \citep{Pescosolido2018WoWMoM,Pescosolido2018AdHocNetworks}.

Consider two nodes, and define them as neighbors if and only if their
distance is less than or equal to a (nominal) maximum transmission
range $r_{\max}^{\text{}}$. In previous works that follow the above
described approach, \emph{in the case that, at the time of a content
request, there are no PCPs within a range $d_{\max}$ from the requesting
node }(i.e., no neighbor has the requested content in its cache),
as soon as the requesting node encounters a PCP, the content is transmitted.
It is clear that, in this case, the transmission takes place at the
maximum transmission range of the devices. Therefore, in a system
with distance-based power control, all the requests that are not fulfilled
at the request time, inherently require the use of the maximum D2D
transmit power. Furthermore, \emph{in the opposite case}, \emph{in
which at the request time there is already a PCP, say at distance}
$r<r_{\max}$, the content delivery requires a transmit power that
may be higher than what would be required if the delivery was postponed
to a later instant, at which the involved (or any other) content provider
could be closer than $r$ to the requesting node.

Motivated by this observation, in this work we propose the following
approach, to define an improved Content Delivery Management System
(CDMS). When a new request arrives, a controller, running, e.g., at
the eNodeB (eNB), exploits knowledge of nodes positions and predicted
motion in the near future (specifically, in the following content
timeout window), to estimate which PCP will be in range of the requesting
node in that timeframe. The content transmission is scheduled with
the PCP that is predicted to be at the minimum distance from the requesting
node, at the point in time when this will happen. In this way, provided
that a distance-dependent transmit power control is in use, the smallest
possible transmit power will be required for that content transmission.
We will show that, using this approach, the energy consumption of
the considered protocol for delay-tolerant application can be considerably
reduced. This work extends our previous work \citep{WWIC2018}, which
provided preliminary simulation results regarding the energy consumption
aspect. With respect to \citep{WWIC2018}, in this work, we provide
an analytical framework which allows to compute the statistics of
the energy consumption of the proposed system, and a performance evaluation
which quantifies, besides the energy consumption, also the spectrum
usage of the proposed system, in comparison with a benchmark plain
cellular system and with the CDMS system proposed by us in \citep{Pescosolido2018WoWMoM,Pescosolido2018AdHocNetworks}.

The paper is organized as follows. We position our work with respect
to the recent research trends in this area in Section~\ref{sec:Related-work}.
In Section~\ref{sec:System-model} we describe our system model,
positioning the proposed CDMS in the framework of a protocol stack
tailored for D2D data offloading protocols. In Section~\ref{sec:Content-Delivery-Management}
we present in detail the proposed Content Delivery Management System
(CDMS) and provide an analytical framework to predict its performance.
In Section~\ref{sec:MAC-and-physical} we describe a possible MAC
(adapted from an existing solution) for an in-band implementation
of the proposed D2D offloading scheme. In Section~\ref{sec:Performance-evaluation},
through extensive system-level simulations, we validate the proposed
analytical framework and evaluate the performance of the proposed
system in terms of the average energy consumption per content delivery,
and average spectrum use, required to satisfy a given system-wise
traffic demand. Finally, Section~\ref{sec:Conclusion} concludes
the paper, summarizing our contribution and most relevant results.

\section{Related work\label{sec:Related-work}}

The use of D2D communications to offload traffic from infrastructure
nodes has been investigated in the recent years by the researchers
of different communities. Works like \citep{Ji2016a,Ji2016b} aim
at investigating scaling laws and network throughput from a fundamental
limits perspective. Works like \citep{Lin2014,Yang2017} (amongst
many others), aim at devising radio resource allocation strategies,
and/or other physical layer parameters, like coding rates and transmit
power levels, assuming that the D2D and/or I2D links to be scheduled
are given as an input to the problem. More specific protocol-oriented
works have appeared in the last years as well. The interested reader
may want to check, e.g., \citep{Rebecchi2015} for an extensive survey.
In these works, the objective is to determine and schedule I2D and
D2D offloading communications as a function of the request patterns
(as opposed to the above mentioned works, in which the links to be
scheduled are an input to the problem). In \citep{Whitbeck2012},
the peculiarity of D2D data offloading for delay-tolerant applications
was first addressed, clarifying the advantages of offloading cellular
traffic from the network infrastructure, and targeting the offloading
efficiency\footnote{In \citep{Whitbeck2012}, the term used to indicate the offloading
efficiency is ``offload ratio''.} as the key performance metric. In \citep{Bruno2014}, the authors
propose a basic CDMS for D2D data offloading an analyze its performance
in a vehicular scenario, investigating the interplay of the content
timeout duration with other system or scenario parameters, in a vehicular
scenario. The presence of multiple contents with different popularity
(which is related to the rate at which a specific content is requested
by the devices) is not considered. In \citep{Rebecchi2016}, in a
scenario in which content delivery mostly relies on D2D-offloading,
a strategy for I2D re-injection of contents in the network is proposed
to mitigate the effect of temporal content starving in a certain areas.
In \citep{Rebecchi2015b}, in the framework of a content dissemination
problem (i.e., when contents need to reach all the nodes, without
having been explicitly requested), the authors propose a mixed I2D-multicast
and D2D-relaying reinforcement-learning-based strategy, which determines
which users should receive the contents through D2D relaying from
a neighboring device or through a direct I2D transmission. The above
mentioned works, although providing interesting insights from the
perspective of offloading efficiency maximization, devote less attention
to performance metrics which are closer to physical quantities, like
energy consumption and spectrum efficiency. Our work is motivated
by the need to take into account such metrics in the system design,
and optimize the design to maximize them. In \citep{Pescosolido2018WoWMoM,Pescosolido2018AdHocNetworks},
we have elaborated a CDMS building on the one presented in \citep{Bruno2014},
and proposed an analytical model to evaluate its performance \citep{Pescosolido2018WoWMoM}.
In \citep{Pescosolido2018AdHocNetworks}, we evaluate the impact,
on the performance evaluation, of using different channel models,\uline{}
showing that simplistic scalar models models\footnote{For instance, deterministic or flat fading path loss models coupled
with an SNR threshold-based packet error modeling.} can lead to high inaccuracy when dealing with performance metrics
tightly related with the physical layer aspects, like energy consumption
or spectrum use. With respect to \citep{Bruno2014}, our works \citep{Pescosolido2018WoWMoM,Pescosolido2018AdHocNetworks}
take into account contents with different popularity, and considers
energy consumption and spectrum occupation, besides offloading efficiency,
as key performance metrics. The analytical model in \citep{Pescosolido2018WoWMoM}
investigates the effect of content popularity and vehicles speed on
the D2D transmit power, provides expressions for the offloading efficiency
and the energy consumption of both I2D and D2D transmissions, and
relies on them to select the best value for the maximum D2D transmission
range. The CDMS considered in \citep{Pescosolido2018WoWMoM}, however,
does not optimize the D2D transmission time, letting the nodes transmit
a requested content as soon as they encounter a node requesting it.
In this work, differently from the above mentioned ones, we leverage
the degree of freedom entailed by delay tolerance by deferring the
D2D transmission instant to the time it will require the lowest power,
thus achieving quite significant performance gains in terms of energy
consumption. We also deem it appropriate to take into account accurate
channel models, since using relatively simplistic models may result
in an inaccurate estimation of the performance gain of a particular
design \citep{Pescosolido2018AdHocNetworks}. Furthermore, we consider
it necessary, when dealing with the type of performance metrics discussed
above, to integrate in the performance evaluation an actual radio
resource management technique. Among the many available, as done in
\citep{Pescosolido2018AdHocNetworks}, we used the solution proposed
in \citep{Yang2017}, adapting it to a multi-cell scenario and to
deal with frequency selective channels.

\section{System model\label{sec:System-model}}

\subsection{Nodes topology, mobility, and content requests\label{subsec:Nodes-topology}}

\begin{figure}[t]
\begin{centering}
\includegraphics[width=1\columnwidth]{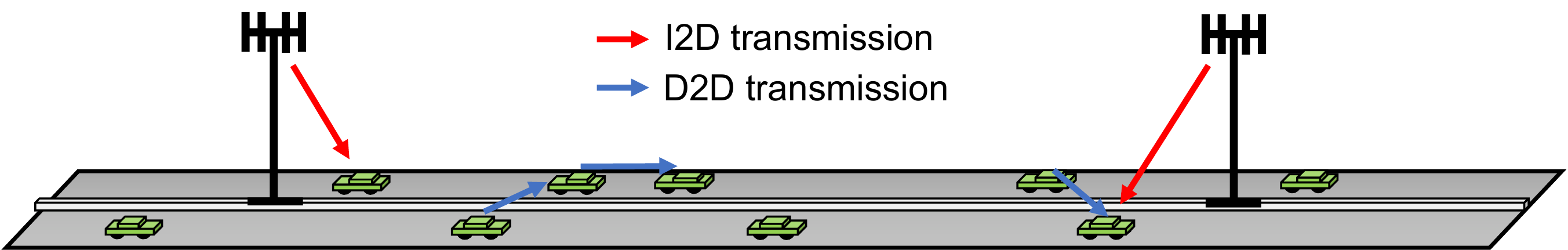}
\par\end{centering}
\caption{Sketch of the considered scenario.}
\label{fig:sketch}
\end{figure}

We consider a Region of Interest (ROI) consisting of a bidirectional
street chunk which vehicle enter, traverse, and exit from both ends,
as shown in Fig.~\ref{fig:sketch}.

Vehicles enter the street according to a given stationary temporal
arrival process, with an average arrival rate of $\lambda_{t}$ vehicles
per second ($\lambda_{t}/2$ vehicles per second on each end). Each
vehicle $n$ traverses the ROI at an average speed which is the sample
of a random variable $V^{*}$ with Probability Density Function (PDF)
$p_{V^{*}}(v)$. We assume that the speed value is bounded by a maximum
speed $v_{\max}$. Each vehicle has onboard a mobile device,
which can be either a human hand-held device or part of the vehicle
equipment. Along the street, a set of eNBs is regularly placed. At
each instant, each device (vehicle) is under the coverage of an eNB.
Each device issues content requests according to a given stationary
content request process with an average content request rate of $\lambda_{Z}$
requests per second, by sending requests messages to the eNB it is
associated to at the request time\footnote{Alternatively, content delivery requests may be originated at a remote
server, intended to specific nodes. For instance, this could be the
case of contents related to specific applications running at many
devices, that a remote server instructs to be delivered to the devices
running it. For the purpose of this work, it is not important which
is the actual origin of the request, since they will be handled in
the same way.}. The specific content being requested is drawn from a content popularity
distribution $P_{Z}(z)$. Similarly to \citep{Whitbeck2012,Bruno2014},
we assume that the content requests can be fulfilled with some delay
tolerance, i.e., they must be served at most within a \emph{content
timeout} $\tau_{c}$, starting at the request instant. A request may
be fulfilled either by a PCP, through a D2D communication,\uline{
}or, if there are no PCPs, by some remote server in the Internet,
using the eNB of a cellular network as the final communication hop\footnote{The problem of placing contents on remote servers is an orthogonal
problem to the one we address, and the location of such servers in
the Internet has no effect either on the algorithm features of the
proposed CDMS or on its performance evaluation.}. We assume that, in such case, the content is directly sent by an
eNB. In this work we assume that, within the content timeout, the
first option (delivery through D2D) is always privileged, and I2D
transmissions are performed only at the end of the content timeout,
if it has expired before any PCP has been found. The rationale is
that, in this way, we maximize the advantage of D2D transmissions
in offloading traffic from the cellular infrastructure, which is one
of the primary goals of any offloading system. Furthermore, to keep
the probability of cache overflow limited,\uline{ }each device
keeps the contents it has received in its cache for a \emph{sharing
timeout} $\tau_{s}$, starting at the content reception instant, making
it available to other nodes encountered by the device which may request
it. At the expiration of the sharing timeout, to avoid an indefinite
increase of the cache occupation, the content is removed from the
cache. Finally, another important parameter of interest is the maximum
nominal\footnote{I.e., computed on the basis on a deterministic channel attenuation
model which relates the distance to the nominal channel gain, see
Section~\ref{sec:MAC-and-physical}.} transmission range of the devices, indicated with \uline{\mbox{$r_{\max}^{\text{(D2D)}}$}.}
Table~\ref{tab:basic_sysparam} summarized the basic scenario parameters
introduced so far.
\begin{table}[h]
\label{tab:basic_sysparam}\caption{Basic system model parameters}

\centering{}%
\begin{tabular}{c|c}
parameter & symbol\tabularnewline
\hline 
\hline 
Vehicles arrival rate & $\lambda_{t}$\tabularnewline
\hline 
Vehicles speed distribution & $p_{V}\left(v\right)$\tabularnewline
\hline 
Maximum speed & \tabularnewline
\hline 
Content request rate & $\lambda_{Z}$\tabularnewline
\hline 
Content popularity distribution & $P_{Z}\left(z\right)$\tabularnewline
\hline 
Content timeout & $\tau_{c}$\tabularnewline
\hline 
Sharing timeout & $\tau_{s}$\tabularnewline
\hline 
maximum nominal D2D transmission range & $r_{\max}^{\text{(D2D)}}$\tabularnewline
\end{tabular}
\end{table}
The assumptions are quite general. For the purpose of performance
evaluation, specific models need to be assumed for the involved random
processes. We leave the description of the specific assumptions used
for our performance evaluation to Section~\ref{sec:Performance-evaluation}.

\subsection{High-level view on D2D offloading control\label{subsec:High-level-view-on}}

In general, D2D-aided data offloading protocols define a strategy
to handle each content request during its lifetime, from the instant
it is taken in charge, to the time the content is finally delivered
to the requesting node. The network infrastructure may be involved
in this process in different ways. At one extreme, the whole process
can be carried out autonomously by the mobile devices, typically operating
out of the cellular network band, e.g. using WiFi-direct or other
similar enabling communication technologies. This approach requires
the frequent execution of neighbor discovery routines, and each node
first seeks to obtain a content of interest directly from the neighbors,
without the need of any control or support from the network infrastructure
elements (such as the eNBs). Only at the approaching of the content
timeout expiration, in what is sometimes called the ``panic zone'',
if the content has not yet been received, the node requires the content
to some remote server via the cellular infrastructure. This approach
has been considered, for instance, in \citep{Bruno2014}. 

\begin{figure}[t]
\begin{centering}
\includegraphics[width=0.9\textwidth]{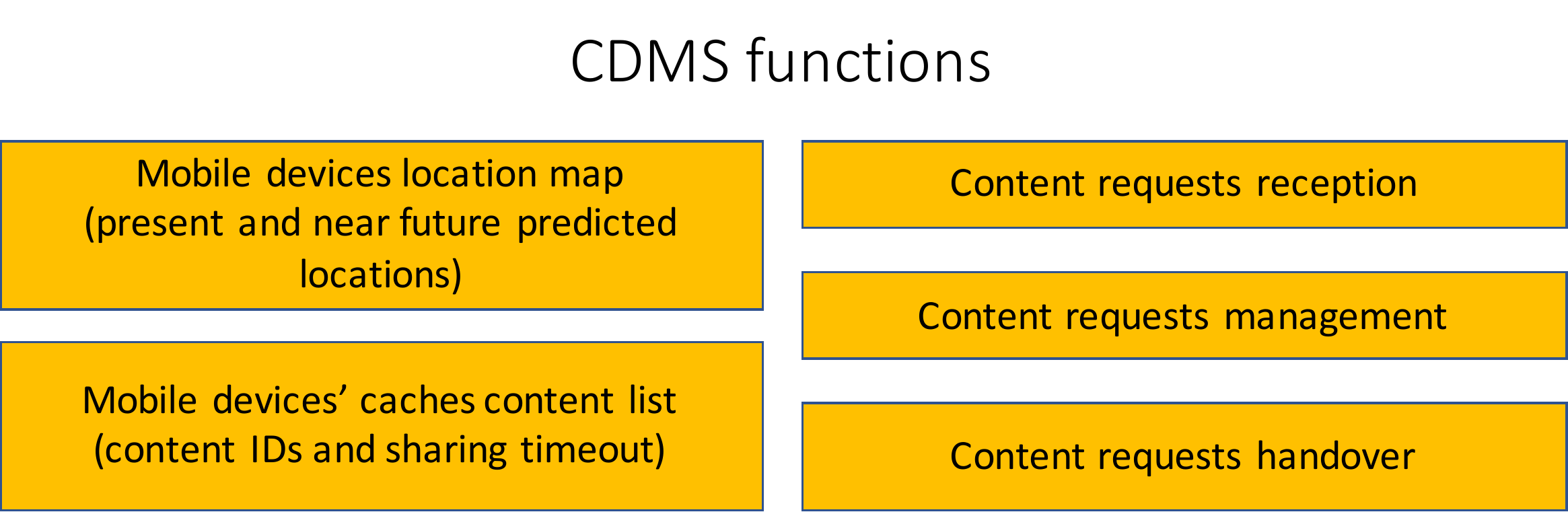}\vspace{-3mm}
\par\end{centering}
\caption{CDMS functions}
\label{fig:CDMS_functions}\vspace{-2mm}
\end{figure}

Alternatively, as proposed in this work, the D2D-aided data offloading
protocol is entirely executed under the supervision of an entity that
we call Content Delivery Management System. The CDMS is a distributed
software agent under the control of the network operator. Most of
its functions are executed at the eNBs. Whenever a content request
is generated by a user, it reaches the CDMS, which is responsible
for handling it from the time it is issued by a device, until its
fulfillment, deciding how and when the content request will be satisfied,
either through D2D or through I2D communications. The main CDMS functions
are summarized in Figure~\ref{fig:CDMS_functions}. 
The left column shows the two functions which provide the required
information for the CDMS to operate. In general, D2D data offloading
may rely on different types of information regarding the presence
nodes in the region of interest, e.g., statistical (node density)
or deterministic information (nodes positions). In this work, we assume
that both the current nodes positions and their predicted trajectories
are available. The right-hand side of Figure~\ref{fig:CDMS_functions}
shows the set of functions for handling the content requests. The
core CDMS function is the content request management, which consists
in the execution of a specific D2D offloading algorithm. The protocol
decides whether a content should be provided to the requesting node
by one of its neighbors or by an eNB, and at what time the transmission
should be performed. In our previous works \citep{Pescosolido2018WoWMoM,Pescosolido2018AdHocNetworks},
this protocol essentially consisted in delivering the content through
D2D as soon as there is an available (i.e., within radio transmission
range) PCP. In this work, we introduce a new strategy for the content
provider selection, which also schedules the optimal instant and position
at which the content provider is supposed to transmit the content
to the requesting device. As we will show, carefully scheduling the
content transmission allows to obtain a considerable performance improvement,
in terms of both energy consumption and radio spectrum use. The details
of the proposed protocol are described in Section~\ref{sec:Content-Delivery-Management}.

In the case that a node, while waiting for a content, moves from one
cell to another, the management procedure associated to that request
is handed over from the eNB currently in charge of it to the adjacent
one. This requires an exchange of information across adjacent eNBs.

Finally, the CDMS relies on a radio resource reuse management scheme
(RRRM) which operates at the MAC layer of the cellular network protocol
stack. For the purposes of this work, we have implemented a scheme
that we have adapted from \citep{Yang2017}, and already used in \citep{Pescosolido2018WoWMoM,Pescosolido2018AdHocNetworks}.
A detailed description of the considered RRRM scheme can be found
in \citep{Pescosolido2018AdHocNetworks} and it is briefly recalled
in Section~\ref{sec:MAC-and-physical}, where we also provide details
on the implementation of physical layer related aspects such as the
channel model and the transmission error model. It is important to
emphasize that without an accurate modeling of such aspects, it would
be difficult to obtain reliable simulation results, in terms of energy
consumption and spectrum use \citep{Pescosolido2018AdHocNetworks}.
The RRRM scheme is responsible for periodically allocating the radio
resources, within the time horizon of short control intervals (with
duration in the order of one second) to the set of D2D and I2D content
transmissions whose transmission time has been determined, with a
coarser time scale, by the scheduling performed at CDMS level.

\begin{figure}[t]
\begin{centering}
\includegraphics[width=0.9\textwidth]{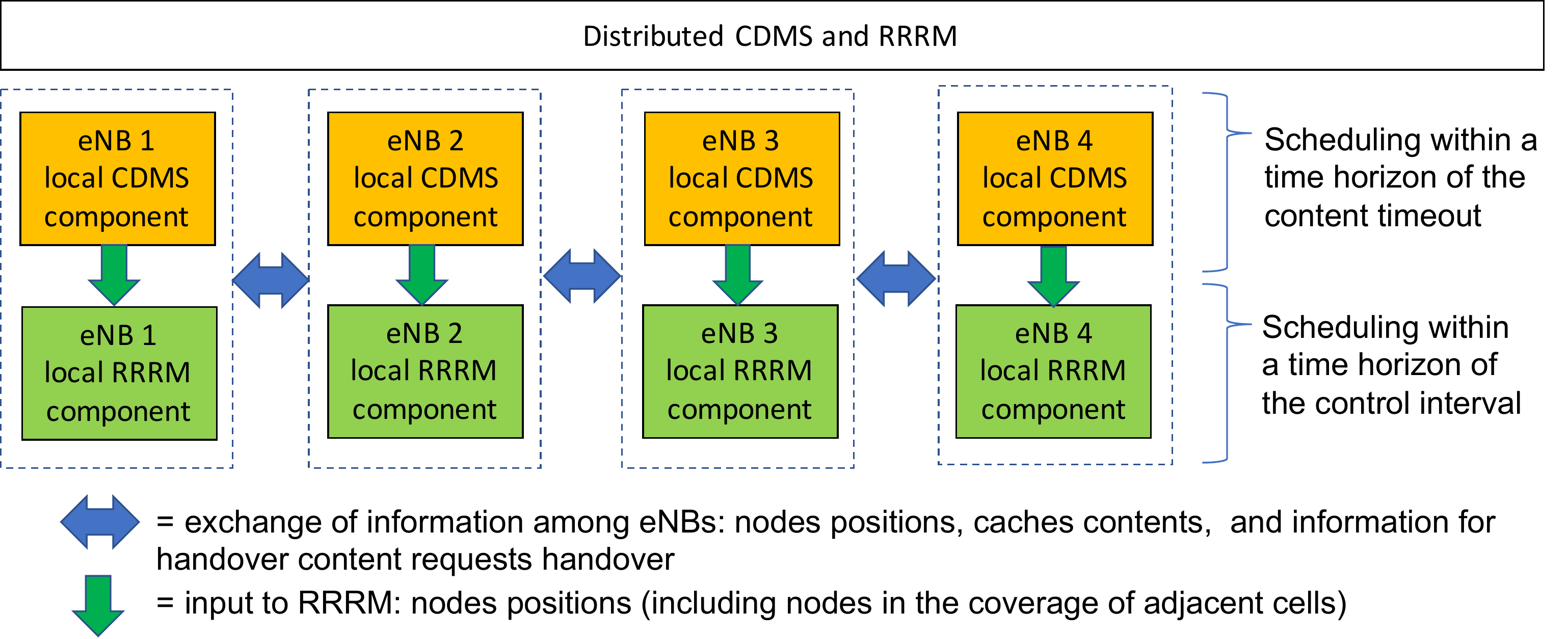}\vspace{-3mm}
\par\end{centering}
\caption{High level abstraction of the distributed CDMS.}
\label{fig:distributed_CDMS}\vspace{-2mm}
\end{figure}

Figure~\ref{fig:distributed_CDMS} provides an high level abstraction
of how the proposed CDMS can be implemented in a distributed way.
The horizontal arrows represent the necessary exchange of information
across adjacent cells. This control information flow would be typically
carried out through high speed fiber connection using, e.g., the X2
interface of 4G and 5G systems. The vertical arrows represent the
information provided to the RRRM component by the CDMS component.
\vspace{-2mm}

\section{Content Delivery Management System with optimized delivery time\label{sec:Content-Delivery-Management}}

For each content request it receives by the mobile devices, the CDMS
executes an algorithm (explained next) which requires that it is aware
of the location and expected trajectory of each node. To this end,
the CDMS acts on a distributed database containing the up-to-date
list of each node's position and an estimation of their trajectories
for the next $\tau_{c}$ seconds. Each device may obtain a running
estimation of its speed and trajectory in the next seconds, either
through the use of GPS or, if it is part of the vehicle electronic
equipment, directly from the speedometer, and send it periodically
to the eNBs. Alternatively, the devices can send the GPS information
only to the eNB, leaving the burden of trajectory estimation to the
CDMS. In general, different combinations are possible, whose details
are outside the scope of this work. In this way, essentially, the
CDMS has a picture of how the network topology will evolve in the
next seconds. In this work, we assume a perfect prediction of the
vehicles' trajectory for an amount of time equal to the content timeout,
leaving the evaluation of the robustness of the system with respect
to trajectory prediction errors to a future work.

Each device $k$ has an internal content cache $\mathcal{C}_{k}$
populated with previously downloaded contents. We assume that, at
any time, the CDMS also has an index of the contents in each node's
cache, and it knows the instants at which each content will be removed
from the node's cache due to the expiration of the associated sharing
timeout. Each eNB keeps the above described information for all the
nodes in its coverage and all the nodes in the adjacent eNBs cells,
see Figure~\ref{fig:distributed_CDMS}. The detailed actions for
the execution of the proposed protocol by the CDMS are provided in
Algorithm~\ref{algo_CDMS_req_handle}. Regarding the requesting node,
all it does after issuing a content request is to wait for the content
to be delivered to it. At the expiration of the content timeout, if
it has not yet received the content by a neighboring device, it will
anyway receive it from an eNB through an I2D transmission.\\
~\\
\begin{algorithm}[H]
\caption{{\small{}Actions taken by CDMS for handling content request $(k,z)$}}
\label{algo_CDMS_req_handle}

\begin{algorithmic} [1]

\item {\footnotesize{}$\mathbf{Upon}$ receiving }\texttt{\footnotesize{}$(k,z)$\_cont\_req }{\footnotesize\par}

\item {\footnotesize{}$\mathbf{set}$ }\texttt{\footnotesize{}$(k,z)$\_served}{\footnotesize{}
= $\mathbf{false}$ }{\footnotesize\par}

\item {\footnotesize{}$\mathbf{set}$ }\texttt{\footnotesize{}$(k,z)$\_content\_timeout}{\footnotesize{} }{\footnotesize\par}

\item {\footnotesize{}$\mathbf{compute}$ the region of interest
$\mathcal{A}^{(k,z)}$ : the area within which all }PCP{\footnotesize{}s
can be located at the request time}{\footnotesize\par}

\item {\footnotesize{}{*}$\mathbf{compute}$ the set of }PCP{\footnotesize{}s
$\mathcal{Q}_{(k,z)}=\left\{ q_{1}^{(k,z)},\ldots,q_{N_{(k,z)}}^{(k,z)}\right\} $
within the area $\mathcal{A}^{(k,z)}$}{\footnotesize\par}

\item {\footnotesize{}$\mathbf{if}$ $\mathcal{Q}_{(k,z)}\neq\emptyset$}{\footnotesize\par}

\item {\footnotesize{}{*}\quad{} $\forall q_{i}\in\mathcal{Q}_{(k,z)}$,
$\mathbf{compute}$ the optimal time and distance $t_{i}^{(k,z)}$
and $\delta_{i}^{(k,z)}$ for delivering content $z$ to device $k$
using the PCP $q_{i}$}{\footnotesize\par}

\item {\footnotesize{}\quad{}$\mathbf{compute}$ $\hat{i}=\arg\min_{i\in\left\{ 1,\ldots,N_{(k,z)}\right\} }\left(\delta_{i}^{(k,z)}\right)$ }{\footnotesize\par}

\item {\footnotesize{}\quad{}$\mathbf{set}$ $\hat{\delta}^{(k,z)}:=\delta_{\hat{i}}^{(k,z)}$}{\footnotesize\par}

\item {\footnotesize{}\quad{}$\mathbf{set}$ $\hat{q}^{(k,z)}:=q_{\hat{i}}^{(k,z)}$
(the selected content provider for delivering content $z$ to node
$k$)}{\footnotesize\par}

\item {\footnotesize{}\quad{}$\mathbf{set}$ $\hat{t}^{(k,z)}:=t_{\hat{i}}^{(k,z)}$
(the selected instant for delivering content $z$ to node $k$ using
$\hat{q}^{(k,z)}$ )}{\footnotesize\par}

\item {\footnotesize{}$\mathbf{else}$}{\footnotesize\par}

\item {\footnotesize{}\quad{}$\mathbf{set}$ $\hat{t}^{(k,z)}:=\tau_{c}$,
$\hat{q}^{(k,z)}:=\mathbf{null}$}{\footnotesize\par}

\item {\footnotesize{}$\mathbf{end\,if}$}{\footnotesize\par}

\item {\footnotesize{}$\mathbf{while}$ }\texttt{\footnotesize{}$t<\hat{t}^{(k,z)}$}{\footnotesize{}
$\mathbf{do}$ ({*} the condition check is performed at every control
interval)}{\footnotesize\par}

\item {\footnotesize{}\quad{}$\mathbf{if}$ }\texttt{\footnotesize{}$\exists r\mid\boldsymbol{x}_{r}\in\mathcal{A}^{(k,z)},q\notin\mathcal{Q}_{(k,z)},\mathcal{C}_{r}\ni z$}{\footnotesize{}
$\mathbf{then}$ }{\footnotesize\par}

\item {\footnotesize{}\quad{}\quad{}$\mathbf{set}$ $q_{\text{new}}^{(k,z)}:=q$}{\footnotesize\par}

\item {\footnotesize{}\quad{}\quad{}$\mathbf{compute}$ the optimal
time and distance $t_{\text{new}}^{(k,z)}$ and $\delta_{\text{new}}^{(k,z)}$
for delivering using $q_{\text{new}}^{(k,z)}$ }{\footnotesize\par}

\item {\footnotesize{}\quad{}\quad{}$\mathbf{if}$ $\delta_{\text{new}}^{(k,z)}<\hat{\delta}^{(k,z)}$
$\mathbf{then}$}{\footnotesize\par}

\item {\footnotesize{}\quad{}\quad{}\quad{}$\mathbf{set}$ $\hat{q}^{(k,z)}:=q_{\text{new}}^{(k,z)}$}{\footnotesize\par}

\item {\footnotesize{}\quad{}\quad{}\quad{}$\mathbf{set}$ $\hat{t}^{(k,z)}:=t_{\text{new}}^{(k,z)}$}{\footnotesize\par}

\item {\footnotesize{}\quad{}\quad{}\quad{}$\mathbf{set}$ }$\hat{\delta}^{(k,z)}:=\delta_{\text{new}}^{(k,z)}$

$\vdots$

\algstore{myalg}
\end{algorithmic}
\end{algorithm}
\begin{algorithm}[t]
$\hspace{1.4cm}\vdots$

\begin{algorithmic}\algrestore{myalg}

\item {\footnotesize{}\quad{}\quad{}$\mathbf{else}$}{\footnotesize\par}

\item {\footnotesize{}\quad{}\quad{}\quad{}$\mathcal{R}_{(k,z)}:=\mathcal{R}_{(k,z)}\cup\left\{ r\right\} $}{\footnotesize\par}

\item {\footnotesize{}\quad{}\quad{}\quad{}$\mathbf{discard}$
$q_{\text{new}}^{(k,z)}$, $t_{\text{new}}^{(k,z)}$, $\delta_{\text{new}}^{(k,z)}$}{\footnotesize\par}

\item {\footnotesize{}\quad{}\quad{}$\mathbf{end\,if}$}{\footnotesize\par}

\item {\footnotesize{}\quad{}\quad{}$\mathbf{repeat}$ steps 17-25
for each $q$ satisfying conditions at step 15}{\footnotesize\par}

\item {\footnotesize{}\quad{}$\mathbf{end\,if}$}{\footnotesize\par}

\item {\footnotesize{}$\mathbf{end\,while}$}{\footnotesize\par}

\item {\footnotesize{}$\mathbf{if}$ $\hat{q}^{(k,z)}\neq\mathbf{null}$
$\mathbf{and}$ }\texttt{\footnotesize{}$(k,z)$\_served}{\footnotesize{}
= $\mathbf{false}$}{\footnotesize\par}

\item {\footnotesize{}\quad{}$\mathbf{while}$ $t\leq\tau_{c}$}{\footnotesize\par}

\item {\footnotesize{}\quad{}\quad{}$\mathbf{trigger}$ transmission
$\hat{q}^{(k,z)}\stackrel{z}{\rightarrow}k$ }{\footnotesize\par}

\item {\footnotesize{}\quad{}\quad{}$\mathbf{while}$ }\texttt{\footnotesize{}$(k,z)$\_}{\footnotesize{}ACK
not received}{\footnotesize\par}

\item {\footnotesize{}\quad{}\quad{}$\mathbf{wait\,for}$ }\texttt{\footnotesize{}$(k,z)$\_}{\footnotesize{}ACK }{\footnotesize\par}

\item {\footnotesize{}\quad{}\quad{}$\mathbf{upon}$ }\texttt{\footnotesize{}$(k,z)$\_}{\footnotesize{}ACK
reception }{\footnotesize\par}

\item {\footnotesize{}\quad{}\quad{}\quad{}$\mathbf{set}$ }\texttt{\footnotesize{}$(k,z)$\_served}{\footnotesize{}
= $\mathbf{true}$ }{\footnotesize\par}

\item {\footnotesize{}\quad{}\quad{}\quad{}$\mathbf{remove}$
$(k,z)$ from $\mathcal{L}_{\text{req}}$ }{\footnotesize\par}

\item {\footnotesize{}\quad{}$\mathbf{end\,while}$ }{\footnotesize\par}

\item {\footnotesize{}$\mathbf{end\,\mathbf{if}}$}{\footnotesize\par}

\item {\footnotesize{}$\mathbf{while}$}\texttt{\footnotesize{} $(k,z)$\_served}{\footnotesize{}
= $\mathbf{false}$}{\footnotesize\par}

\item {\footnotesize{}\quad{}$\mathbf{send}$ }\texttt{\footnotesize{}$z$}{\footnotesize{}
to }\texttt{\footnotesize{}$k$ }{\footnotesize{}from eNB }{\footnotesize\par}

\item {\footnotesize{}\quad{}$\mathbf{wait\,for}$ }\texttt{\footnotesize{}$(k,z)$\_}{\footnotesize{}ACK }{\footnotesize\par}

\item {\footnotesize{}\quad{}$\mathbf{upon}$ }\texttt{\footnotesize{}$(k,z)$\_}{\footnotesize{}ACK
reception }{\footnotesize\par}

\item {\footnotesize{}\quad{}\quad{}$\mathbf{set}$ }\texttt{\footnotesize{}$(k,z)$\_served}{\footnotesize{}
= $\mathbf{true}$ }{\footnotesize\par}

\item {\footnotesize{}\quad{}\quad{}$\mathbf{remove}$ $(k,z)$
from $\mathcal{L}_{\text{req}}$ }{\footnotesize\par}

\item {\footnotesize{}$\mathbf{end\,while}$ }{\footnotesize\par}

\item {\footnotesize{}$\mathbf{Cancel}$ }\texttt{\footnotesize{}$(k,z)$\_content\_timeout}{\footnotesize{} }{\footnotesize\par}

\end{algorithmic}

\vspace{5mm}
\end{algorithm}
Essentially, on a coarse timescale, with respect to a given content
request, the requesting node and the proposed CDMS act as follows.
Upon receiving a content request from a node within its coverage,
the eNB performs the following operations:
\begin{enumerate}
\item It determines the region within which PCPs for the considered request
can be located. In practice, the region is determined by the maximum
speed parameter $v_{\max}$, the content timeout $\tau_{c}$, and
the maximum D2D transmission range $r_{\max}^{\text{(D2D)}}$. These
parameters are system parameters known to the CDMS and which determine
the set of PCPs that the requesting node is supposed to encounter
before the content timeout for the request expires. (steps 4-5) 
\item It compares the estimated trajectory of the requesting node for the
next $\tau_{c}$ seconds, with those of all the nodes that have the
requested content in their caches. For each PCP, it compares the expiration
instant of the sharing timeout for the requested content with the
expiration time of the content timeout associated to the request.
If the sharing timeout will expire before the content timeout, the
estimated trajectory of the PCP is considered only up to the expiration
instant of the sharing timeout. (steps 6-7) 
\item On the basis of the trajectories of all the PCPs, it computes (i)
which provider will achieve the shortest distance from requesting
device, (ii) the value of such distance, and (iii) the instant at
which the two nodes are going to find themselves that close to each
other. The provider with the shortest prospective distance is selected
as the one who will transmit the content to the requesting node. (step
8) 
\item It schedules the transmission of the content from the selected content
provider to the requesting node at the time instant in which the two
nodes will be at their shortest distance (compatible with the expiration
of the content and sharing timeout). (steps 9-14) 
\item Before the scheduled transmission instant arrives, the CDMS, with
respect to the considered content request, keeps track of any device
other than the selected content provider which (i) is not included
in the initial set of PCPs and (ii) is supposed to encounter the requesting
before the expiration of the content timeout. If any such node receives
the same requested content in this period, the CDMS computes the shortest
distance it will reach from the requesting node. If this new shortest
distance is found to be shorter than the originally computed shortest
distance, the content delivery is rescheduled to be performed by the
newly found PCP, at the (new) time instant it will find itself at
the newly found shortest distance. (steps 15-29)
\item At the scheduled transmit time, trigger the transmission as per the
result of the assignment of the transmission to a PCP or to an eNB,
and re-trigger it until an ACK is received or the content timeout
expires. (steps 30-39)
\item At the expiration of the content timeout, if the content has not been
received yet, transmit the content from the eNB under which the requesting
device is located. (steps 40-47)\vspace{-1mm}
\end{enumerate}
The operations described above are executed, in practice, in discrete-time,
with control intervals of duration typically much lower than the content
timeout. For instance, the content timeout can be in the order of
one minute, and the control interval duration is in the order of 1
second. We consider a typical multi-carrier system, with control intervals
determined by the organization of the radio resources onto frames,
each one corresponding to a rectangular time-frequency grid of Physical
Resource Blocks. For instance, considering an LTE-like MAC, a control
interval could be mapped to a frame, i.e., it would last one second.
The scheduled content delivery instants are hence computed in terms
of number of control intervals, and mapped to the future control intervals.
The content deliveries scheduled by the CDMS within the time horizon
of the content timeout, will contribute, in the control interval corresponding
to the prescribed delivery time, to the input to the Radio Resource
Reuse (RRR) and allocation scheme described in Section~\ref{sec:MAC-and-physical}.

\section{Analytical model\label{sec:Analytical-model}}

\begin{figure}
\begin{centering}
\includegraphics[width=1\columnwidth]{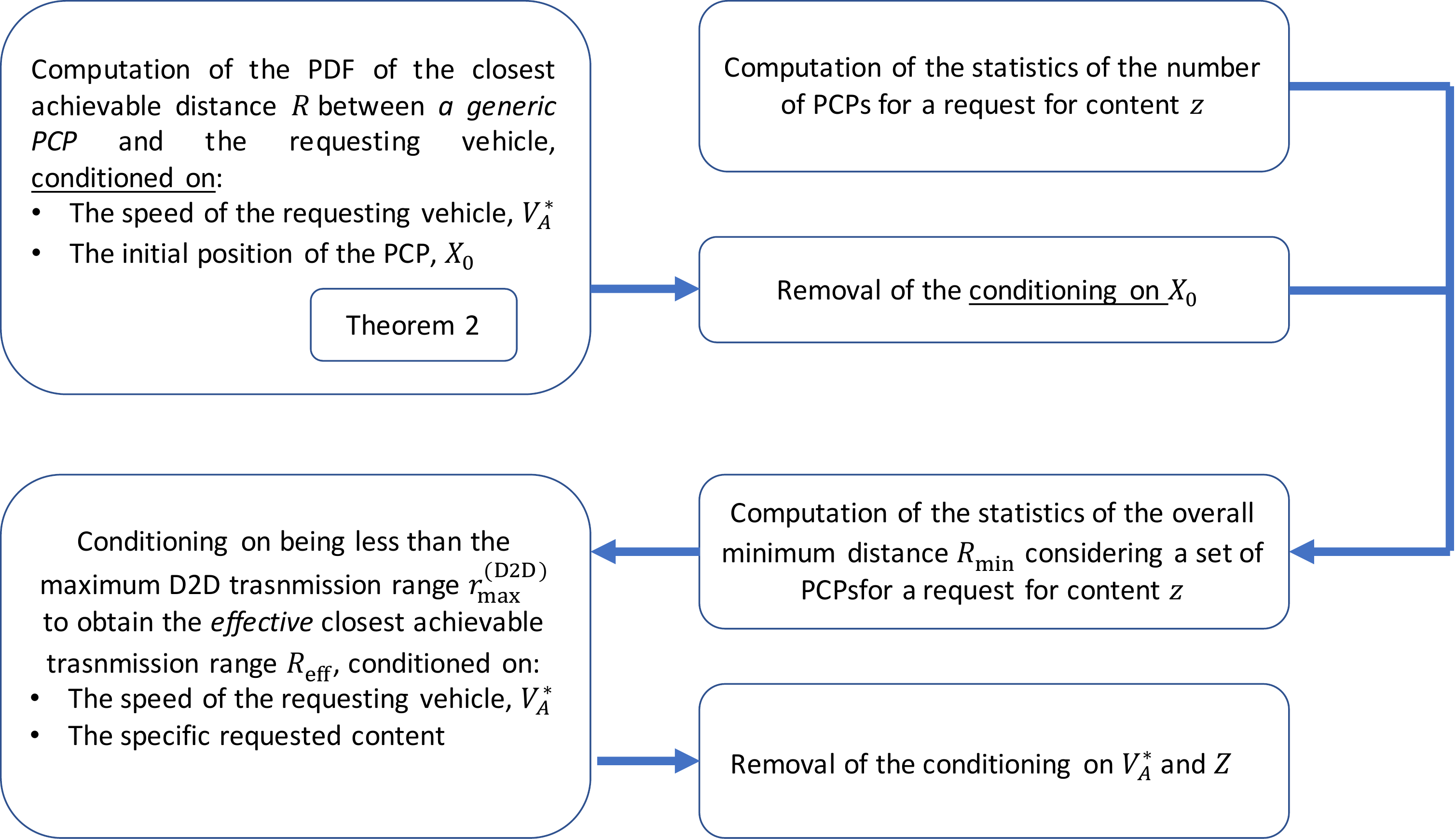}
\par\end{centering}
\caption{Summary of the derivations}
\label{fig:summary_derivations}
\end{figure}

In this section, we provide an analytical model for computing the
statistics of the D2D transmission distance, and the associated energy
consumption of mobile devices, when the CDMS described in Section
4 is in operation in the scenario described in Section~\ref{sec:System-model}.
In the rest of this section, we represent the nodes positions in the
street chunk as a unidimensional Homogeneous Spatial Poisson Point
Process (HSPPP), i.e., we only consider the spatial dimension along
the street median axis. For our derivations, we will use analytical
results obtained in our previous paper \citep{Pescosolido2018WoWMoM},
which are briefly summarized in the following Subsection~\ref{subsec:Preliminary-results}.
In Subsection~\ref{subsec:Analytical-model-distance} we compute
the statistics of the D2D transmission distance resulting from the
use of the proposed CDMS, and use them to compute the statistics of
the associated energy consumption in Subsection~\ref{subsec:Analytical-model-energy}.
The derivations in Subsection~\ref{subsec:Analytical-model-distance}
will follow the line of reasoning represented in Figure~\ref{fig:summary_derivations}.
Before starting, it will be useful to introduce the following notation:
The symbol $p_{X}\left(\cdot\right)$ and $P_{Z}\left(\cdot\right)$
indicate the PDF and the Probability Mass Function (PMF) of continuous
and discrete random variables, respectively. $F_{X}\left(\cdot\right)$
is used to represent the cumulative distribution function (CDF) of
a random variable $X$, for both continuous and discrete random variables\footnote{For discrete random variables, the CDF is a staircase function.}.
The math blackboard expression $\mathbb{P}\left(\cdot\right)$ indicates
the probability of the event enclosed in the parentheses. The function
$u_{[a,b]}\left(x\right)$ represents the rectangular function defined
as $u_{[a,b]}(x)=\begin{cases}
1 & x\in[a,b]\\
0 & x\notin[a,b]
\end{cases}$. In case the domain interval is open at one or both edges, the notation
$u_{(a,b]}(x)$ and $u_{(a,b)}(x)$ will be used. Setting one of the
extremes to infinity, the same notation indicates the step functions
equal to unity for values of $x$ larger than or equal to $a$, and
zero otherwise, $u_{[a,+\infty)}(x)$, or equal to unity for values
of $x$ less than or equal to $b$, and zero otherwise, $u_{(-\infty,b]}(x)$.
The function $u_{0}\left(x\right)$ indicates the Dirac pulse function.
The operator $\circ$ used between two functions, as in $f\left(\cdot\right)\circ g\left(\cdot\right)$,
represents the convolution operator, i.e., $f\left(x\right)\circ g\left(x\right)=\smallint_{-\infty}^{\infty}f\left(x'\right)g\left(x'-x\right)dx'$.

\subsection{Preliminary results\label{subsec:Preliminary-results}}

Let us assume that vehicles enter the street according to a Homogeneous
Temporal Poisson Point Process with a rate $\lambda_{t}$ vehicles
per second (see Section~\ref{subsec:Nodes-topology}) and each vehicle
traverses the street at a constant speed $v$, which is independent
sample of a random variable $V^{*}$ with PDF $p_{V^{*}}\left(v\right)$,
and the direction of motion is incorporated in the sign of $v$. The
the following hold true \citep[Lemma 2]{Pescosolido2018WoWMoM}:
\begin{lyxlist}{00.00.0000}
\item [{(i)}] The positions of the nodes along the street is a HSPPP with
linear density 
\begin{equation}
\rho=\int_{-\infty}^{\infty}\frac{1}{\left|v\right|}\lambda_{t}p_{V^{*}}(v)dv.\label{eq:overall_spatial_density}
\end{equation}
\item [{(ii)}] In the special case of uniformly distributed speeds, i.e.,
assuming
\begin{equation}
p_{V^{*}}\left(v\right)=\frac{1}{2}\left(v_{\max}-v_{\min}\right)u_{[-v_{\max},-v_{\min}]}\left(v\right)+\frac{1}{2}\left(v_{\max}-v_{\min}\right)u_{[v_{\min},v_{\max}]}\left(v\right),\label{eq:PDF_speed}
\end{equation}
the linear density of vehicles present in the street at a given instant
is\texttt{}
\begin{equation}
\rho=\lambda_{t}\left(\ln v_{\max}-\ln v_{\min}\right)/(v_{\max}-v_{\min}).\label{eq:overall_spatial_density_special}
\end{equation}
\end{lyxlist}
Furthermore, under the assumption that content requests arrive according
to a HTPPP with interarrival rate $\lambda_{Z}$ and that the requested
contents of different requesting nodes and across different requests
are i.i.d. random variables with PMF $P_{Z}\left(z\right)$ representing
the content popularity, we have that \citep[Lemma 3]{Pescosolido2018WoWMoM}
\begin{lyxlist}{00.00.0000}
\item [{(iii)}] The temporal process of arrival of requests for a specific
content $z$ is a HTPPP with interarrival rate\vspace{-2mm}
\begin{equation}
\lambda_{z}=P_{Z}(z)\lambda_{Z}.\label{eq:HTPPP_z_rate}
\end{equation}
\item [{(iv)}] The positions of nodes having content $z$ in their caches
at any given time instant is a HSPPP with linear density satisfying
the tight lower bound\footnote{Note that, as discussed in \citep{Pescosolido2018WoWMoM}, under the
assumption that $\tau_{s}\gg\tau_{c}$, the approximation~\eqref{eq:SPPP-geo-content-distribution-2}
is quite accurate.}\vspace{-2mm}
\begin{equation}
\rho_{z}\gtrsim\rho\left(1-e^{-\lambda_{z}(\tau_{s}-\tau_{c})}\right).\label{eq:SPPP-geo-content-distribution-2}
\end{equation}
\end{lyxlist}
A futher result we will need is the probability that a given request
is for a content that \emph{is not} already cached at the device requesting
it or; in other words, the probability that the request is ``non-repeated''.
This probability is given by $\mathbb{P}\left(\mathrm{NR}\right)=\sum_{z}P_{Z}\left(z\right)\mathbb{P}\left(\mathcal{C}\not\ni z\right)$,
where $\mathcal{C}$ is the set of contents in cache of the requesting
node at the request time. Finally, and the probability that the requested
content is $z$, conditioned on the request being non-repeated, is
given by \citep[Lemma 4]{Pescosolido2018WoWMoM}\vspace{-2mm}
\begin{equation}
p_{Z}\left(z\mid\mathrm{NR}\right)=\frac{Pr\left(Z=z\right)Pr\left(\mathcal{C}\not\ni z\right)}{\sum_{z\in\mathcal{L}}Pr\left(Z=z\right)Pr\left(\mathcal{C}\not\ni z\right)}.
\end{equation}
The probability that the request is non-repeated is the probability
that the request fulfillment will require a transmission, either from
an eNB or from a mobile device (vehicle).\vspace{-2mm}

\subsection{Analytical model for the optimal D2D transmission distance\label{subsec:Analytical-model-distance}}

We start considering a device requesting a content $z$ at a given
instant $t_{0}$, which is onboard a vehicle denoted with letter A,
and a PCP for that request which is onboard a vehicle B. We indicate
with $V_{A}^{*}$ and $V_{B}^{*}$ the random speeds at which the
two vehicles are moving, and with $v_{a}^{*}$ and $v_{b}^{*}$, their
respective realizations. $V_{A}^{*}$ and $V_{B}^{*}$ are i.i.d.
and distributed according to a PDF $p_{V^{*}}\left(v\right)$. We
incorporate the marching direction in the speed value, associating
positive speed values to one direction and negative values to the
opposite one. For simplicity, we assume that the absolute (i.e., unsigned)
values of the speeds of vehicles marching in the two opposite directions
are distributed in the same way. Since $V^{*}$ is defined as the
\emph{signed} speed value, this assumptions entails that $p_{V^{*}}\left(v\right)$
is symmetric around 0.

We introduce the random variable representing the relative speed between
the two vehicles $V=V_{B}^{*}-V_{A}^{*}$. The PDF of the relative
speed $V$, conditioned on $V_{A}^{*}$, is given by
\begin{equation}
p_{V\mid V_{A}^{*}}\left(v\mid v_{a}^{*}\right)=p_{V^{*}}\left(v+v_{a}^{*}\right).\label{eq:PDF_V_B_conditioned}
\end{equation}
We can assume, without loss of generality, that $v_{a}^{*}$ is positive\footnote{If $v_{a}^{*}$ was negative, all the following derivations would
still be valid by redefining the sign of both $V_{A}^{*}$ and $V_{B}^{*}$.}. 

Consider now the direction of motion of A and the half-line originating
at A \emph{and extending in its motion direction}, and assume that
vehicle B is on this half-line\footnote{The possibility that vehicle B is in the remaining half-line will
be considered later on.}. With the above definition of $V$ and assumption on the location
of B, it holds that $v<0$ if the two vehicles are getting closer
to each other, $v>0$ if their distance is increasing, and $v=0$
if the distance between the two vehicles is constant in time (since
they proceed at the same speed $v^{*}$)\footnote{Conversely, assuming that vehicle B is in the opposite half-line (the
half-line behind A), $v>0$ if the vehicles are getting closer to
each other, and $v<0$ if they are getting farther.\label{fn:relative_speed_B_behind}}. The PDF of the relative speed between a PCP is the starting point
to compute an approximate analytical expression for the PDF of the
transmission range from which the eventually selected content provider
will transmit the content to the requesting device. Before starting
with the derivation of the approximate PDF, we first prove the following
result on the maximum time limit within which a PCP should transmit
the content (in case it was selected).

\begin{lem}\label{lem:PHI}Consider two devices and A and B and assume
that device A requests a content $z$ at $t_{0}$ and that $z$ is
present in device B's contents cache. Assume that the content timeout
duration, $\tau_{c}$ is lower than the sharing timeout, $\tau_{s}$.
Then the effective time limit within which vehicle B should transmit
the content to the requesting device A, is a random variable $\Phi$
with the following PDF
\begin{equation}
p_{\Phi}\left(\phi\right)=\frac{1}{\tau_{s}}u_{[0,\tau_{c})}\left(\phi\right)+\left(1-\frac{\tau_{c}}{\tau_{s}}\right)u_{0}\left(\phi-\tau_{c}\right),\label{eq:time limit PDF}
\end{equation}
 and average value
\begin{align}
\bar{\Phi} & =\tau_{c}-\frac{\tau_{c}^{2}}{2\tau_{s}}.
\end{align}

\end{lem}

\begin{pf}Let $\Phi'$ be a random variable representing the amount
of time left, at $t=t_{0}$, before the expiration of the sharing
timeout for content $z$ in vehicle B's cache. At the expiration
of the sharing timeout, the content will be deleted from vehicle B's
cache. Since the request time $t_{0}$ is independent from the time
vehicle B has (previously) obtained the content, we can claim that
$\Phi'$ is uniformly distributed over the interval $[0,\tau_{s}]$,
i.e., $p_{\Phi'}\left(\phi\right)=\frac{1}{\tau_{s}}u_{[0,\tau_{s}]}\left(\phi\right)$.
At the same time, the content timeout duration (which is a deterministic
quantity) can be seen as a random variable whose PDF just includes
a probability mass concentrated at $\tau_{c}$. To keep the same notation,
indicating this variable with $\Phi''$, we have $p_{\Phi''}\left(\phi\right)=u_{0}\left(\phi-\tau_{c}\right)$.
The effective time limit, within which vehicle B could transmit content
$z$ to vehicle A, is determined by the first expiring timeout, among
the content timeout and the sharing timeout. This time limit is, therefore,
a new random variable defined as $\Phi=\min\left(\Phi',\Phi''\right)$.
It is easy to check that the corresponding PDF is given by \eqref{eq:time limit PDF}.\hfill{}$\blacksquare$

\end{pf}

The effective time limit is the superposition of a rectangular function
of size $\tau_{c}$ weighted by $1/\tau_{s}$ and a probability mass
$\left(1-\tau_{c}/\tau_{s}\right)$ concentrated at $\phi=\tau_{c}$.
The first term represents the event that the sharing timeout expires
before the content timeout. Its probability is given by $\tau_{c}/\tau_{s}$.
If this is the case, device B will need to transmit the content \emph{before}
the expiration of the content timeout. The second term is the probability
that the content timeout expires after the sharing timeout. In this
case, device B can wait until $t=t_{0}+\tau_{c}$ to transmit the
content. Note that the introduced random variable $\Phi$ represents
the effective time \emph{limit }within which device B can transmit
the content, and \emph{not} the instant at which it will eventually
do so.

We now proceed with the derivation of the PDF of the closest transmission
distance that a generic PCP can achieve within its time limit $\Phi$.
We start considering a coordinate system still with earth, and with
origin at the location of vehicle A at the request time, and indicate
the random position of vehicle B at the request time with $X_{0}$.
We consider now \emph{a coordinate system integral with vehicle A's
motion}, with origin coincident with the (time-varying) location of
vehicle A in the former coordinate system, and with the positive semi-axis
of the position variable corresponding to the half-line ahead of the
motion. We indicate with $x_{0}$ the realization of $X_{0}$, and
observe that the position of B at the request time has the same value,
$x_{0}$, in both coordinate systems. Setting, without loss of generality,
$t_{0}=0$, the \emph{relative} trajectory\footnote{Here the term ``relative trajectory'' has the meaning that the trajectory
refers to the coordinate system integral with vehicle A's motion.} of vehicle B with respect to vehicle A is given by
\begin{equation}
x\left(t\right)=x_{0}+vt.\label{eq:relative_trajectory}
\end{equation}
We indicate the best time and relative position of vehicle B, with
respect to vehicle A, to eventuallyhave the PCP transmit the content
the requesting device with $t^{*}$ and $x^{*}$. Given the trajectory
\eqref{eq:relative_trajectory}, the best time and position for transmission
are the those at which the distance between B and A is minimal, within
the time limit $\Phi$. Intuition suggests that three cases are possible:
\begin{enumerate}
\item \label{enu:Vehicle-B-is-1}Vehicle B is moving away from vehicle A,
i.e., it moves in the same direction and with an absolute speed larger
than or equal to vehicle A's speed. In this case the optimal instant
and position to transmit are just $t^{*}=0$ and $x^{*}=x_{0}$, since
$x(t)$ increases with time, and transmitting the content later would
require more and more energy.
\item \label{enu:Vehicle-B-is-2}Vehicle B is either moving in the opposite
direction of vehicle A's direction, or it is moving in the same direction
with a lower speed, \emph{but} the two vehicles will \emph{not} get
to a zero distance\footnote{It is worth recalling that we are considering only one spatial dimension.
A distance equal to zero between two vehicles represents, in practice,
an overtaking between the two vehicles, if they are moving in the
same direction, or their crossing across each other, if they are moving
on two lanes of the street that have opposite direction.}. In this case, the optimal time and position are given by $t^{*}=\phi$
and $x^{*}=x\left(t^{*}\right)=x_{0}+v\phi$, respectively (where
$\phi$ is the realization of the above defined time limit random
variable $\Phi$).
\item \label{enu:Vehicle-B-is-3}Vehicle B is either moving in the opposite
direction of vehicle A's direction, or it is moving in the same direction
with a lower speed, \emph{and} the two vehicles are going to find
themselves at the same location before the time limit. In this case,
the optimal position is obviously $x^{*}=0$ and the optimal time
is $t^{*}=-x_{0}/v$. Note that the minus sign is coherent with the
convention that, for a vehicle in the half-line ahead of vehicle A's
motion, if the two vehicles are getting closer to each other, the
relative speed $v$ is negative, and hence $t^{*}$ is a positive
quantity.
\end{enumerate}
We indicate the closest distance that the PCP can achieve from the
requesting device with $R$. We indicate the PDF of $R$, conditioned
on the initial position $X_{0}$ of the PCP, and on the speed $V_{A}^{*}$
of the requesting device, with $p_{R\mid X_{0},V_{A}^{*}}\left(r\mid x_{0},v_{a}^{*}\right)$,
and characterize the PDF through the following

\begin{thm}\label{thm:1}

Consider a device onboard a vehicle A requesting a content $z$ at
time $t_{0}$ and that $z$ is present in the cache of a device onboard
a vehicle B, which is therefore a PCP for that request. Consider a
unidimensional coordinate system still with earth, with origin at
the position of vehicle A at the request time, and with positive axis
corresponding to the motion direction of vehicle A. Let the two i.i.d.
random variables $V_{A}^{*}$ and $V_{B}^{*}$, with common PDF $p_{V^{*}}\left(v^{*}\right)$,
represent the absolute, signed speed of vehicle A and B, respectively,
and let the relative speed of vehicle B with respect to vehicle A
be defined as $V=V_{B}^{*}-V_{A}^{*}$. Let $f\left(v,v_{a}^{*}\right)\triangleq p_{V\mid V_{A}^{*}}\left(v\mid v_{a}^{*}\right)=p_{V^{*}}\left(v+v_{a}^{*}\right)$.
Let $X_{0}$ denote be a random variable representing the position
of vehicle B at the request time, in the so defined coordinate system.
Let $R$ denote the closest distance that vehicle B can achieve, within
a time limit $\Phi$ distributed as in Lemma~\ref{lem:PHI}.

Assume that $X_{0}$ is in the positive axis, i.e., vehicle B on the
half-line ahead of vehicle A in its motion direction. Then, the PDF
of $R$ can be written as\small{
\begin{align}
 & \hspace{-3mm}p_{R\mid X_{0}^{+},V_{A}^{*}}\left(r\mid x_{0},v_{a}^{*}\right)=\label{eq:D2D_distance_single_content_provider_POS}\\
= & +u_{(0,\infty)}\left(x_{0}\right)u_{0}\left(r-\left|x_{0}\right|\right)\int_{0}^{\infty}f\left(v,v_{a}^{*}\right)dv\nonumber \\
 & +u_{(0,\infty)}\left(x_{0}\right)u_{(0,\left|x_{0}\right|)}\left(r\right)\left(\frac{1}{\tau_{s}}\int_{-\infty}^{\left(r-x_{0}\right)/\tau_{c}}f\left(v,v_{a}^{*}\right)\frac{1}{\left|v\right|}dv+\left(\frac{1}{\tau_{c}}-\frac{1}{\tau_{s}}\right)f\left(\frac{r-x_{0}}{\tau_{c}},v_{a}^{*}\right)\right)\hspace{-7mm}\nonumber \\
 & +u_{(0,\infty)}\left(x_{0}\right)u_{0}\left(r\right)\left(\int_{-\infty}^{-x_{0}/\tau_{c}}f\left(v,v_{a}^{*}\right)dv-\frac{\left|x_{0}\right|}{\tau_{s}}\int_{-\infty}^{-x_{0}/\tau_{c}}f\left(v,v_{a}^{*}\right)\frac{1}{\left|v\right|}dv\right).\nonumber 
\end{align}
}

Assume now that $X_{0}$ is in the negative axis, i.e., vehicle B
on the half-line behind vehicle A in its motion direction. Then, the
PDF of $R$ can be written as \small{
\begin{align}
 & \hspace{-3mm}p_{R\mid X_{0}^{-},V_{A}^{*}}\left(r\mid x_{0},v_{a}^{*}\right)=\label{eq:D2D_distance_single_content_provider_NEG}\\
= & u_{(-\infty,0)}\left(x_{0}\right)u_{0}\left(r-\left|x_{0}\right|\right)\int_{-\infty}^{0}f\left(v,v_{a}^{*}\right)dv\nonumber \\
 & +u_{(-\infty,0)}\left(x_{0}\right)u_{(0,\left|x_{0}\right|)}\left(r\right)\left(\frac{1}{\tau_{s}}\int_{\left(\left|x_{0}\right|-r\right)/\tau_{c}}^{\infty}f\left(v,v_{a}^{*}\right)\frac{1}{\left|v\right|}dv+\left(\frac{1}{\tau_{c}}-\frac{1}{\tau_{s}}\right)f\left(\frac{\left|x_{0}\right|-r}{\tau_{c}},v_{a}^{*}\right)\right)\nonumber \\
 & +u_{(-\infty,0)}\left(x_{0}\right)u_{0}\left(r\right)\left(\int_{-x_{0}/\tau_{c}}^{\infty}f\left(v,v_{a}^{*}\right)dv-\frac{\left|x_{0}\right|}{\tau_{s}}\int_{-x_{0}/\tau_{c}}^{\infty}f\left(v,v_{a}^{*}\right)\frac{1}{\left|v\right|}dv\right)\nonumber 
\end{align}

Finally, if the $X_{0}=0$, $R$ is deterministically equal to 0.

Overall, the PDF of $R$ is given by

\begin{align}
p_{R\mid X_{0},V_{A}^{*}}\left(r\mid x_{0},v_{a}^{*}\right)= & u_{(-\infty,0)}\left(x_{0}\right)p_{R\mid X_{0}^{-},V_{A}^{*}}\left(r\mid x_{0},v_{a}^{*}\right)\label{eq:Theorem1}\\
 & +u_{(0,\infty)}\left(x_{0}\right)p_{R\mid X_{0}^{+},V_{A}^{*}}\left(r\mid x_{0},v_{a}^{*}\right)+u_{0}\left(r\right)u_{0}\left(x_{0}\right)\nonumber 
\end{align}
}\end{thm}

\begin{pf}

See Appendix~\ref{subsec:APP_Theroem_1}\hfill{}$\blacksquare$

\end{pf}

\subsubsection{Closest D2D distance from a random number of potential content providers}

As described in Section~\ref{sec:Content-Delivery-Management}, for
each request, the CDMS computes the trajectories of \emph{a set} of
PCPs, and selects the best one according to the minimum of the shortest
distances from the requesting device that they can achieve within
their respective time limits. Such shortest distances are determined
by the expiration of the content timeout (common to all) or the respective
sharing timeouts (which are specific for each PCP, and distributed
according to \eqref{eq:time limit PDF}).

The set of devices eligible to transmit the content is the result
of the spatial point process of the positions, at the request time,
of the devices that have the requested content $z$ in their caches.
This process, according to our assumption, as recalled in Subsection~\ref{subsec:Preliminary-results},
is a HSPPP completely characterized by its linear density, $\rho_{z}$,
which is given by~\eqref{eq:SPPP-geo-content-distribution-2}.

Let $r_{\max}^{\text{(D2D)}}$ be the maximum D2D transmission range,
defined as a system parameter, and consider a coordinate system still
with earth, with origin at the position of vehicle A at the request
time. It is straightforward to show that
\begin{lyxlist}{00.00.0000}
\item [{(i)}] Conditioned on $V_{A}$, a vehicle B \emph{behind} vehicle
A at the request time (and with the desired content in its cache)
has a chance to come within a distance from vehicle A lower than or
equal to the maximum transmission range $r_{\max}^{\text{(D2D)}}$
if, at the request time, it is located in the interval $[-X_{\text{\ensuremath{\inf}}}\left(v_{a}^{*}\right),0]$,
with $X_{\text{\ensuremath{\inf}}}\left(v_{a}^{*}\right)=r_{\max}^{\text{(D2D)}}+\left(v_{\max}-v_{a}^{*}\right)\tau_{c}.$
\item [{(ii)}] Conditioned on $V_{A}$, a vehicle B \emph{ahead of} vehicle
A at the request time (and with the desired content in its cache)
has a chance to come within a distance from vehicle A lower than or
equal to the maximum transmission range $r_{\max}^{\text{(D2D)}}$
if, at the request time, it is located in the interval $[0,X_{\sup}\left(v_{a}^{*}\right)]$,
with $X_{\sup}\left(v_{a}^{*}\right)=r_{\max}^{\text{(D2D)}}+\left(v_{\max}-v_{a}^{*}\right)\tau_{c}$.
\end{lyxlist}
We recognize that the two spatial boundaries $X_{\text{\ensuremath{\inf}}}\left(v_{a}^{*}\right)$
and $X_{\sup}\left(v_{a}^{*}\right)$ have the same expression. Defining
\[
X_{\text{\ensuremath{\lim}}}\left(v_{a}^{*}\right)\triangleq r_{\max}^{\text{(D2D)}}+\left(v_{\max}-v_{a}^{*}\right)\tau_{c},
\]
we have that the street chunk corresponding to the spatial interval
$[-X_{\text{\ensuremath{\lim}}}\left(v_{a}^{*}\right),X_{\text{\ensuremath{\lim}}}\left(v_{a}^{*}\right)]$
is the region in which any PCP can be located at the request time.

For the properties of HSPPPs, the initial position of the PCP in this
region, $X_{0}$, conditioned on vehicle A's speed, is uniformly distributed
in the interval $[-X_{\text{\ensuremath{\lim}}}\left(v_{a}^{*}\right),X_{\text{\ensuremath{\lim}}}\left(v_{a}^{*}\right)]$,
i.e.,
\begin{equation}
p_{X_{0}}\left(x_{0}\right)=\frac{1}{2X_{\text{\ensuremath{\lim}}}\left(v_{a}^{*}\right)}u_{[-X_{\text{\ensuremath{\lim}}}\left(v_{a}^{*}\right),X_{\text{\ensuremath{\lim}}}\left(v_{a}^{*}\right)]}\left(x_{0}\right).\label{eq:PDF_X0}
\end{equation}

Removing the conditioning on $X_{0}$ from \eqref{eq:D2D_distance_single_content_provider_POS},
we obtain that the closest distance achievable by a PCP for a given
content request, conditioned on the speed of the requesting vehicle,
$V_{A}$, is distributed as
\begin{align}
p_{R\mid V_{A}^{*}}\left(r\mid v_{a}^{*}\right)= & \frac{1}{2X_{\text{\ensuremath{\lim}}}\left(v_{a}^{*}\right)}\int_{-X_{\sup}\left(v_{a}^{*}\right)}^{X_{\text{inf}}\left(v_{a}^{*}\right)}p_{R\mid X_{0},V_{A}^{*}}\left(r\mid x_{0},v_{a}^{*}\right)dx_{0},\label{eq:PDF_D2D_distance_single_content_provider_2}
\end{align}
where $p_{R\mid X_{0},V_{A}^{*}}\left(r\mid x_{0},v_{a}^{*}\right)$
is given by \eqref{eq:Theorem1}. Note that \eqref{eq:PDF_D2D_distance_single_content_provider_2}
does not depend on the specific content $z$. The specific content
$z$, instead, comes into play in the following of our derivation.

Considering a content request, indicating with $Z$ the random variable
representing the requested content, we can state that

\begin{lem}The number of devices with content $z$ in their cache,
positioned within the region $[-X_{\text{\ensuremath{\lim}}}\left(v_{a}^{*}\right),X_{\text{\ensuremath{\lim}}}\left(v_{a}^{*}\right)]$
centered at the position of the requesting device at the request time
(i.e., the number of PCPs for that content request), a Poisson random
variable $N_{\text{PCP}}\left(v_{a}^{*};z\right)$ with mean
\begin{align}
\overline{N}_{\text{PCP}}\left(v_{a}^{*};z\right) & =\rho_{z}2X_{\text{\ensuremath{\lim}}}\left(v_{a}^{*}\right)\label{eq:expected_number_of_PCP_cond_Z}
\end{align}
and PMF
\begin{equation}
P_{N_{\text{PCP}}\mid V_{a}^{*},Z}\left(n\mid v_{a}^{*},z\right)=e^{-\overline{N}_{\text{PCP}}\left(v_{a}^{*};z\right)}\frac{\overline{N}_{\text{PCP}}\left(v_{a}^{*};z\right)^{n}}{n!},\label{eq:Poisson_V*_A}
\end{equation}
where we have explicitly indicated the dependence on the variables
$V_{A}^{*}$ and $Z$.\end{lem}\begin{pf}The result comes straightforward
from well known properties of homogeneous Poisson point processes.\hfill{}$\blacksquare$\end{pf}

It is worth pointing out that since the PMF, evaluated at $n=0$,
is the probability that there are no PCPs in the eligibility region,
it coincides with the probability that the content request will not
be offloaded. Therefore, indicating the probability of offloading
conditioned on a specific content $z$, and on a given speed $v_{a}^{*}$
of the vehicle with onboard the requesting device, with $\mathbb{P}\left(\text{off}\mid z,v_{a}^{*}\right)$,
and the probability of sending the content using an eNB as $\mathbb{P}\left(\text{non-off}\mid z,v_{a}^{*}\right)$,
we can write\begin{subequations}
\begin{align}
\mathbb{P}\left(\text{non-off}\mid z,v_{a}^{*}\right) & =e^{-\overline{N}_{\text{PCP}}\left(v_{a}^{*};z\right)},\label{eq:P_NON-OFF_COND_Z_VA}\\
\mathbb{P}\left(\text{off}\mid z,v_{a}^{*}\right) & =1-e^{-\overline{N}_{\text{PCP}}\left(v_{a}^{*};z\right)}..\label{eq:P_OFFLOADING_COND_Z_VA}
\end{align}
\end{subequations}

We now proceed by computing the best achievable transmission range
resulting from the overall set of PCPs. To each PCP within the set,
we can associate random variables of the kind $X_{0}$ (initial position)
and $\Phi$ (PCP-specific effective time limit for eventually sending
the content), resulting in two sets $X_{0}^{(1)},\ldots,X_{0}^{\left(N_{\text{PCP}}\left(v_{a}^{*};z\right)\right)}$
and $\Phi{}_{1},\ldots,\Phi{}_{N_{\text{PCP}}\left(v_{a}^{*};z\right)}$.
The random variables in both sets are i.i.d. with common distribution
\eqref{eq:PDF_X0} and \eqref{eq:time limit PDF}, respectively. Each
pair $\left[X_{0}^{(n)},\Phi{}_{n}\right],\,n\in\left\{ 1,\ldots,N_{\text{PCP}}\left(v_{a}^{*};z\right)\right\} ,$
refers to a different PCP, and determines a new random variable $R_{n}$,
corresponding to the closest achievable distance of the $n$-th PCP,
which, conditionally on the requesting vehicle speed, is distributed
with PDF \eqref{eq:PDF_D2D_distance_single_content_provider_2}.
By construction, the random variables in the new set $\left\{ R_{1},\ldots,R_{N_{\text{PCP}}\left(v_{a}^{*};z\right)}\right\} $
are conditionally independent and identical distributed.

According to the proposed CDMS operation, the device that would eventually
be selected to transmit the content to vehicle A is the one with the
smallest prospective minimum distance in the set $\left\{ R_{1},\ldots,R_{N_{\text{PCP}}\left(v_{a}^{*};z\right)}\right\} $.
We indicate this overall minimum distance as
\[
R_{\min}=\min\left(R_{1},\ldots,R_{N}\right).
\]
Since the random variables $R_{1},\ldots,R_{N}$ are conditionally
i.i.d., using the well known property that the CDF of the minimum
among a set of i.i.d. random variables with common CDF $F\left(r\right)$
is given by $F_{\min}\left(r\right)=1-\left(1-F\left(r\right)\right)^{N},$
and introducing the conditional CDF of $R$ as
\[
F_{R\mid V_{A}^{*}}\left(r\mid v_{a}^{*}\right)=\smallint_{0}^{r}p_{R\mid V_{A}^{*}}\left(r'\mid v_{a}^{*}\right)dr',
\]
we obtain the conditional CDF of $R_{\min}$ (with conditioning random
variables $V_{A}^{*}$ and $N_{\text{PCP}}\left(v_{a}^{*}/z\right)$
as
\[
F_{R_{\min}\mid V_{A}^{*},\overline{N}_{\text{PCP}}\left(v_{a}^{*};z\right)}\left(r\mid v_{a}^{*},n\right)=1-\left(1-F_{R\mid V_{A}^{*}}\left(r\mid v_{a}^{*}\right)\right)^{n},
\]
and the corresponding PDF as
\begin{align}
p_{R_{\min}\mid V_{A}^{*},N_{\text{PCP}}\left(v_{a}^{*};z\right)}\left(r\mid v_{a}^{*},n\right) & =\frac{d}{dr}F_{R_{\min}\mid V_{A}^{*},N}\left(r\mid v_{a}^{*},n\right)\label{eq:D2D_distance_N_content_providers}\\
 & =-n\left(1-F_{R\mid V_{A}^{*}}\left(r\mid v_{a}^{*}\right)\right)^{n-1}\frac{d}{dr}\left(1-F_{R\mid V_{A}^{*}}\left(r\mid v_{a}^{*}\right)\right)\nonumber \\
 & =n\left(1-F_{R\mid V_{A}^{*}}\left(r\mid v_{a}^{*}\right)\right)^{n-1}p_{R\mid V_{A}^{*}}\left(r\mid v_{a}^{*}\right).\nonumber 
\end{align}

We now observe that, since the content will be actually delivered
through a D2D transmission \emph{only if} the closest distance will
be lower than or equal the maximum nominal D2D transmission rage $r_{\max}^{\text{(D2D)}}$,
the \emph{effective} transmission distance, conditioned on $V_{A}^{*}$,
results from conditioning $R_{\min}$ to being lower than or equal
to $r_{\max}^{\text{(D2D)}}$. We indicate this \emph{effective} transmission
distance as $R_{\text{eff}}$. Its PDF is related to the PDF of the
closest distance achieved by the set of PCPs (whose number, here indicated
with $N$, is determined by \eqref{eq:D2D_distance_N_content_providers})
through
\begin{equation}
p_{R_{\text{eff}}^{*}\mid V_{A}^{*},N,Z}\left(r\mid v_{a}^{*},n,z\right)=\frac{p_{R^{*}\mid V_{A}^{*},N,Z}\left(r\mid v_{a}^{*},n,z\right)u_{[0,r_{\max}^{\text{(D2D)}}]}\left(r\right)}{F_{R^{*}\mid V_{A}^{*},N}\left(r_{\max}^{\text{(D2D)}}\mid v_{a}^{*},n\right)},\label{eq:D2D_distance_N_content_providers_2}
\end{equation}
where $p_{R^{*}\mid V_{A}^{*},N,Z}\left(r\mid v_{a}^{*},n,z\right)$
is given by \eqref{subsec:D2D_distance_single_content_provider},
but we have made it explicit its dependence on the specific requested
content $z$, which comes into play through \eqref{eq:expected_number_of_PCP_cond_Z}.

Combining \eqref{eq:Poisson_V*_A} and \eqref{eq:D2D_distance_N_content_providers_2},
we obtain the following PDF of the effective D2D transmission distance
for the considered content $z$, conditioned, now, only on $V_{A}^{*}$
and the content itself
\begin{equation}
p_{R_{\text{eff}}^{*}\mid V_{A}^{*},Z}\left(r\mid v_{a}^{*},z\right)=\frac{\sum_{n=1}^{\infty}P_{N_{\text{PCP}}\mid V_{a}^{*},Z}\left(n;z\right)p_{R_{\text{eff}}^{*}\mid V_{A}^{*},N_{\inf}}\left(r\mid v_{a}^{*},n\right)}{1-e^{-\bar{N}_{\text{PCP}}\left(v_{a}^{*};z\right)}}.\label{eq:PDF_R_Eff_COND_VA_Z}
\end{equation}
The final step to obtain the PDF of the effective D2D transmission
distance $R_{\text{eff}}^{*}$ is to average out the dependency on
$V_{A}^{*}$ and $Z$. In doing this, we must keep in mind that all
the derivations in this section have built on the convention of taking
vehicle A's motion direction as a reference for defining the positive
and negative axis of the coordinate system. Therefore, in the considered
system, the speed of vehicle A is, by construction, always positive.
In other words, the marginal PDF which needs to be used to compute
the unconditional PDF of $R_{\text{eff}}$ for a given content $z$
is $p_{V_{A}^{*}}^{+}(v)=p_{V^{*}}(v)+p_{V^{*}}(-v)$, which, under
the symmetry assumption on $p_{V^{*}}(v)$, becomes $p_{V_{A}^{*}}^{+}(v)=2p_{V^{*}}(v)$.
Further removing the conditioning on the requested content $z$, we
obtain the final, unconditional PDF of the optimal D2D transmission
range. This result is stated in the following theorem

In conclusion, the PDF of the effective D2D transmission range for
a request of content $z$ is provided by the following

\begin{thm}Consider a content request issued by a device onboard
a vehicle moving at constant (unsigned) speed which is a realization
of a random variable $V_{A}^{*}$ with PDF $p_{V_{a}^{*}}^{+}(v)$,
and assume that the specific requested content, $z$, is the realization
of a discrete random variable $Z$ representing the content popularity,
with realizations in a content library $\mathcal{Z}$ and PMF $P_{Z}\left(z\right)$.
Let the assumptions on the vehicle arrival process and content request
process made in Subsection~\ref{subsec:Preliminary-results} hold.
Let $\rho_{z}$ be the linear density of devices with the desired
content $z$ in their caches resulting from \ref{eq:SPPP-geo-content-distribution-2}.
Then:
\begin{lyxlist}{00.00.0000}
\item [{(i)}] The PDF of the distance $R_{\text{eff}}$ from which the
PCP that would eventually send the content to the requesting device,
conditioned on the specific content $z$, is given by 
\begin{align}
p_{R_{\text{eff}}\mid Z}\left(r\mid z\right) & =\smallint_{0}^{\infty}p_{R_{\text{eff}}^{*}\mid V_{A}^{*},Z}\left(r\mid v_{a}^{*},z\right)p_{V_{a}^{*}}^{+}(v)dv\label{eq:PDF_R_EFF_COND_Z}\\
 & =\smallint_{0}^{\infty}p_{V_{a}^{*}}^{+}(v)\sum_{n=1}^{\infty}P_{N_{\text{PCP}}\mid V_{a}^{*}}\left(n;z\right)p_{R_{\text{eff}}^{*}\mid V_{A}^{*},N_{\inf},Z}\left(r\mid v,n,z\right)dv\nonumber \\
 & =\smallint_{0}^{\infty}p_{V_{A}^{*}}^{+}(v)\sum_{n=1}^{\infty}e^{-\overline{N}_{\text{PCP}}\left(v_{a}^{*};z\right)}\frac{\overline{N}_{\text{PCP}}\left(v_{a}^{*};z\right)^{n}}{n!}p_{R_{\text{eff}}^{*}\mid V_{A}^{*},N_{\inf},Z}\left(r\mid v,n,z\right)dv\nonumber \\
\text{with: } & \text{\ensuremath{\overline{N}_{\text{PCP}}\left(v_{a}^{*};z\right)}}=\rho_{z}\left(2r_{\max}^{\text{(D2D)}}+\left(v_{\max}+v_{\min}-2v_{a}^{*}\right)\tau_{c}\right),\nonumber 
\end{align}
where $p_{R_{\text{eff}}^{*}\mid V_{A}^{*},N_{\text{PCP}},Z}\left(r\mid v,n,z\right)$
is given in \eqref{eq:D2D_distance_N_content_providers_2}.
\end{lyxlist}
and
\begin{lyxlist}{00.00.0000}
\item [{(ii)}] The unconditional PDF of the minimum transmission range
$R_{\text{eff}}$ is 
\end{lyxlist}
\begin{align}
p_{R_{\text{eff}}}\left(r\right)= & \sum_{z\in\mathcal{Z}}\left(p_{Z}\left(z\mid\mathrm{NR}\right)\int_{0}^{\infty}p_{V_{A}^{*}}^{+}\left(v_{a}^{*}\right)\vphantom{\frac{\sum_{n=1}^{\infty}P_{N_{\text{PCP}}\mid V_{a}^{*},Z}\left(n;z\right)p_{R_{\text{eff}}^{*}\mid V_{A}^{*},N_{\inf}}\left(r\mid v_{a}^{*},n\right)}{1-e^{-\overline{N}_{\text{PCP}}\left(v_{a}^{*};z\right)}}}\right.\label{eq:PDF_R_FINAL}\\
 & \cdot\left.\frac{\sum_{n=1}^{\infty}P_{N_{\text{PCP}}\mid V_{a}^{*},Z}\left(n;z\right)p_{R_{\text{eff}}^{*}\mid V_{A}^{*},N_{\inf}}\left(r\mid v_{a}^{*},n\right)}{1-e^{-\overline{N}_{\text{PCP}}\left(v_{a}^{*};z\right)}}dv_{a}^{*}\right)\nonumber 
\end{align}
where $\mathcal{Z}$ is the content library, and $p_{Z}\left(z\mid\mathrm{NR}\right)$
is the content probability, conditioned to the fact the the content
is not already in the cache of the requesting device.\end{thm}\begin{pf}The
two expressions of $p_{R_{\text{eff}}\mid Z}\left(r\mid z\right)$
and $p_{R_{\text{eff}}}\left(r\right)$ are simply obtained by averaging
out the conditioning random variables $V_{A}^{*}$ and $Z$ from the
conditional PDF~\eqref{eq:PDF_R_Eff_COND_VA_Z}. The denominator
on the right-hand side of \eqref{eq:PDF_R_FINAL} is the probability
of offloading the content, see~\eqref{eq:P_OFFLOADING_COND_Z_VA}.\hfill{}$\blacksquare$\end{pf}

\subsection{Analytical model for the energy consumption\label{subsec:Analytical-model-energy}}

To determine the energy consumption (due to the radio transmissions)
induced on both the network infrastructure nodes and the devices by
our CDMS, it is necessary to specify how the transmit power is set.
In this work, we assume that both cellular communications (I2D) and
D2D ones rely on a power control mechanism, which relates the transmit
power to the distance between transmitter and receiver. More specifically,
the system relies on a nominal channel gain function of transmission
range, $g\left(r\right)$. Based on this function (and on standard
physical layer parameters related to modulation and coding) it is
able to determine the transmit power required to achieve a desired
radio link reliability\footnote{For D2D communications, since the CDMS is aware of the position of
the nodes at the transmission time, it can communicate the power to
use to the PCP responsible for the content delivery. }. More details on these aspects are provided in Section~\ref{sec:MAC-and-physical}
and Appendix~\ref{Appendix:Transmit-power-setting-and-error-model}.

We indicate with $g_{\text{I2D}}\left(r\right)$ and $g_{\text{D2D}}\left(r\right)$
two nominal channel gain functions, related to I2D and D2D transmissions,
respectively\footnote{We distinguish between two different functions because the path loss
behavior, as a function of distance, is different, see e.g. \citep{METIS_chanmod}}, and with $\mathcal{E}\left(g\right)$ the function that relates
the energy to the nominal channel gain\footnote{The dependemce of $\mathcal{E}\left(g\right)$ on the transmit power
and the content size has been omitted to simplify the notation.}. Furthermore, we indicate with $p_{R}^{\text{(I2D)}}(r)$ the PDF
of the transmission range for the \emph{cellular} tranmissions, and
with $r_{\max}^{\text{(I2D)}}$ the coverage of eNBs (i.e, the cell
radius). Under the assumption that the nodes positions in time are
a HSPPP, using basic HSPPP properties, it is straightforward to show
that $p_{R}^{\text{(I2D)}}(r)=\frac{1}{r_{\max}^{\text{(I2D)}}}u_{[0,r_{\max}^{\text{(I2D)}}]}\left(r\right)$,
which does not depend on either $z$ or $v_{a}^{*}$. Thus, the average
energy consumption associated to a content transmission performed
by an eNB is
\begin{equation}
\overline{E}_{\text{I2D}}=\frac{1}{r_{\max}^{\text{(I2D)}}}\int_{0}^{r_{\max}^{\text{(I2D)}}}\mathcal{E}\left(g_{\text{I2D}}\left(r\right)\right)dr.
\end{equation}

Furthermore, the probabity that a content delivery is not offloaded
is given by (see~\eqref{eq:P_NON-OFF_COND_Z_VA})
\begin{align}
\mathbb{P}\left(\text{non-off}\right) & =\sum_{z\in\mathcal{Z}}\left(P_{Z}\left(z\mid\text{NR}\right)\int_{0}^{\infty}\text{\ensuremath{\mathbb{P}\left(\text{non-off}\mid z,v_{a}^{*}\right)}}p_{V_{A}^{*}}^{+}\left(v_{a}^{*}\right)dv_{a}^{*}\right)\\
 & =\sum_{z\in\mathcal{Z}}\left(P_{Z}\left(z\mid\text{NR}\right)\int_{0}^{\infty}e^{-\overline{N}_{\text{PCP}}\left(v_{a}^{*};z\right)}p_{V_{A}^{*}}^{+}\left(v_{a}^{*}\right)dv_{a}^{*}\right),\nonumber 
\end{align}

For D2D trasnmissions, it is straightforward to show that
\begin{align}
\overline{E}_{D2D} & =\sum_{z\in\mathcal{Z}}\left(P_{Z}\left(z\mid\text{NR}\right)\int_{0}^{r_{\max}^{\text{(D2D)}}}\mathcal{E}\left(g_{\text{D2D}}\left(r\right)\right)\vphantom{\frac{\mathbb{P}\left(\text{off}\mid z,v_{a}^{*}\right)}{1-\mathbb{P}\left(\text{non-off}\right)}}\right.\\
 & \left.\int_{0}^{\infty}\frac{\mathbb{P}\left(\text{off}\mid z,v_{a}^{*}\right)}{1-\mathbb{P}\left(\text{non-off}\right)}p_{R_{\text{eff}}^{*}\mid V_{A}^{*},Z}\left(r\mid v_{a}^{*},z\right)p_{V_{A}^{*}}^{+}\left(v_{a}^{*}\right)dv_{a}^{*}dr\right)\nonumber 
\end{align}
where $\text{\ensuremath{p_{R_{\text{eff}}^{*}\mid Z}\left(r\right)}}$
is given by \eqref{eq:PDF_R_EFF_COND_Z}.

Finally, the overall average energy consumption for delivering a content
is
\begin{align}
\overline{E}= & \mathbb{P}\left(\text{non-off}\right)\overline{E}_{\text{I2D}}+\left(1-\mathbb{P}\left(\text{non-off}\right)\right)\overline{E}_{\text{D2D}}\\
= & \mathbb{P}\left(\text{non-off}\right)\overline{E}_{\text{I2D}}\nonumber \\
 & +\int_{0}^{r_{\max}^{\text{(D2D)}}}\mathcal{E}\left(g_{\text{D2D}}\left(r\right)\right)\text{\ensuremath{\int_{0}^{\infty}}\ensuremath{\mathbb{P}\left(\text{off}\mid z,v_{a}^{*}\right)}}p_{V_{A}^{*}}^{+}\left(v_{a}^{*}\right)p_{R_{\text{eff}}^{*}\mid V_{A}^{*},Z}\left(r\mid v_{a}^{*},z\right)dv_{a}^{*}dr\nonumber 
\end{align}
where $p_{R_{\text{eff}}^{*}\mid V_{A}^{*},Z}\left(r\mid v_{a}^{*},z\right)$
is given by~\eqref{eq:PDF_R_Eff_COND_VA_Z}, $\mathbb{P}\left(\text{off}\mid z,v_{a}^{*}\right)$
by \eqref{eq:P_OFFLOADING_COND_Z_VA}, and $r_{\max}^{\text{(D2D)}}$
is the maximum transmission range of the devices.

\section{MAC and physical layer implementation\label{sec:MAC-and-physical}\vspace{-2mm}
}

In evaluating the performance of the proposed CDMS, we considered
it important to use a sufficiently detailed and realistic implementation
of medium access control and radio resource management layers, which
takes into account the physical layer aspects that have an considerable
impact on the energy consumption and interference among concurrent
transmission. As shown in our recent work \citep{Pescosolido2018AdHocNetworks},
failing to do so may result in a high degree of inaccuracy of the
results. The physical layer aspects taken into account are the multipath
frequency selective fading of the radio channels, spatially correlated
lognormal shadowing, and interference across simultaneous transmissions.
For the RRRM component, we use the same solution presented in \citep{Pescosolido2018AdHocNetworks},
which is also compatible for being used with the CDMS proposed in
this work. In the following, we summarize the main features of the
RRRM component, whereas the description of the physical layer and
channel models we used, the transmit power settings, and how modeled
transmission errors are left to Appendix~\ref{Appendix:Transmit-power-setting-and-error-model}.

We have considered a multi-carrier system in which the radio resources
are organized in a time-frequency grid of Physical Resource Blocks
(PRBs) of fixed bandwidth $w$ and duration $\tau$. Concurrent D2D
and I2D transmissions are allowed to spatially reuse the PRBs in a
very flexible way\footnote{Our RRR implementation follows the approach of the \emph{resource-sharing
oriented} scheme proposed in \citep{Yang2017}. We have modified the
algorithms in \citep{Yang2017} to use different transmit power levels
across concurrent links, and including multiple eNBs and spatial frequency
reuse for I2D communications (besides D2D ones) in the design, which
allows to run the RRR scheme across multiple cells. Additionally,
it is worth mentioning that the solution proposed in \citep{Yang2017}
is evaluated under a flat fading channel assumption, the implementation
of both the RRR scheme includes frequency selective channels. Further
details on the considered channel model are provided in .}. Specifically, we have followed the approach of the \emph{resource-sharing
oriented} scheme proposed in \citep{Yang2017}, modifying the algorithms
in \citep{Yang2017} to use different transmit power levels across
concurrent links, and including multiple eNBs and spatial frequency
reuse for I2D communications (besides D2D ones) in the design, which
allows to run the RRR scheme across multiple cells\footnote{It is worth mentioning that the solution proposed in \citep{Yang2017}
is evaluated under a flat fading channel assumption, whereas our implementation
includes frequency selective channels.}.

Time is organized in control intervals. In each control interval,
a set of ID2 and D2D links have to be scheduled for transmission.
The set of I2D and D2D links to schedule in each control interval
is determined by the CDMS according to the procedure described in
Section~\ref{sec:Content-Delivery-Management}. Radio Resource allocation
is performed by a distributed RRR agent residing at the eNBs. We assume
that the position of each device is known to the RRR agent, and hence,
it can compute the distance between any node pair.

The RRR agent, taking in input the distance $r$ between the transmitter
and receiver of each link to be scheduled, computes the transmit power
of each link. The transmit power is computed to guarantee that the
channel capacity (which is a random quantity determined by fading
and interference) supports the transfer of the desired amount of information
with an outage probability $P_{e}\ll1$. More details on the transmit
power setting are provided in Subsection~\ref{Appendix:Transmit-power-setting-and-error-model}.

The set of links is partitioned\footnote{The RRR set partitioning algorithm is similar to \citep[Algorithm 1]{Yang2017}.}
into RRR sets in order to satisfy a set of cross-interference mitigation
constraints. The constraints are computed using an estimation of the
interference across links obtained by computing the nominal channel
gain $g$ between any link transmitter and any link receiver among
the set of links to be scheduled. A suitable amount of PRBs is assigned
to each RRR set. This amount is a function of the number and size
of the contents that have to be transmitted by each link in the RRR
set. D2D links in the same RRR set can use the same radio resources,
since their belonging to the same set stands for the fact that their
cross-interference is sufficiently low not to compromise the communications.
I2D links originating from the same eNB are assigned radio resources
in an exclusive way, selected as a portion of the pool of PRBs assigned
to the RRR set they have been included in. In its portion of PRBs,
however, each I2D link is subject to the interference coming from
the D2D links included in the same RRR set. Finally, I2D links originating
from different eNBs, that are included in the same RRR set, can be
assigned the same portion of PRBs within the pool of PRBs assigned
to that RRR set. If the RRR set partitioning and consequent PRBs allocation
to each RRR set, due to the cross-interference constraints and to
the limited number of PRBs in a control interval, prevent to accomodate
the transmission of all the data required by any of the links, the
data to be transmitted are pruned until reaching a feasible amount.
The pruned transmissions will be rescheduled in the next control interval.
Pruning is performed giving a higher scheduling priority to content
deliveries related to requests whose content timeout is closer to
expire. Therefore, I2D communications have a higher priority then
D2D ones, since they are by design related to content requests whose
timeout has already expired. If, due to pruning, the content timeout
of any content request expires, the corresponding delivery is redirected
to be performed by an eNB.

\section{Performance evaluation\label{sec:Performance-evaluation}}

We evaluate the perfomance of the proposed CDMS using both the analytical
model and simulation results. We describe the considered scenario
in Subsection~\ref{subsec:Scenario-description}, and validate the
theoretical model and draw some conclusions based on it in Subsection~\ref{subsec:Analytical-model-validation}.
Extensive simulations results and further comments are provided in
Subsection~\ref{subsec:Simulation-results}.

\subsection{Scenario description\label{subsec:Scenario-description}}

We considered a two-lane street chunk of length 3 Km and width 20
m. The two lanes correspond to opposite marching directions. Six eNBs
are placed at the horizontal coordinates of 0, 600, 1200, 1800, 2400,
and 3000 m, respectively, at the center of the street (see Fig.~1).
The eNB antenna height is 10 m. These numbers are in line with the
``Urban Micro'' scenario \citep{METIS_chanmod}.

The distance between the median axis of the two lanes is 10 m. This
is also the closest distance a vehicle can get to any vehicle marching
in the opposite direction. We modeled the vehicles arrival as a HTPPP.
In all the simulations whose results are, the vehicle arrival rate
was kept fixed at $\lambda_{t}=1/3$ vehicles per second. Similarly,
we used a HTPPP for modeling the request arrival process of each node,
and kept the content request rate per device fixed at $\lambda_{Z}=1/6$
requests per second (10 requests per minute). The content requests
processes of different devices were set to be statistically independent.
The selected content popularity distribution was a Zipf distribution
with parameter $\alpha=1.1$, i.e., $p_{Z}(z)\sim\frac{1}{\zeta(\alpha)}z^{-\alpha},$
truncated to a library size of $10^{4}$ contents. The sharing timeout
was also fixed and equal to $\tau_{s}=600$ seconds. The content size
was fixed and equal to a payload of 432~kB, which we assumed to be
encoded in a packet of 540~kB using a FEC coding rate $\beta=0.8$.
The MAC parameters we used (see Subsection~\ref{sec:MAC-and-physical})
are as follows: each control interval lasts one second, and is divided
in time slots of duration $0.5\,\text{ms}$. Each PRB lasts for 1
time slot and has width $180\,\text{KHz}$. In each PRB bandwidth,
there are 12 subcarriers, the overall system bandwidth is 10.8~MHz,
and in each control interval, a maximum of 120000 PRBs could be allocated
to concurrent I2D and D2D transmissions (possibly spatially reusing
the same PRBs across non-interfering links (see Subsection~\ref{sec:MAC-and-physical}).

\subsubsection{Simulation settings, performance metrics and benchmarks}

To evaluate the performance of the proposed system, validate the analytical
model, we used a custom simulator written in Matlab\footnote{The reason to use a custom simulator, as opposed to classic network
simulators like ns-3 or OMNET++, is to obtain a fine grain control
on implementation of the physical layer aspects, while retaining an
acceptable level of scalability, and using a state of the art channel
model able to reproduce the effects of frequency selective fading.}. The same simulator has been used for our previous works \citep{Pescosolido2018WoWMoM,Pescosolido2018AdHocNetworks,WWIC2018}.
The simulator implements both the CDMS layer and the RRRM layer described
in Sections~\ref{sec:Content-Delivery-Management} and~\ref{sec:MAC-and-physical},
and all the considered aspects of channel, interference, and transmission
error models (see Section~\ref{sec:MAC-and-physical} and Appendix~\ref{Appendix:Transmit-power-setting-and-error-model}).
More details can be found in \citep{Pescosolido2018AdHocNetworks}.

The simulation results are organized in four different sets, each
one obtained by letting a system parameter vary while keeping the
rest of the parameters fixed. We focus on three parameters: the speed
range $[v_{\min},v_{\max}]$ in which each vehicle's speed falls,
the content timeout $\tau_{c},$ and the maximum D2D transmission
distance $r_{\max}^{\text{(D2D)}}$ (we performed two sets of simulations
with varying speed range, using two different fixed values of $\tau_{c}$
and the same value for $r_{\max}^{\text{(D2D)}}$). For each value
of the varying system parameter, we run 10 independent i.i.d simulations,
each lasting 1 hour, reinitializing the random number generator seed
with the same state at the beginning of each batch of 10 simulations.
Each simulation is initialized with a random number of vehicles, positions,
speeds, and cache content of each node according to the results of
our previous work \citep{Pescosolido2018WoWMoM}, in which we computed
the steady state average number of vehicles and cache content distribution.
In each simulation, we used a different independent realization of
the whole set of random components of the channels between any two
points in the grid, and between any eNB and any point in the grid.

We evaluate the performance of the proposed systems using the following
benchmarks:
\begin{lyxlist}{00.00.0000}
\item [{A)}] Plain cellular system with 6 eNBs, numbered eNB1, eNB2,...,
eNB6, following the order of their location. The frequency reuse pattern
of length 3. The set of PRBs in each control interval is partitioned
in three subsets of equal size, and each subset in the partition is
assigned to the eNBs in the subsets \{eNB1,eNB4\}, \{eNB2,eNB5\},
\{eNB2,eNB6\}. Essentially, in each control interval, a PRB can be
used exclusively by one base station within the exclusive spectrum
use regions \{eNB1,eNB2, eNB3\}.
\item [{B)}] CDMS presented in \citep{Pescosolido2018WoWMoM,Pescosolido2018AdHocNetworks},
in which the D2D transmission can occur under the following circumstances:
\begin{itemize}
\item Immediately after the request, if there is at least one PCP within
a distance $r_{\max}^{\text{(D2D)}}$ to the requesting device. In
this case, the closest PCP is selected, and the transmission distance
is the same distance the two devices are from each other at the request
time.
\item During the content timeout, if no PCP is within a range $r_{\max}^{\text{(D2D)}}$
to the requesting device at the request time. In this case, the first
PCP which comes at a distance $r_{\max}^{\text{(D2D)}}$ to the requesting
device is selected for delivering the content, and it does so at the
time its being in-range is detected, therefore transmitting at the
maximum distance.
\end{itemize}
\end{lyxlist}
The performance metrics considered in this work are
\begin{itemize}
\item Offloading efficiency
\item Average energy consumption per content delivery, considering both
I2D and D2D transmissions
\item Average energy consumption per content delivered considering only
D2D transmissions
\item Average spectrum occupation percentage (computed an area equal to
the exclusive spectrum use regions): a PRB is counted as being used
if it used by at least one transmission within an exclusive spectrum
use region of the cellular system. Clearly, for the benchmark cellular
system, the average spectrum occupation percentage coincides with
the ratio between the offered traffic and the traffic that the network
is able to support without being saturated. For D2D offloading schemes,
in which PRBs are spatially reused, the average spectrum occupation
percentage is expected to be less.
\end{itemize}

\subsection{Analytical model validation and performance trends\label{subsec:Analytical-model-validation}}

\begin{figure}[t]
\centering{}\includegraphics[width=0.79\columnwidth]{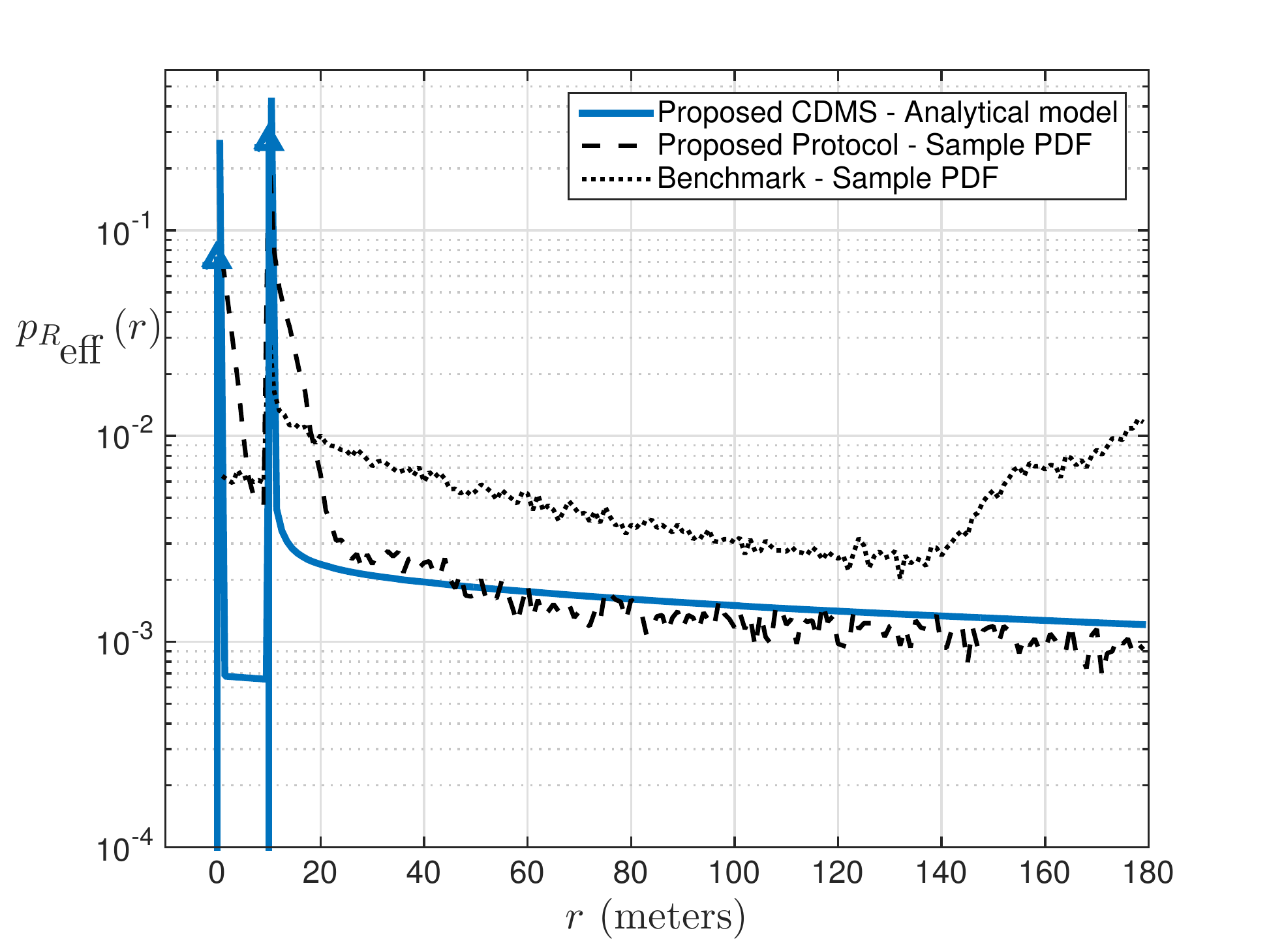}\caption{PDF of the D2D transmission distance $R_{\text{eff}}$}
\label{fig:model_validation}
\end{figure}

We validate our analytical model by comparing the statistics of the
D2D transmission distance computed with it, with the sample PDF obtained
in the simulations. In doing this, we also comment on the major difference,
in terms of D2D transmission distance, between the proposed CDMS and
the benchmark CDMS. Figure~\ref{fig:model_validation} shows the
PDF of the D2D transmission range computed using the analytical model
(solid line), and the sample PDFs obtained with the simulations running
the proposed CDMS (dashed line) and the benchmark CDMS (dotted line).
The system parameters are $\tau_{c}=20\text{ s}$, $r_{\max}^{\text{(D2D)}}=180\,\text{m}$,
and $[v_{\min},v_{\max}]=[6,16]\,\text{m/s}$. In plotting the theoretical
PDF we reintroduced the presence of the spatial dimension transversal
to the street median axis. Defining $r_{y}$ as the distance between
the median axes of the two street lanes, we have that the effective
distance, taking into account both spatial dimensions, is given by
$\widetilde{R}_{\text{eff}}=\sqrt{R_{\text{eff}}^{2}-r_{y}^{2}}$
if the selected PCP is in the opposite street lane with respect to
the requesting device, and $\widetilde{R}_{\text{eff}}=R_{\text{eff}}$
otherwise\footnote{Using standard tools it can be shown that $p_{\widetilde{R}_{\text{eff}}}\left(r\right)=P_{0}u_{0}\left(r\right)+P_{1}u_{0}\left(r-r_{y}\right)+P_{1}p_{\widetilde{R}_{\text{eff}}}\left(\sqrt{r^{2}-r_{y}^{2}}\right)r/\sqrt{r^{2}-r_{y}^{2}},$where
$P_{0}$ and $P_{1}$ are constants corresponding to the probabilities
that $R_{\text{eff}}=0$ conditioned on the fact that the selected
PCP moves in the opposite direction as the requesting device ($P_{0}$)
or in the same direction ($P_{1}$). $P_{0}$ and $P_{1}$ can be
computed using the same techniques used in Section~\ref{sec:Analytical-model}.}. The theoretical model presents two Dirac pulses at $r=0$ and $r=10$,
respectively, which account for the fraction of D2D transmissions
that are performed by the PCP from the sweet spot $R_{\text{eff}}=0$,
i.e., either $\widetilde{R}_{\text{eff}}=0$ or $\widetilde{R}_{\text{eff}}=r_{y}$.

It can be seen that the sample PDF closely follows the tail of the
theoretical PDF, and the trends at small values of the transmission
distance are similar as well. The mismatch in the area of the theoretical
probability masses (which are absent from the sample PDF), is explained
by the spatio-temporal sampling effect represented by the RRRM implementation.
In practice, with the actual implementation of the RRRM component,
the CDMS is able to determine the transmission instant ony with a
precision equal to the control interval duration, which, being in
the order of one second, entails a dispersion of the theoretical proability
mass around an interval of few meters (depending on the speed). We
can conclude that the proposed model is sufficiently accurate, since
it reproduces the tail behaivor, and allows to quantify the percentage
of transmissions that is performed at a very short range, e.g, less
than 20~m, which is given by the overall probability mass at $R_{\text{eff}}=0$
(i.e., either $\widetilde{R}_{\text{eff}}=0$ or $\widetilde{R}_{\text{eff}}=r_{y}$).
 Finally, the figure also shows how much effective is the proposed
CDMS in concentrating the probability mass towards short distances,
with respect to the benchmark CDMS. 

\begin{figure}[t]
\centering{}\includegraphics[width=1\columnwidth]{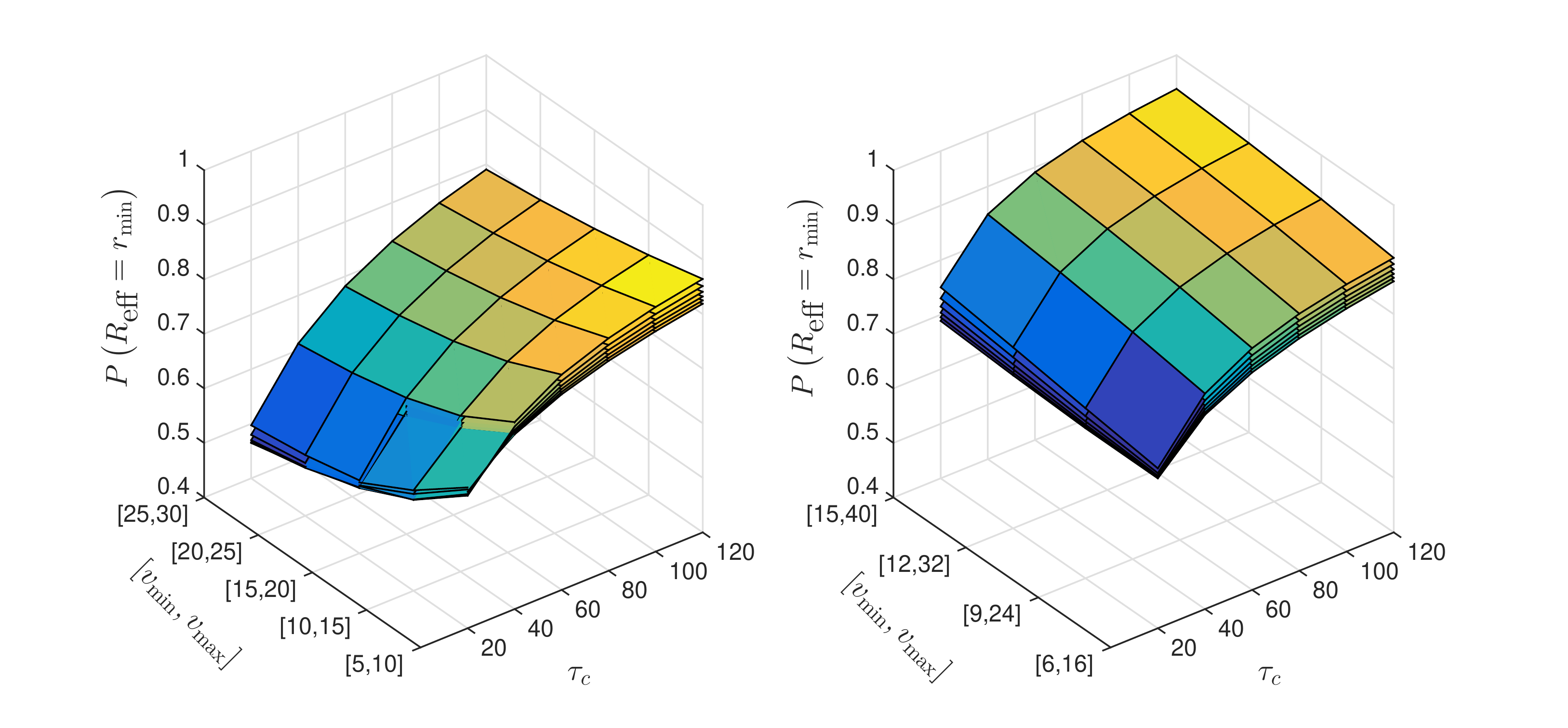}\caption{Probability of D2D transmission at the minimum range}
\label{fig:prob_minimum_D2D_range}
\end{figure}

The surface plots in Figure~\ref{fig:prob_minimum_D2D_range} show
the value of the probability mass at $R_{\text{eff}}=0$, i.e., the
probability that the D2D transmission will be performed at very short
distance (\emph{virtually} equal to zero in case of PCP moving in
the same direction of the requesting vehicle, and $r_{y}$ otherwise),
for different values of the system parameters. We used the PDF \eqref{eq:PDF_speed}.
The horizontal axes correspond to speed range $[v_{\min},v_{\max}]$
and content timeout $\tau_{c}$. Different surfaces correspond to
different values of $r_{\max}^{\text{(D2D)}}$, with surfaces at lower
heights corresponding to higher values of $r_{\max}^{\text{(D2D)}}$,
ranging from 80 to 140~m in steps of 20~m. The difference between
the left and right plots is in the variation of the speed range. In
the left had side surface plot, $v_{\min}$ and $v_{\max}$ are increased
while keeping their difference constant, and the speeds are narrowed
in a 5 m/s interval. In the right hand side, when increasing the speed
range, also the difference between $v_{\max}$ and $v_{\min}$ increases.
From the plots we can observe that, for all configurations of speed
range and maximum D2D transmission range, increasing the content timeout
has a significant impact in terms of probability of transmission near
the closest feasible achievable distance. Increasing the maximum transmission
range $r_{\max}^{\text{(D2D)}}$ (different surfaces layered on top
of each other) results in a moderate decrease of the probability of
short range transmission (the height of the surfaces decreases). Finally,
an interesting aspect is that, increasing $v_{\max}$ and $v_{\min}$
at the same rate (left hand side surfaces) results in a descrease
of the probability of D2D transmission with the PCP close to best
overall spot, whereas, increasing $v_{\max}$ while also increasing
$v_{\min}$, but at a lower rate, i.e., widening the difference $v_{\min}-v_{\max}$,
results in an increase of the probability of D2D trasnmission with
the selected PCP close to the best place.

\subsection{Simulation results and performance evaluation\label{subsec:Simulation-results}}

In the following, we review and comment on the results of our simulations
analyzing different aspects. Each figure displays a specific performance
metric obtained by letting one system parameters vary, and keeping
the other ones fixed. To generate the vehicles speed in inpu to the
simulator, we used the PDF \eqref{eq:PDF_speed}. The considered parameters
are the content timeout $\tau_{c}$, the speed range $[v_{\min},v_{\max}]$,
and the maximum nominal transmission range for D2D communications
$r_{\max}^{\text{(D2D)}}$. The sharing timeout was set to 600~s.The
remaining system parameters, kept fixed as well, are shown in Table~\ref{tab:sysparam}.
\begin{table}[t]
\caption{System parameters used for performance evaluation}
\vspace*{-3mm}
 \label{tab:sysparam}\begin{small}%
\begin{tabular*}{1\columnwidth}{@{\extracolsep{\fill}}l|c|c}
\hspace*{3cm}System parameter & Symbol & value\tabularnewline
\hline 
\hline 
Speed range & $[v_{\min},v_{\max}]$ & variable\tabularnewline
\hline 
Vehicles arrival rate (new vehicles per minute)  & $\lambda_{V}$ & 20 per minute\tabularnewline
\hline 
Node density & $\rho$ & variable (see Section~\ref{subsec:Preliminary-results})\tabularnewline
\hline 
Content requests per minute (for each vehicle) & $\lambda_{C}$ & 6 req. per minute\tabularnewline
\hline 
Content payload size & $L$ & $432\,\text{kB}$\tabularnewline
\hline 
Coded packet size & $L/\beta$ & $540\,\text{kB}$\tabularnewline
\hline 
Zipf distribution parameter for the & \multirow{2}{*}{$\alpha$} & \multirow{2}{*}{1.1}\tabularnewline
content popularity &  & \tabularnewline
\hline 
Content timeout & $\tau_{c}$ & variable\tabularnewline
\hline 
Sharing timeout & $\tau_{s}$ & $600\text{ s}$\tabularnewline
\hline 
Center frequency of the system band & $f_{0}$ & $2.3\text{ GHz}$\tabularnewline
\hline 
System bandwidth & $W$ & $10.8\,\text{MHz}$\tabularnewline
\hline 
control interval duration & $T$ & $1\text{ s}$\tabularnewline
\hline 
PRB duration & $\tau$ & $0.5\,\text{ms}$ \tabularnewline
\hline 
PRB bandwidth & $w$ & $180\,\text{KHz}$\tabularnewline
\hline 
Number of subcarriers per PRB & $K_{sc}$ & $12$\tabularnewline
\hline 
Subcarrier spacing & $w_{c}$ & $15$ $\text{KHz}$\tabularnewline
\hline 
Noise power spectral density & $\mathcal{N}_{0}$ & $-174~\text{dBm/Hz}$\tabularnewline
\hline 
Receiver noise figure & $F$ & $10\,\text{dB}$\tabularnewline
\hline 
Link margin & $M$ & see Section 6.2\tabularnewline
\hline 
Forward error correction coding rate & $\beta$ & 4/5\tabularnewline
\hline 
Transmit spectral efficiency (see Appendix )  & $e$ & 6\tabularnewline
\end{tabular*}\end{small}\vspace*{-5mm}
 
\end{table}

\subsubsection{Offloading efficiency}

\begin{figure*}[!t]
\begin{centering}
\subfloat[{With different values of $\tau_{c}$ (and fixed parameters $r_{\max}^{\text{(D2D)}}=100$~m
and $[v_{\min},v_{\max}]=[9,24]$~m/s)}]{\begin{centering}
\includegraphics[width=0.45\columnwidth]{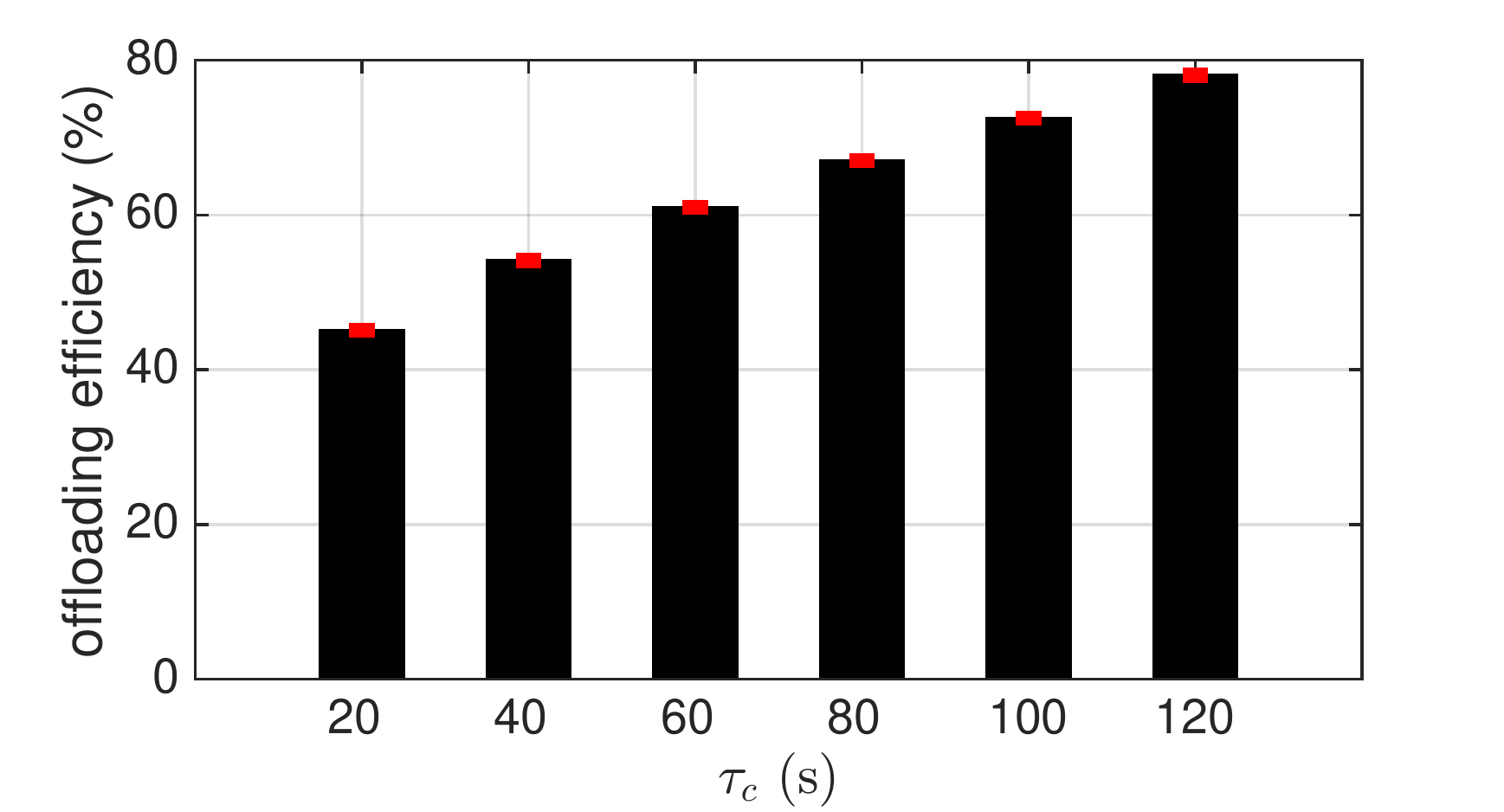}
\par\end{centering}
}\hspace{0.04\columnwidth}\subfloat[{With different values of $r_{\max}^{\text{(D2D)}}$ (and fixed parameters
$\tau_{c}=20$~s and $[v_{\min},v_{\max}]=[9,24]$~m/s)}]{\begin{centering}
\includegraphics[width=0.45\columnwidth]{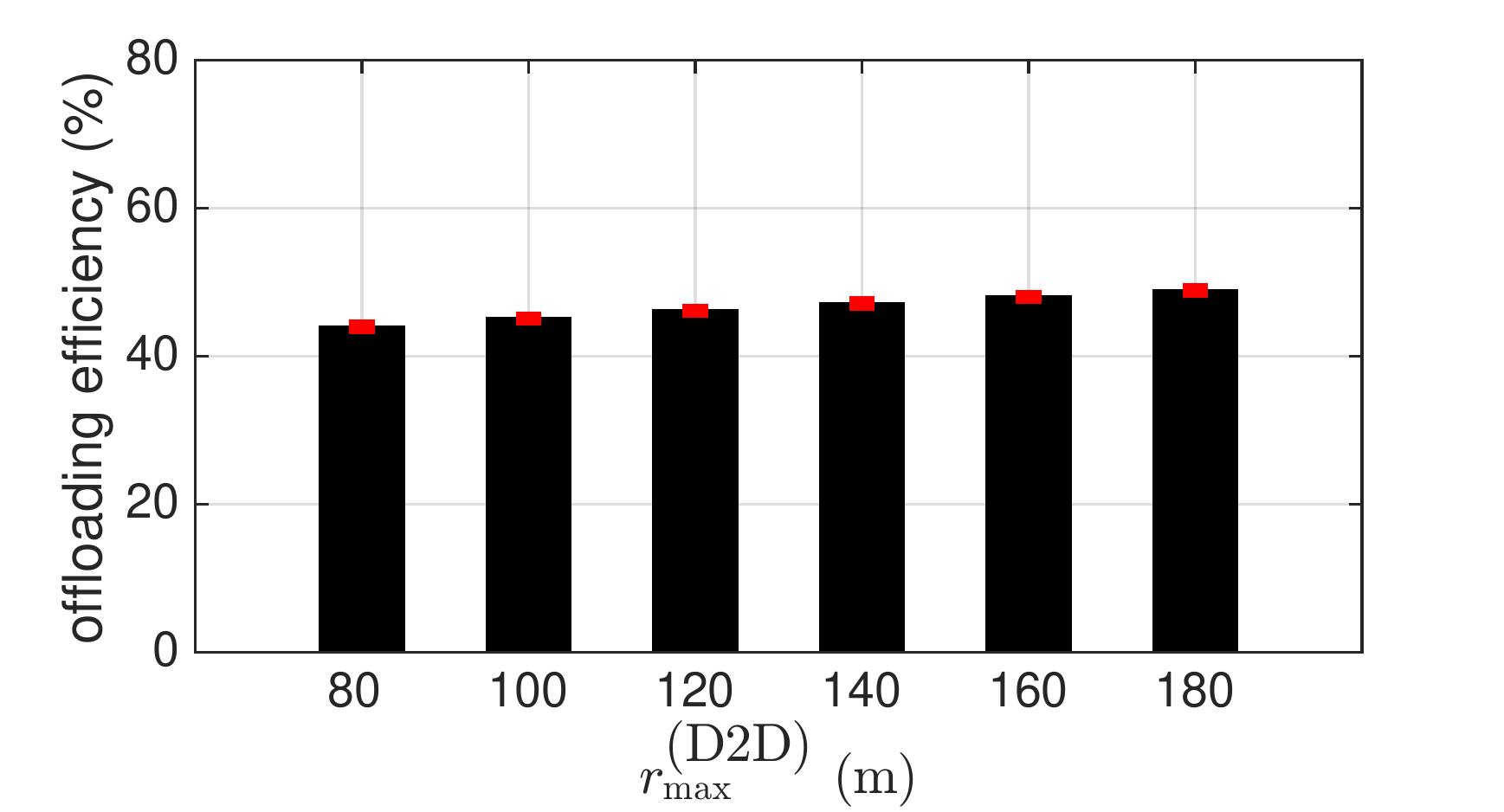} 
\par\end{centering}
}
\par\end{centering}
\centering{}\subfloat[{With different values of $[v_{\min},v_{\max}]$ (and fixed parameters
$\tau_{c}=20\text{ s}$ and $r_{\max}^{\text{(D2D)}}=100\text{ m}$)}]{\begin{centering}
\includegraphics[width=0.45\textwidth]{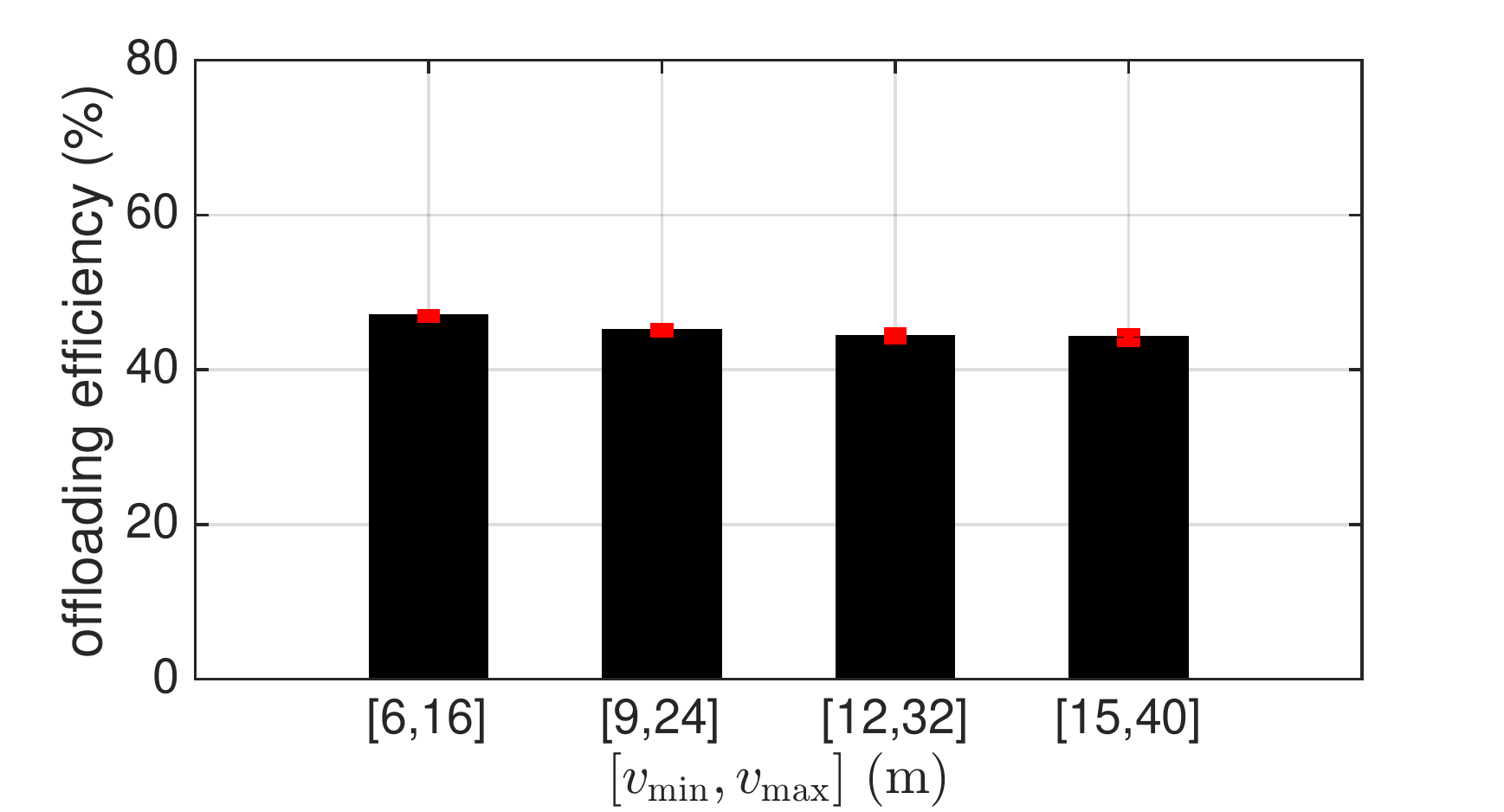} 
\par\end{centering}
}\hspace{0.04\columnwidth}\subfloat[{With different values of $[v_{\min},v_{\max}]$ (and fixed parameters
$\tau_{c}=60\text{ s}$ and $r_{\max}^{\text{(D2D)}}=100\text{ m}$)}]{\begin{centering}
\includegraphics[width=0.45\textwidth]{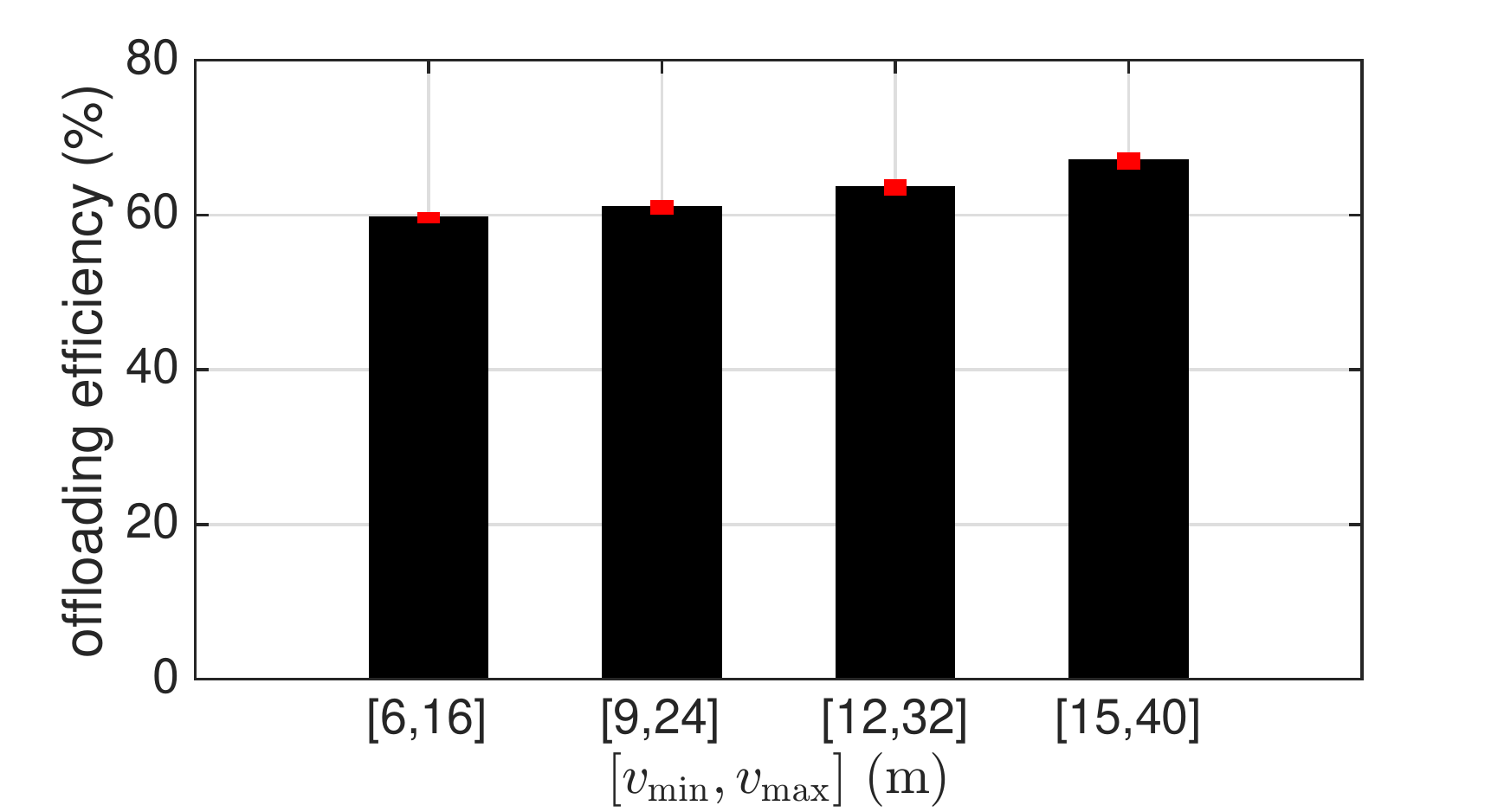} 
\par\end{centering}
}\vspace*{-2mm}
 \caption{Offloading efficiency}
\label{FIG_RES_1_offloading_efficiency} \vspace*{-2mm}
 
\end{figure*}

In Figure~\ref{FIG_RES_1_offloading_efficiency} we plot the results
obtained in terms of offloading efficiency of the considered D2D offloading
system (with 95\% confidence intervals). The offloading efficiency
tends to increase significantly with the duration of the content timeout,
while varying the other parameters yields a moderate effect. Regarding
the offloading efficiency of the benchmark CDMS, it can be shown that,
by construction, it is the same as the proposed scheme, hence it is
not showed in the figure.

\subsubsection{Energy consumption}

\begin{figure*}[!t]
\begin{centering}
\subfloat[{With different values of $\tau_{c}$ (and fixed parameters $r_{\max}^{\text{(D2D)}}=100$~m
and $[v_{\min},v_{\max}]=[9,24]$~m/s)}]{\begin{centering}
\includegraphics[width=0.45\columnwidth]{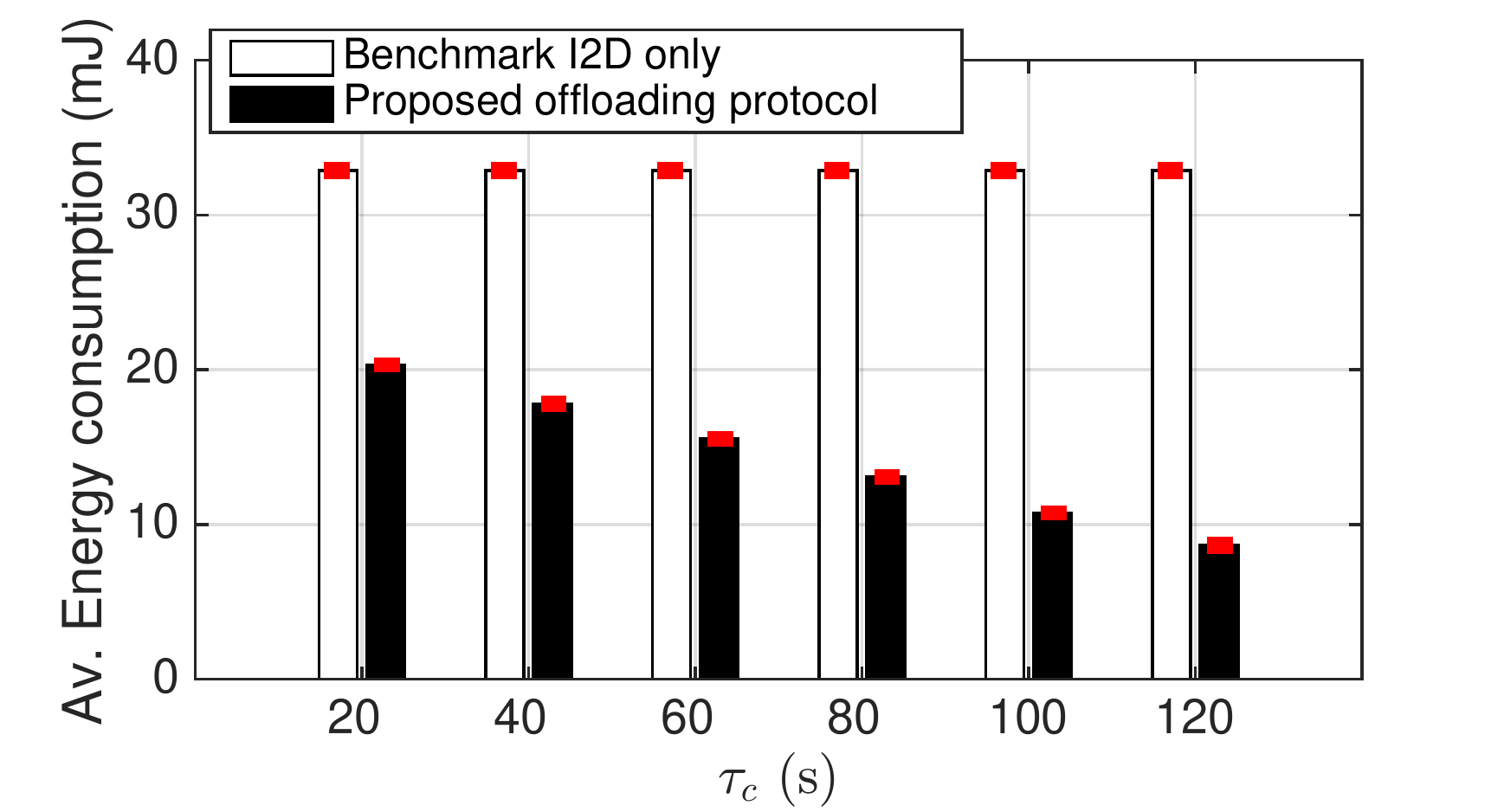}
\par\end{centering}
}\hspace{0.04\columnwidth}\subfloat[{With different values of $r_{\max}^{\text{(D2D)}}$ (and fixed parameters
$\tau_{c}=20$~s and $[v_{\min},v_{\max}]=[9,24]$~m/s)}]{\begin{centering}
\includegraphics[width=0.45\columnwidth]{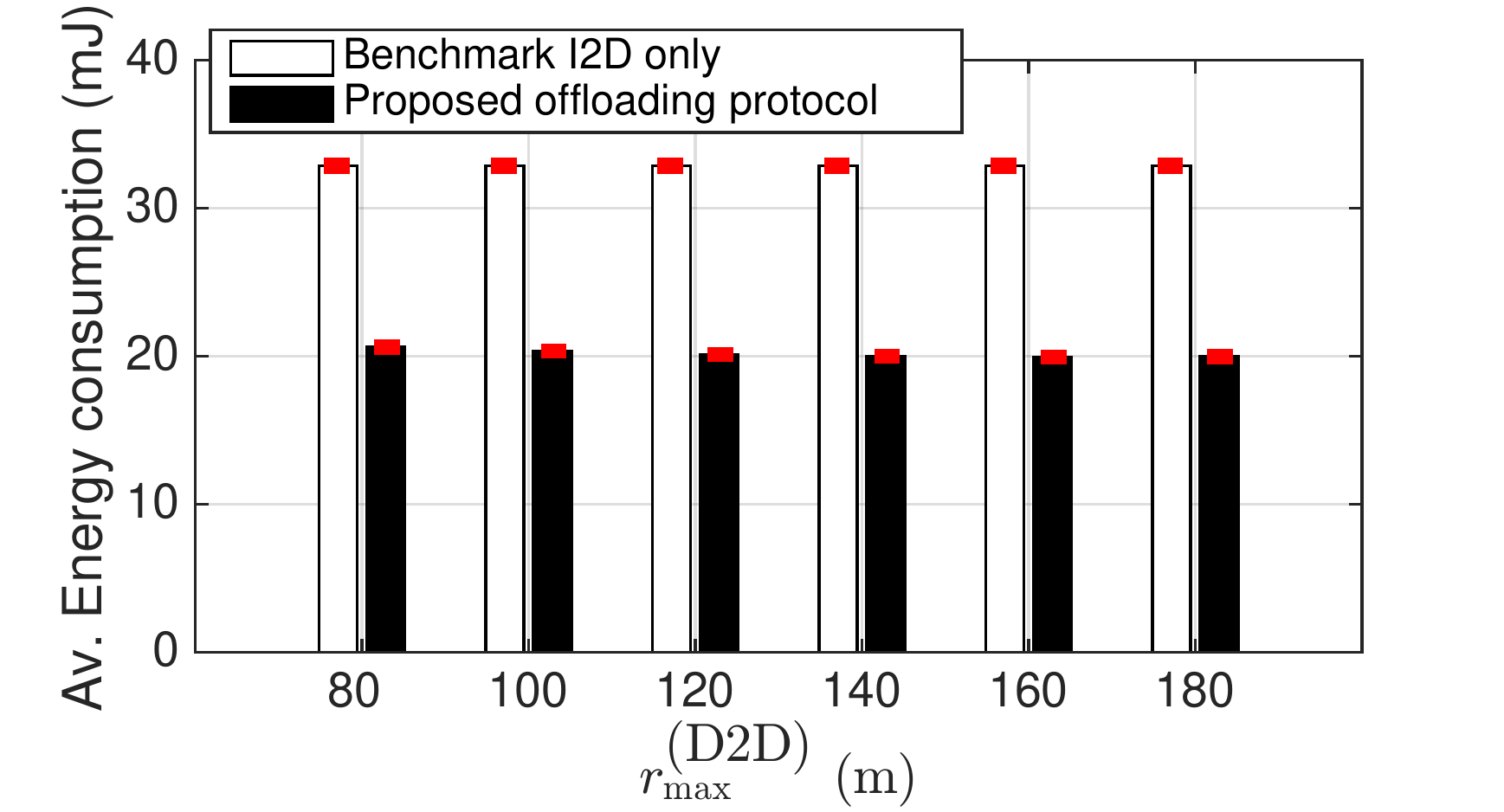} 
\par\end{centering}
}
\par\end{centering}
\centering{}\subfloat[{With different values of $[v_{\min},v_{\max}]$ (and fixed parameters
$\tau_{c}=20\text{ s}$ and $r_{\max}^{\text{(D2D)}}=100\text{ m}$)}]{\begin{centering}
\includegraphics[width=0.45\textwidth]{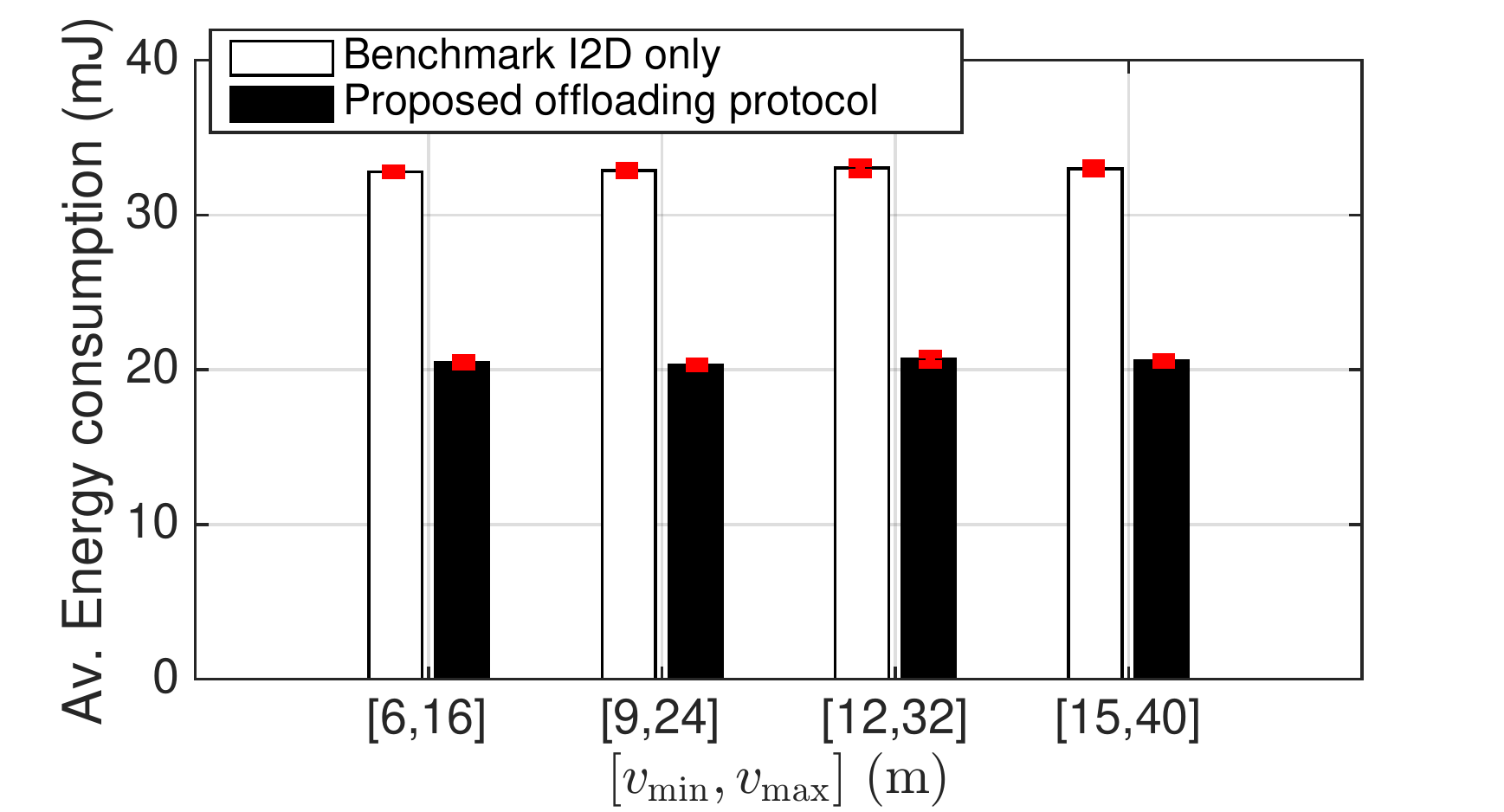} 
\par\end{centering}
}\hspace{0.04\columnwidth}\subfloat[{With different values of $[v_{\min},v_{\max}]$ (and fixed parameters
$\tau_{c}=60\text{ s}$ and $r_{\max}^{\text{(D2D)}}=100\text{ m}$)}]{\begin{centering}
\includegraphics[width=0.45\textwidth]{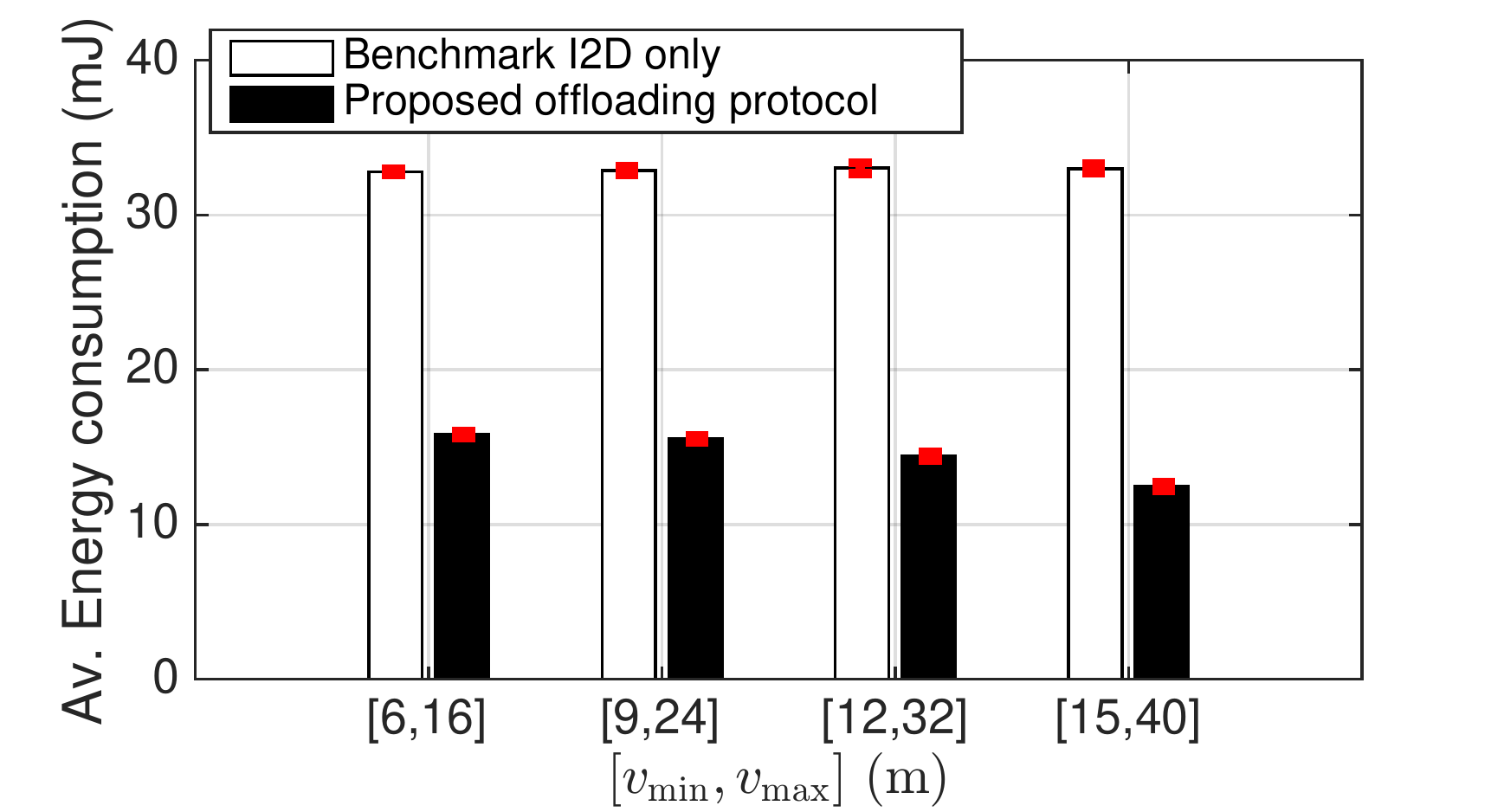} 
\par\end{centering}
}\vspace*{-2mm}
 \caption{Average energy consumption per delivered content}
\label{FIG_RES_2_Energy_consumption} \vspace*{-2mm}
 
\end{figure*}

Figure~\ref{FIG_RES_2_Energy_consumption} shows the energy consumed
on average (with confidence intervals) to deliver a content by the
proposed CDMS and the plain cellular scheme. The average is performed
\emph{on the overall set of both I2D and D2D transmissions} (only
I2D ones for the benchmark cellular scheme). It can be seen that the
proposed CDMS yields a considerable improvement of this performance
metric with respect to the plain cellular system. Using the proposed
CDMS yields a performance gain (i.e., a reduction) of at least $13\text{ mJ}$
per content, and up to $25\text{ mJ}$, over the benchmark plain cellular
protocol.

\begin{figure*}[!t]
\begin{centering}
\subfloat[{With different values of $\tau_{c}$ (and fixed parameters $r_{\max}^{\text{(D2D)}}=100$~m
and $[v_{\min},v_{\max}]=[9,24]$~m/s)}]{\begin{centering}
\includegraphics[width=0.45\columnwidth]{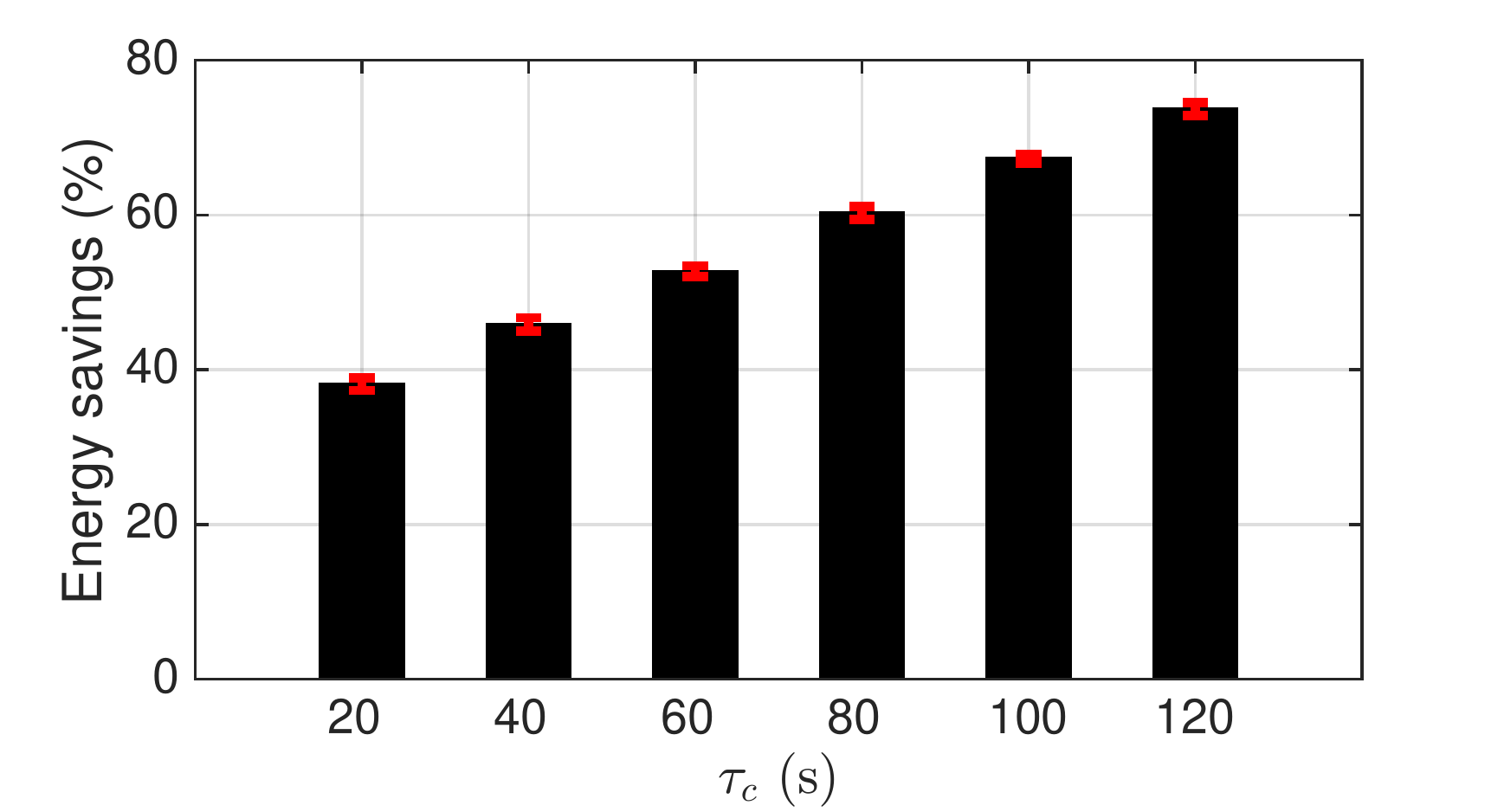}
\par\end{centering}
}\hspace{0.04\columnwidth}\subfloat[{With different values of $r_{\max}^{\text{(D2D)}}$ (and fixed parameters
$\tau_{c}=20$~s and $[v_{\min},v_{\max}]=[9,24]$~m/s)}]{\begin{centering}
\includegraphics[width=0.45\columnwidth]{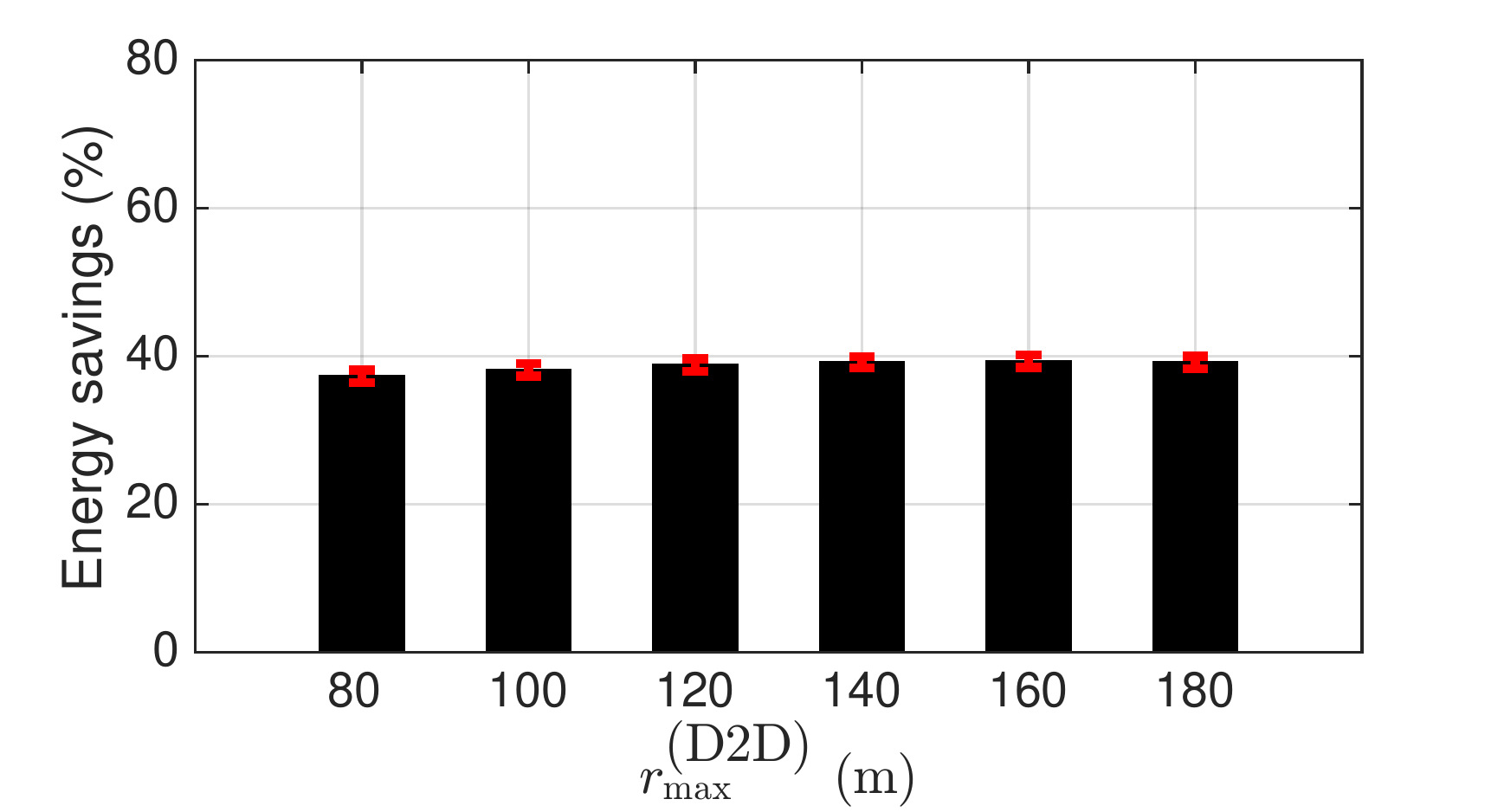} 
\par\end{centering}
}
\par\end{centering}
\centering{}\subfloat[{With different values of $[v_{\min},v_{\max}]$ (and fixed parameters
$\tau_{c}=20\text{ s}$ and $r_{\max}^{\text{(D2D)}}=100\text{ m}$)}]{\begin{centering}
\includegraphics[width=0.45\textwidth]{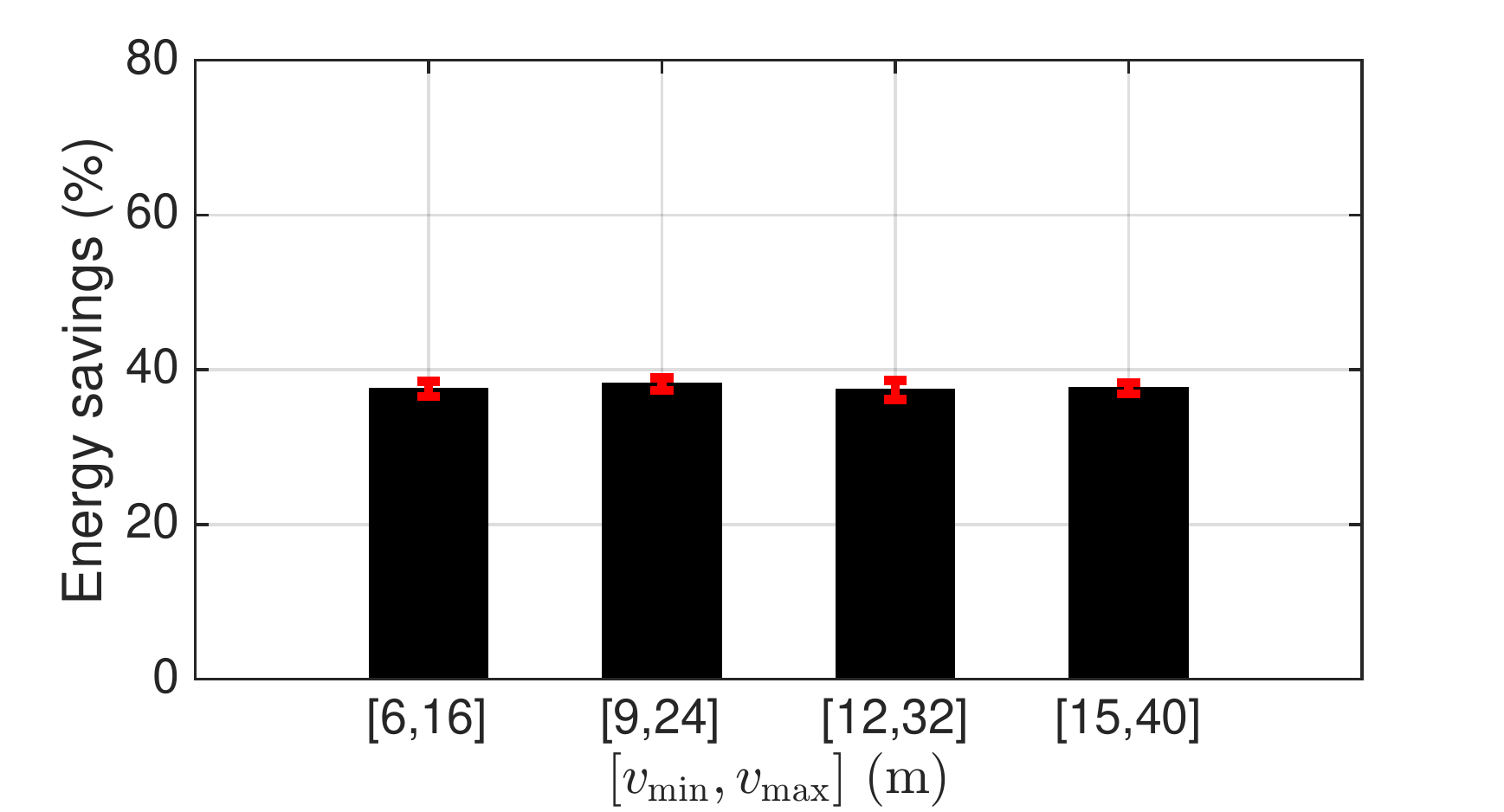} 
\par\end{centering}
}\hspace{0.04\columnwidth}\subfloat[{With different values of $[v_{\min},v_{\max}]$ (and fixed parameters
$\tau_{c}=60\text{ s}$ and $r_{\max}^{\text{(D2D)}}=100\text{ m}$)}]{\begin{centering}
\includegraphics[width=0.45\textwidth]{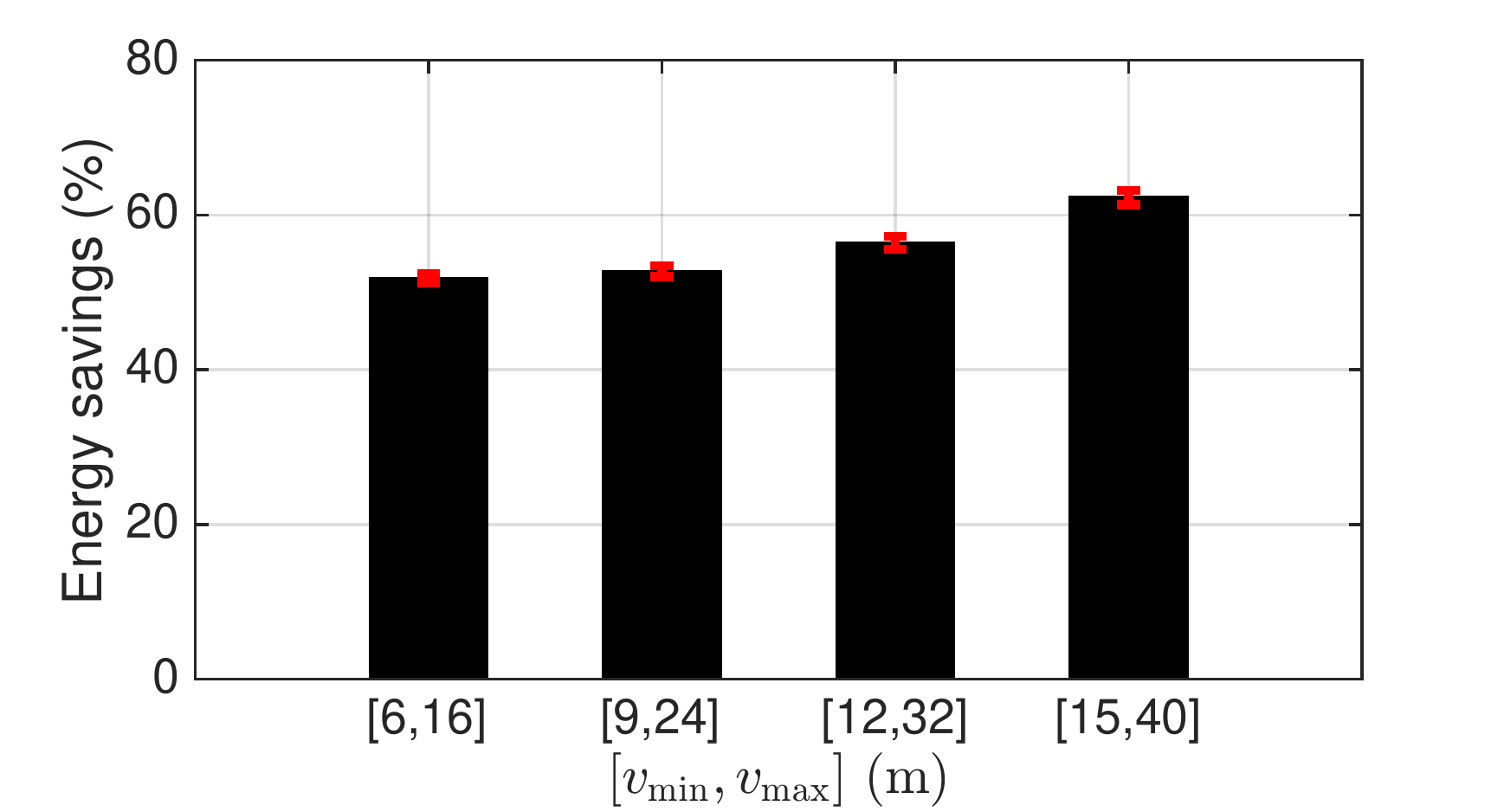} 
\par\end{centering}
}\vspace*{-2mm}
 \caption{Average energy consumption reduction percentage with respect to the
benchmark plain cellular scheme}
\label{FIG_RES_3_ENERGY_GAIN_AB_vs_0} \vspace*{-2mm}
 
\end{figure*}

The same comparison, in terms of percentage reduction of the energy
consumption, is provided in Figure~\ref{FIG_RES_3_ENERGY_GAIN_AB_vs_0}.
The reduction is in the order of 30-40\% in the worst cases, up to
60\% with a speed range $[15,40]\,\text{m/s}$ and a content timeout
of $60\,\text{s}$ (Subfigure~\ref{FIG_RES_3_ENERGY_GAIN_AB_vs_0}d),
and 77\% with a speed range of $[9,24]\,\text{m/s}$ and content timeout
$20\text{ s}$.
\\

\begin{figure*}[!t]
\begin{centering}
\subfloat[{With different values of $\tau_{c}$ (and fixed parameters $r_{\max}^{\text{(D2D)}}=100$~m
and $[v_{\min},v_{\max}]=[9,24]$~m/s)}]{\begin{centering}
\includegraphics[width=0.45\columnwidth]{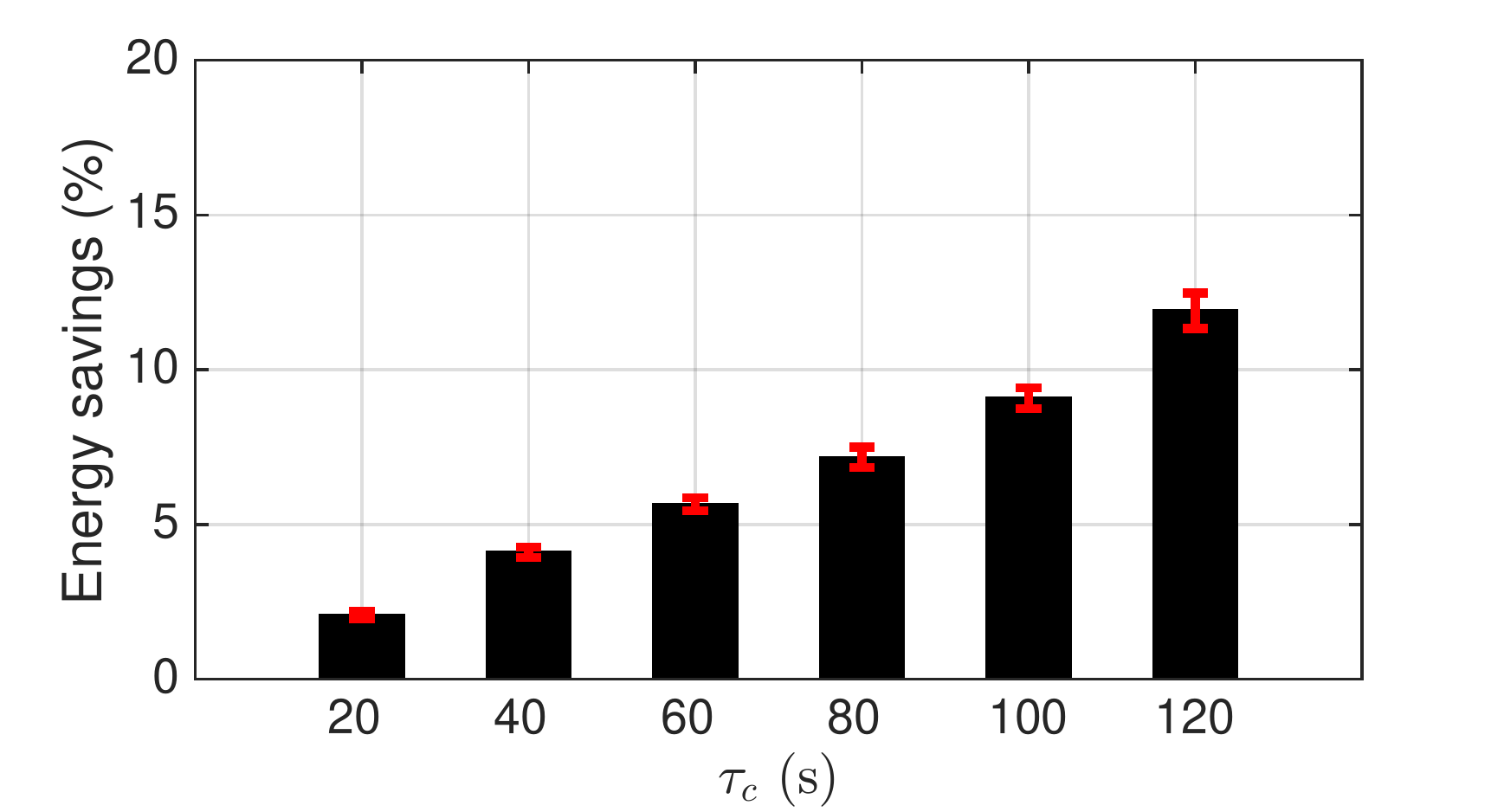}
\par\end{centering}
}\hspace{0.04\columnwidth}\subfloat[{With different values of $r_{\max}^{\text{(D2D)}}$ (and fixed parameters
$\tau_{c}=20$~s and $[v_{\min},v_{\max}]=[9,24]$~m/s)}]{\begin{centering}
\includegraphics[width=0.45\columnwidth]{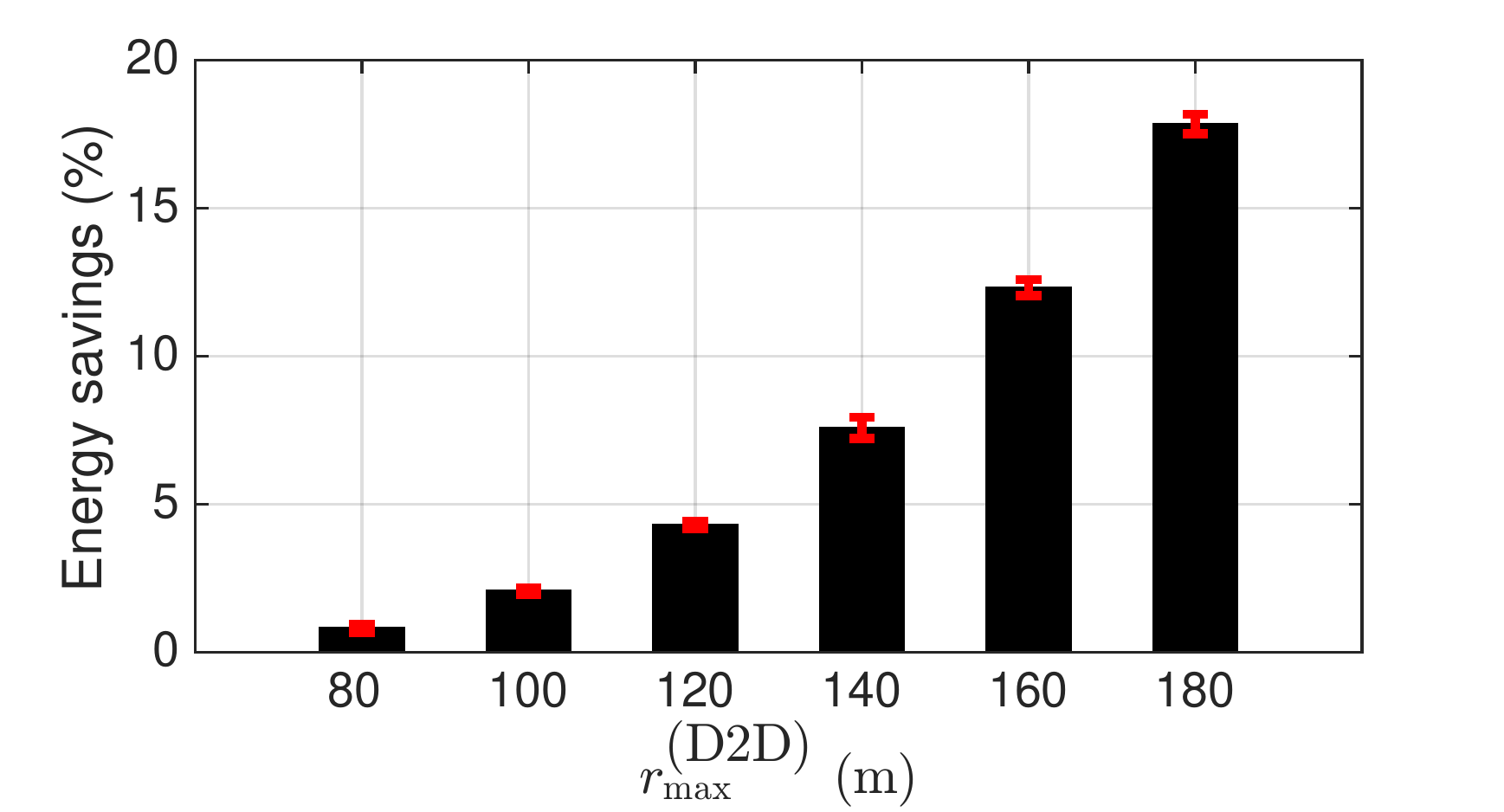} 
\par\end{centering}
}
\par\end{centering}
\centering{}\subfloat[{With different values of $[v_{\min},v_{\max}]$ (and fixed parameters
$\tau_{c}=20\text{ s}$ and $r_{\max}^{\text{(D2D)}}=100\text{ m}$)}]{\begin{centering}
\includegraphics[width=0.45\textwidth]{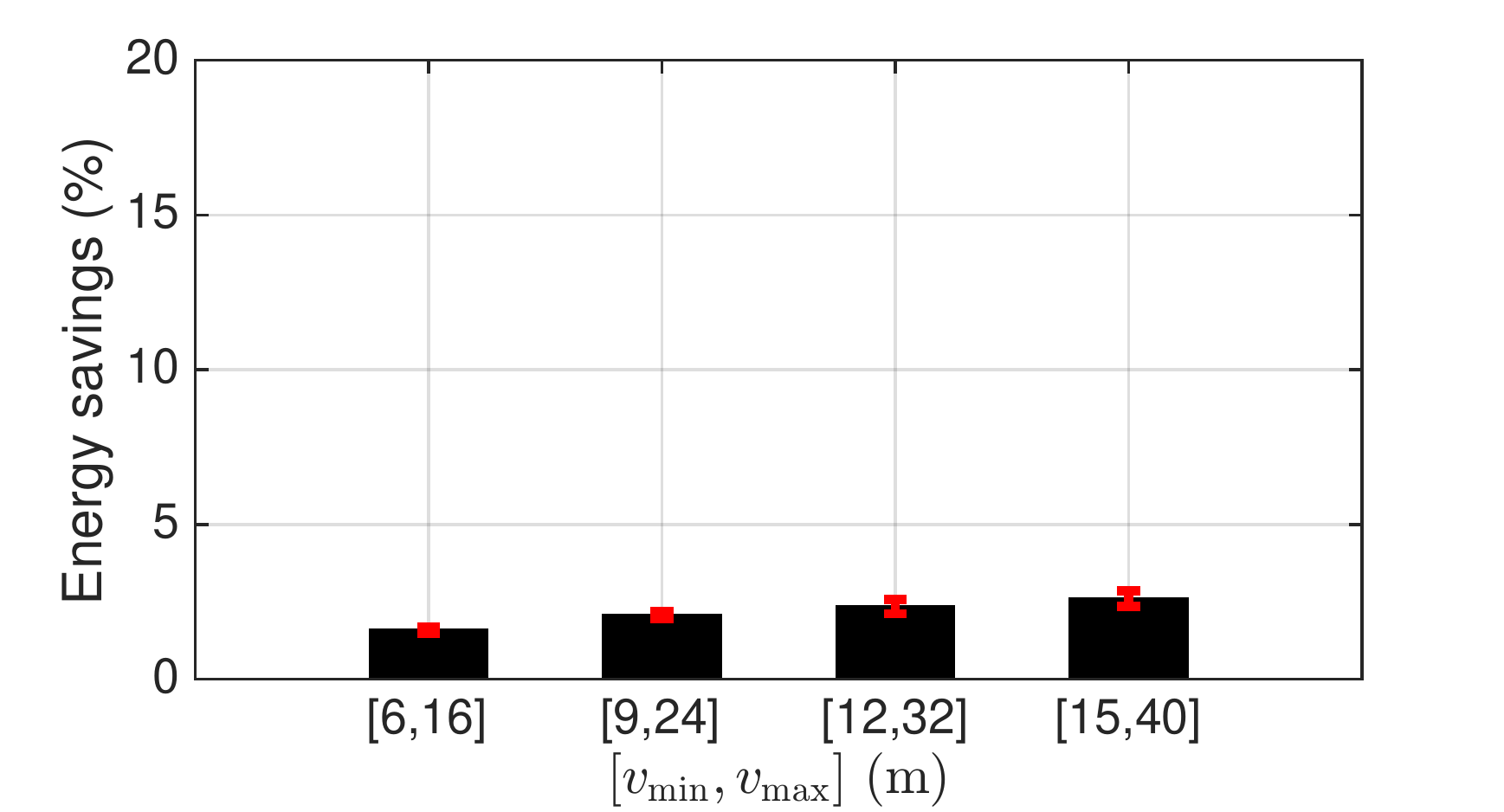} 
\par\end{centering}
}\hspace{0.04\columnwidth}\subfloat[{With different values of $[v_{\min},v_{\max}]$ (and fixed parameters
$\tau_{c}=60\text{ s}$ and $r_{\max}^{\text{(D2D)}}=100\text{ m}$)}]{\begin{centering}
\includegraphics[width=0.45\textwidth]{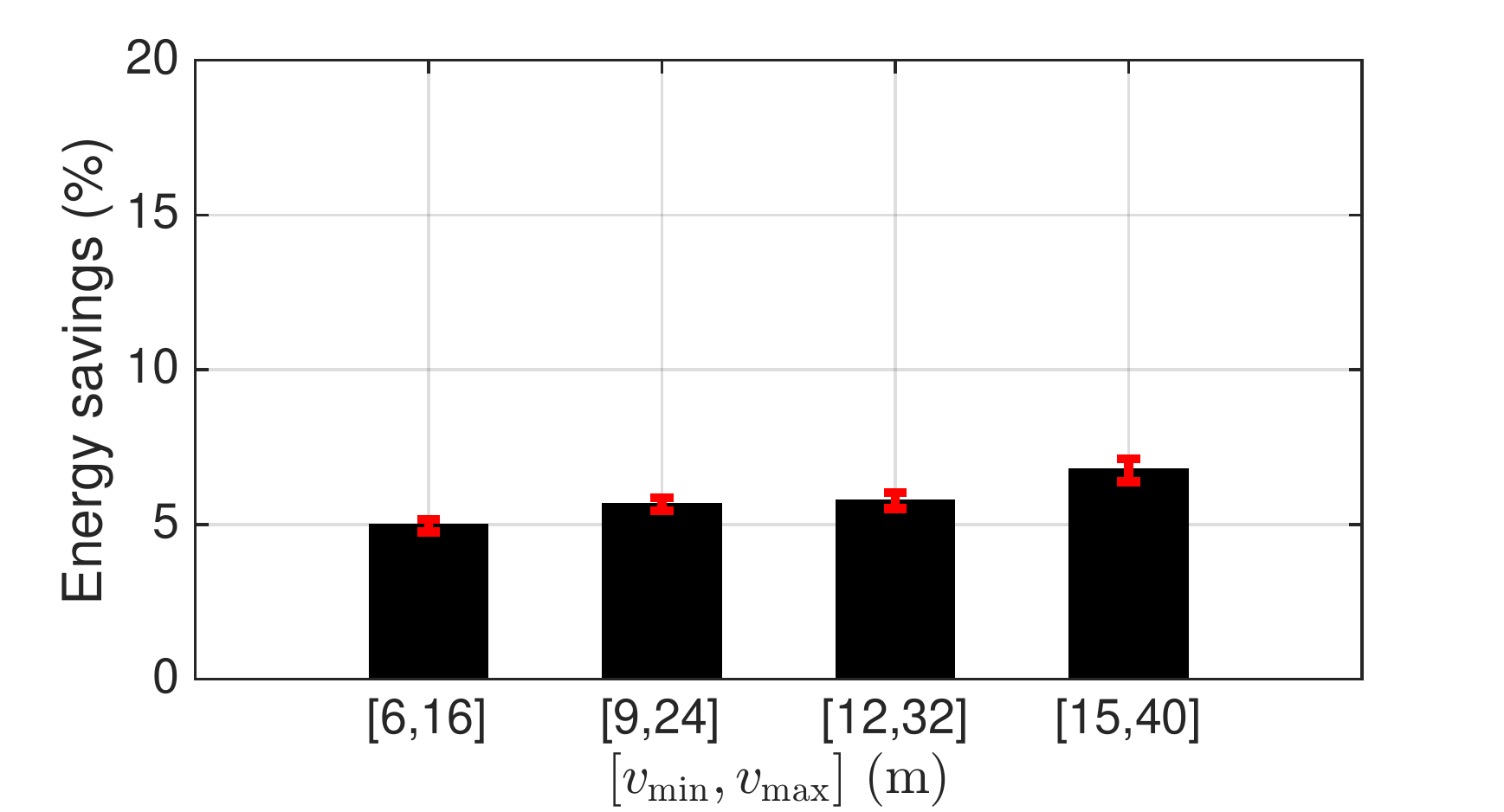} 
\par\end{centering}
}\vspace*{-2mm}
 \caption{System-level (I2D +D2D comms) energy consumption reduction percentage
of the proposed CDMS with respect to the benchmark CDMS}
\label{FIG_RES_4_ENERGY_GAIN_B_vs_A} \vspace*{-2mm}
 
\end{figure*}

\begin{figure*}[!t]
\begin{centering}
\subfloat[{With different values of $\tau_{c}$ (and fixed parameters $r_{\max}^{\text{(D2D)}}=100$~m
and $[v_{\min},v_{\max}]=[9,24]$~m/s)}]{\begin{centering}
\includegraphics[width=0.45\columnwidth]{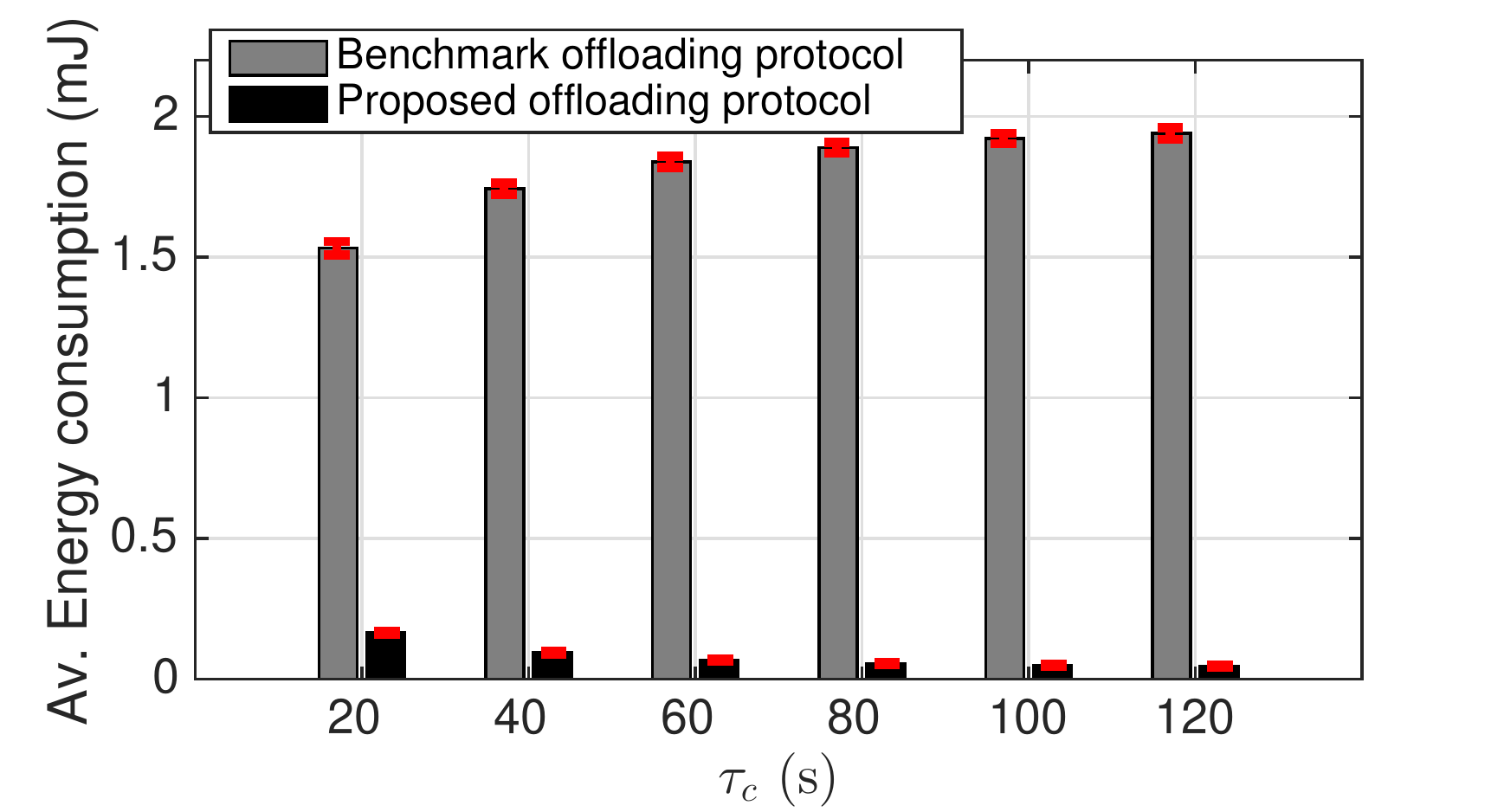}
\par\end{centering}
}\hspace{0.04\columnwidth}\subfloat[{With different values of $r_{\max}^{\text{(D2D)}}$ (and fixed parameters
$\tau_{c}=20$~s and $[v_{\min},v_{\max}]=[9,24]$~m/s)}]{\begin{centering}
\includegraphics[width=0.45\columnwidth]{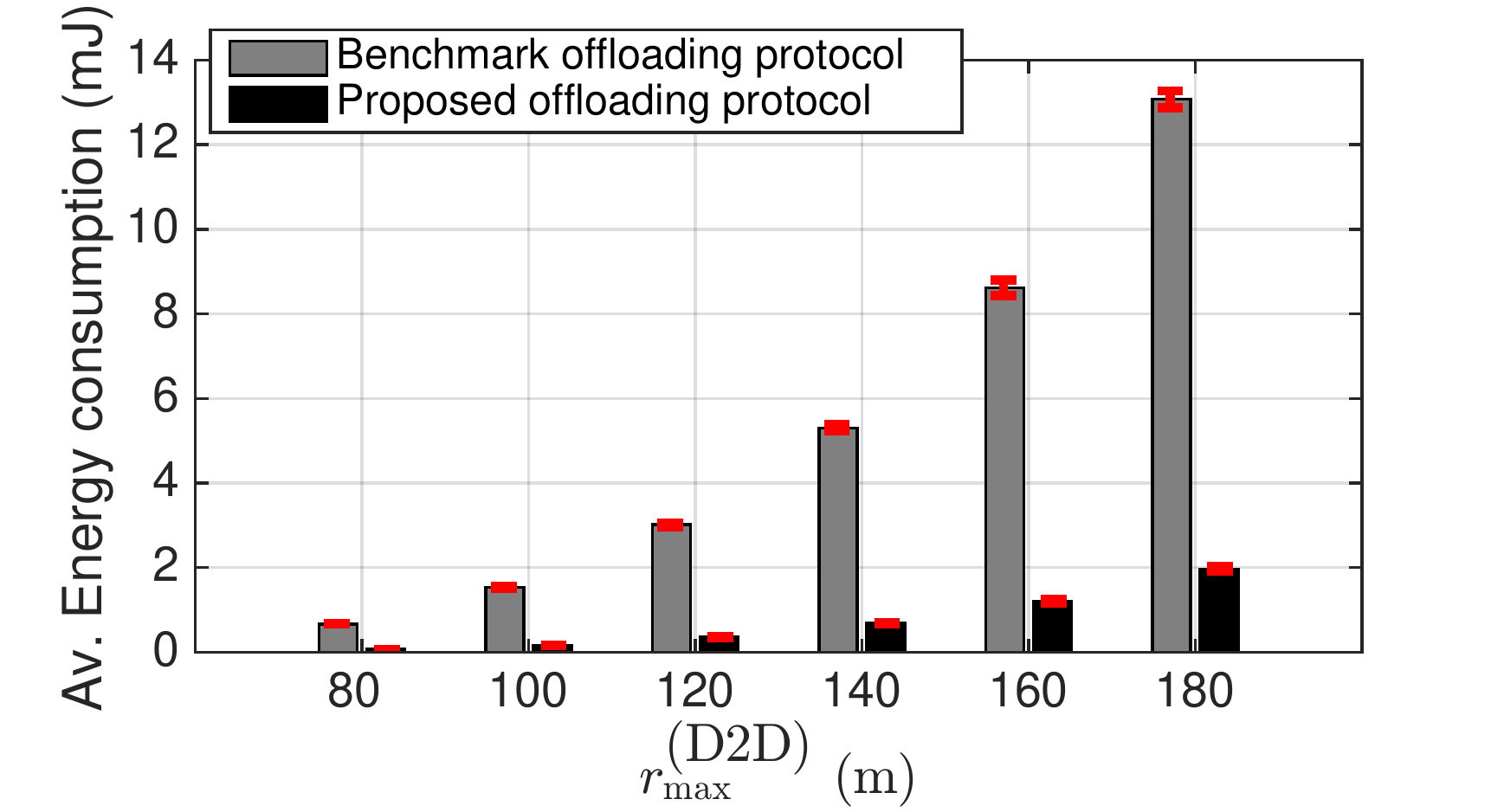} 
\par\end{centering}
}
\par\end{centering}
\centering{}\subfloat[{With different values of $[v_{\min},v_{\max}]$ (and fixed parameters
$\tau_{c}=20\text{ s}$ and $r_{\max}^{\text{(D2D)}}=100\text{ m}$)}]{\begin{centering}
\includegraphics[width=0.45\textwidth]{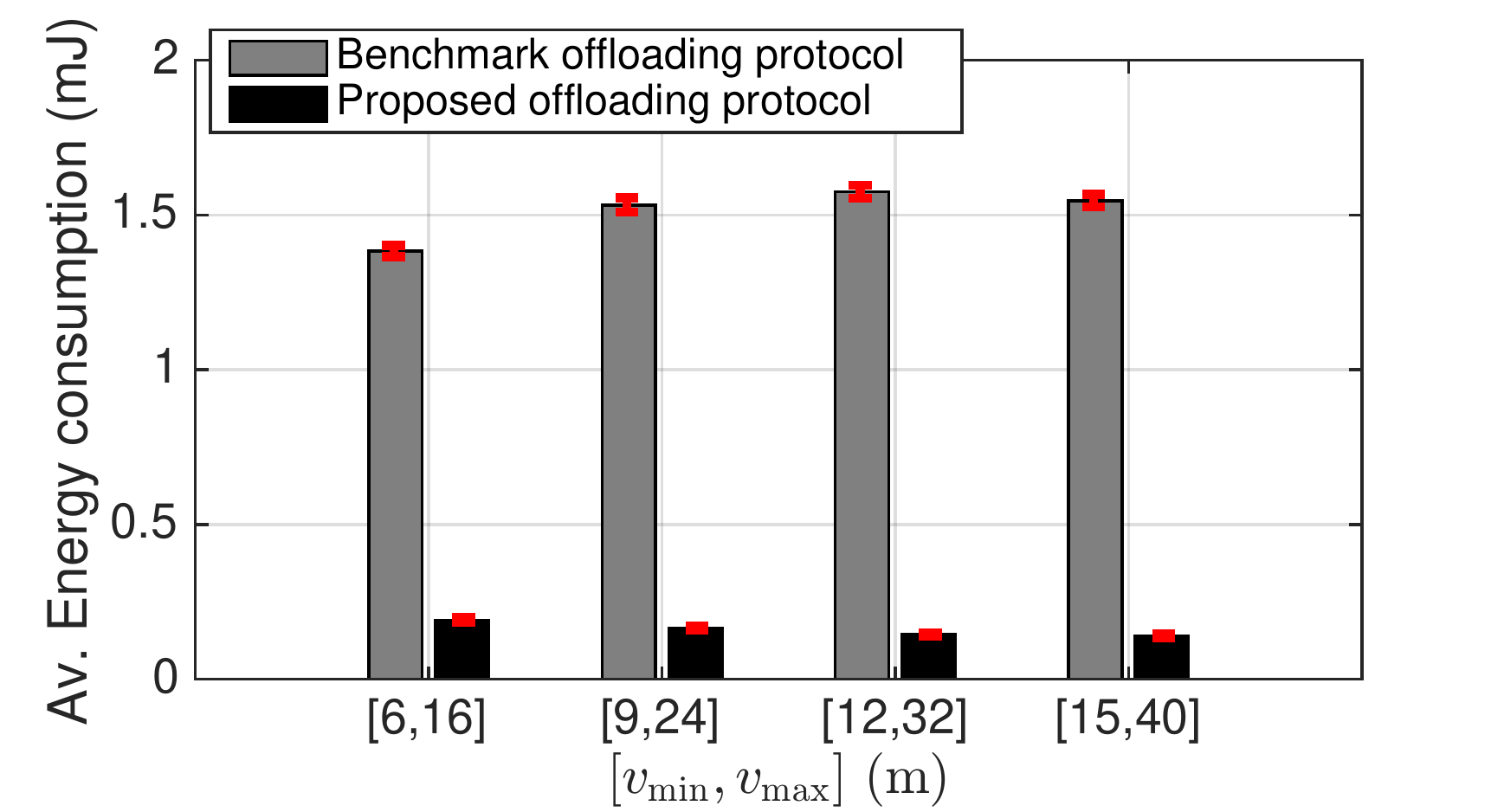} 
\par\end{centering}
}\hspace{0.04\columnwidth}\subfloat[{With different values of $[v_{\min},v_{\max}]$ (and fixed parameters
$\tau_{c}=60\text{ s}$ and $r_{\max}^{\text{(D2D)}}=100\text{ m}$)}]{\begin{centering}
\includegraphics[width=0.45\textwidth]{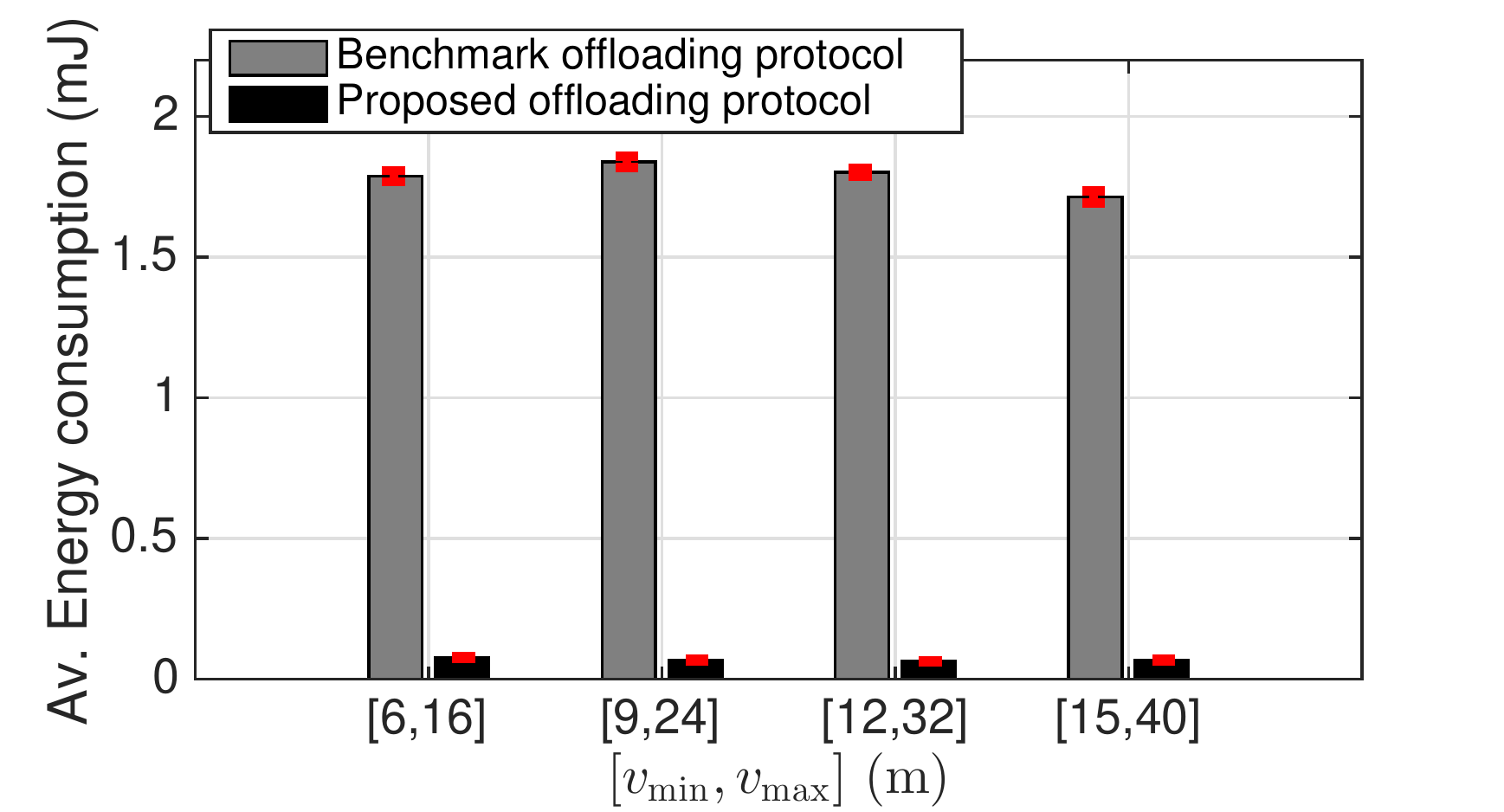} 
\par\end{centering}
}\vspace*{-2mm}
 \caption{Average energy consumption per delivered content, \emph{including
D2D communications only} - proposed CDMS and benchmark CDMS}
\label{FIG_RES_5_Energy_consumption_D2D}\vspace*{-2mm}
 
\end{figure*}

\begin{figure*}[!t]
\begin{centering}
\subfloat[{With different values of $\tau_{c}$ (and fixed parameters $r_{\max}^{\text{(D2D)}}=100$~m
and $[v_{\min},v_{\max}]=[9,24]$~m/s)}]{\begin{centering}
\includegraphics[width=0.45\columnwidth]{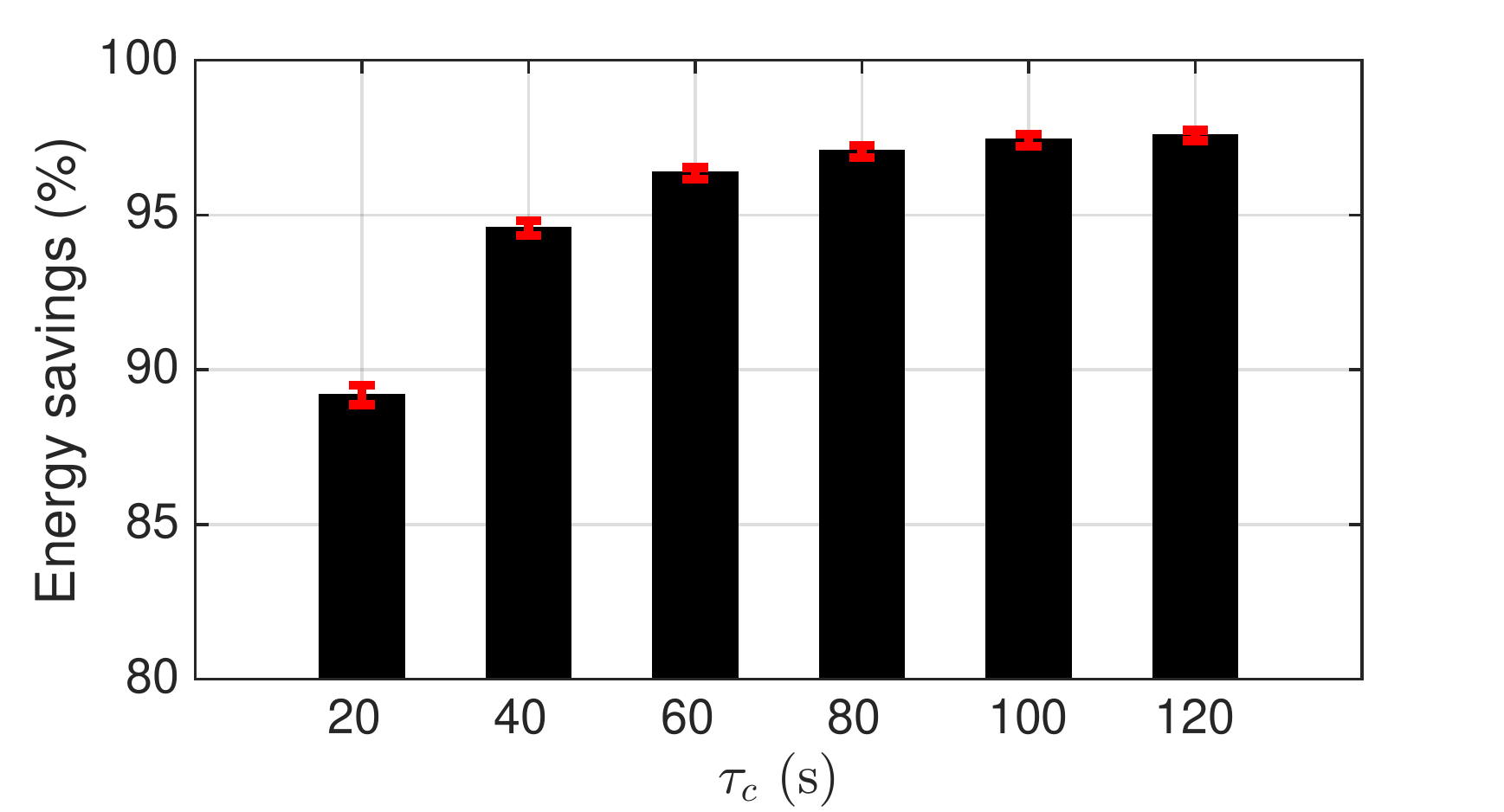}
\par\end{centering}
}\hspace{0.04\columnwidth}\subfloat[{With different values of $r_{\max}^{\text{(D2D)}}$ (and fixed parameters
$\tau_{c}=20$~s and $[v_{\min},v_{\max}]=[9,24]$~m/s)}]{\begin{centering}
\includegraphics[width=0.45\columnwidth]{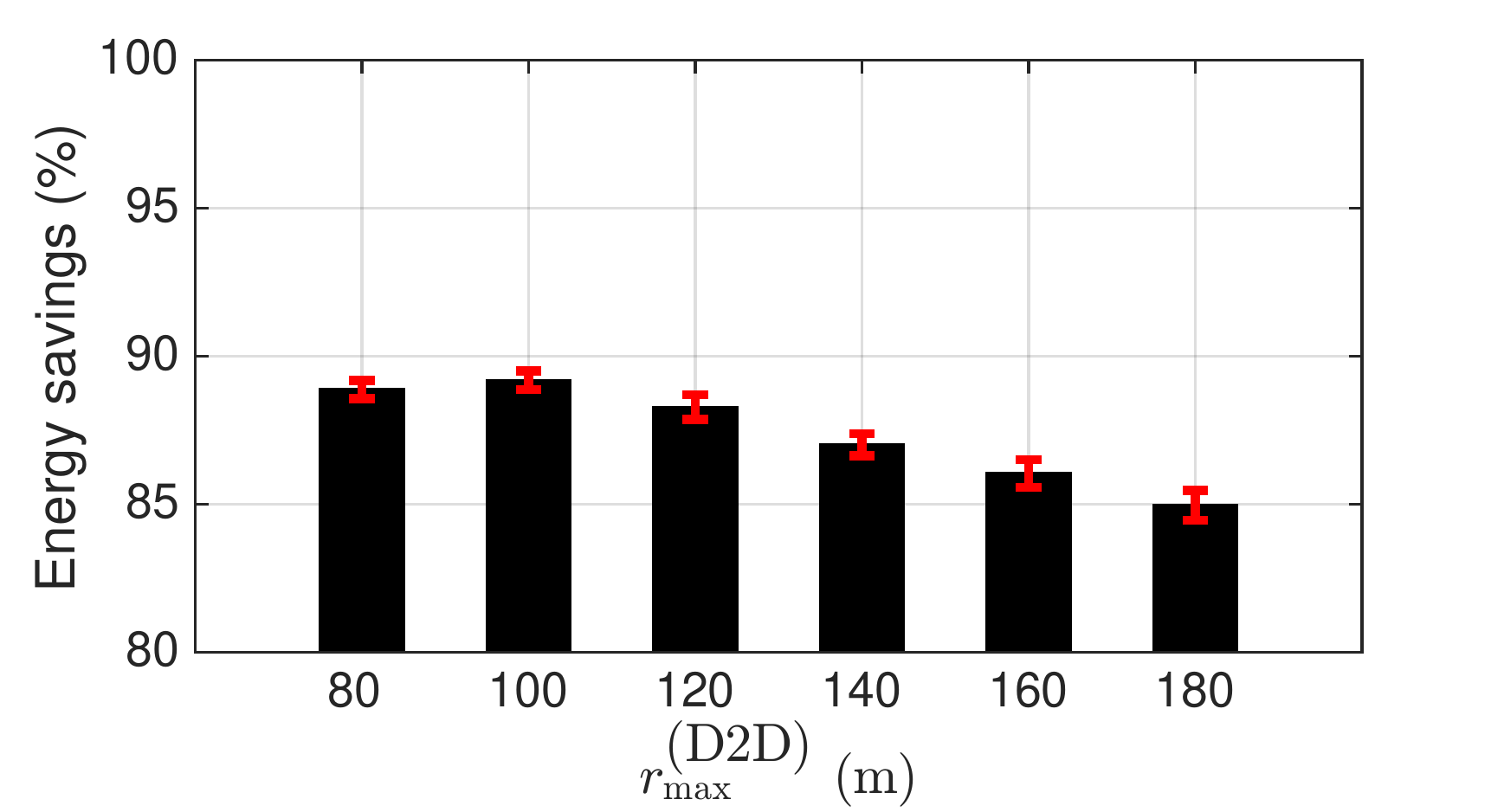} 
\par\end{centering}
}
\par\end{centering}
\centering{}\subfloat[{With different values of $[v_{\min},v_{\max}]$ (and fixed parameters
$\tau_{c}=20\text{ s}$ and $r_{\max}^{\text{(D2D)}}=100\text{ m}$)}]{\begin{centering}
\includegraphics[width=0.45\textwidth]{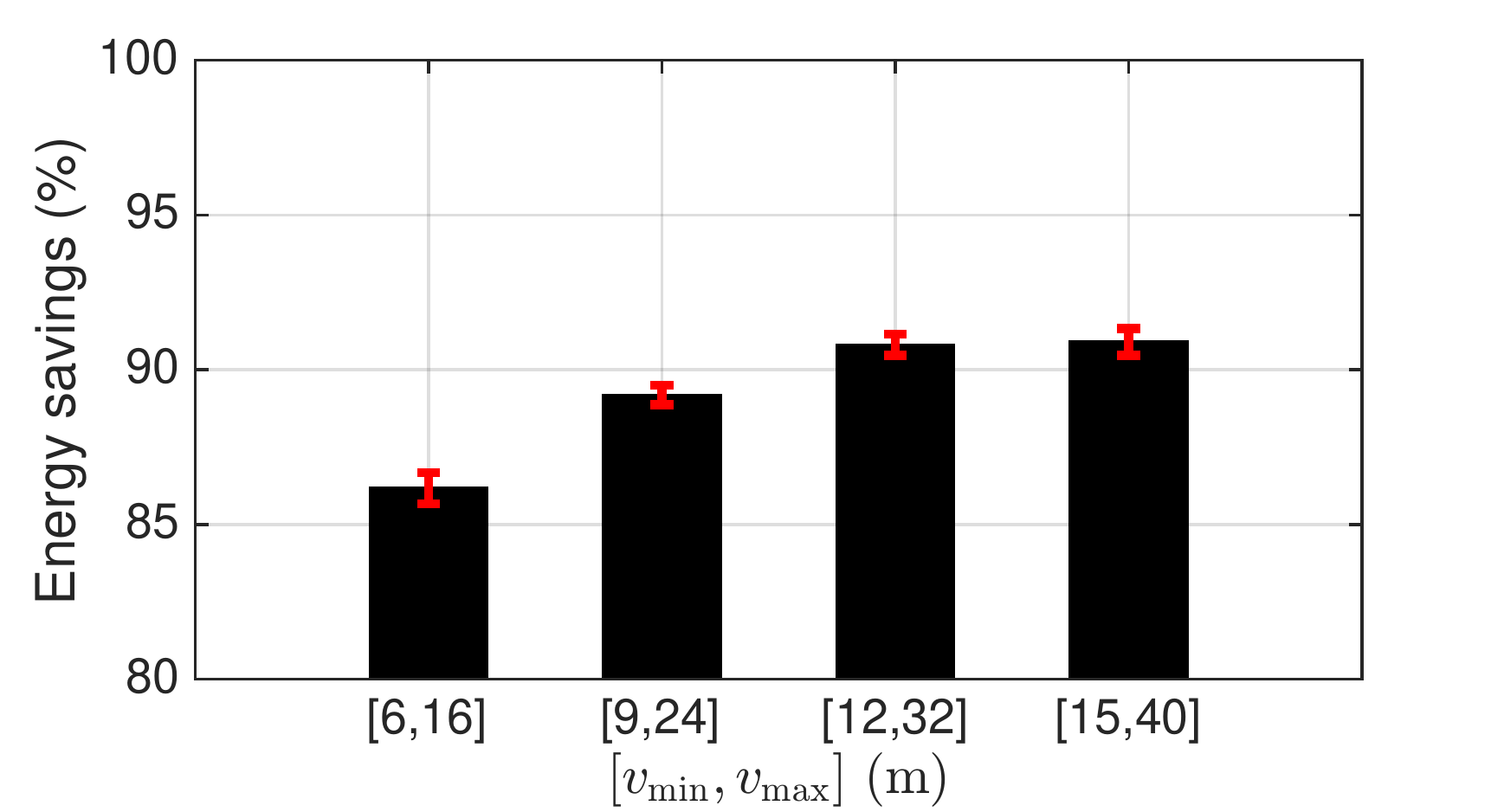} 
\par\end{centering}
}\hspace{0.04\columnwidth}\subfloat[{With different values of $[v_{\min},v_{\max}]$ (and fixed parameters
$\tau_{c}=60\text{ s}$ and $r_{\max}^{\text{(D2D)}}=100\text{ m}$)}]{\begin{centering}
\includegraphics[width=0.45\textwidth]{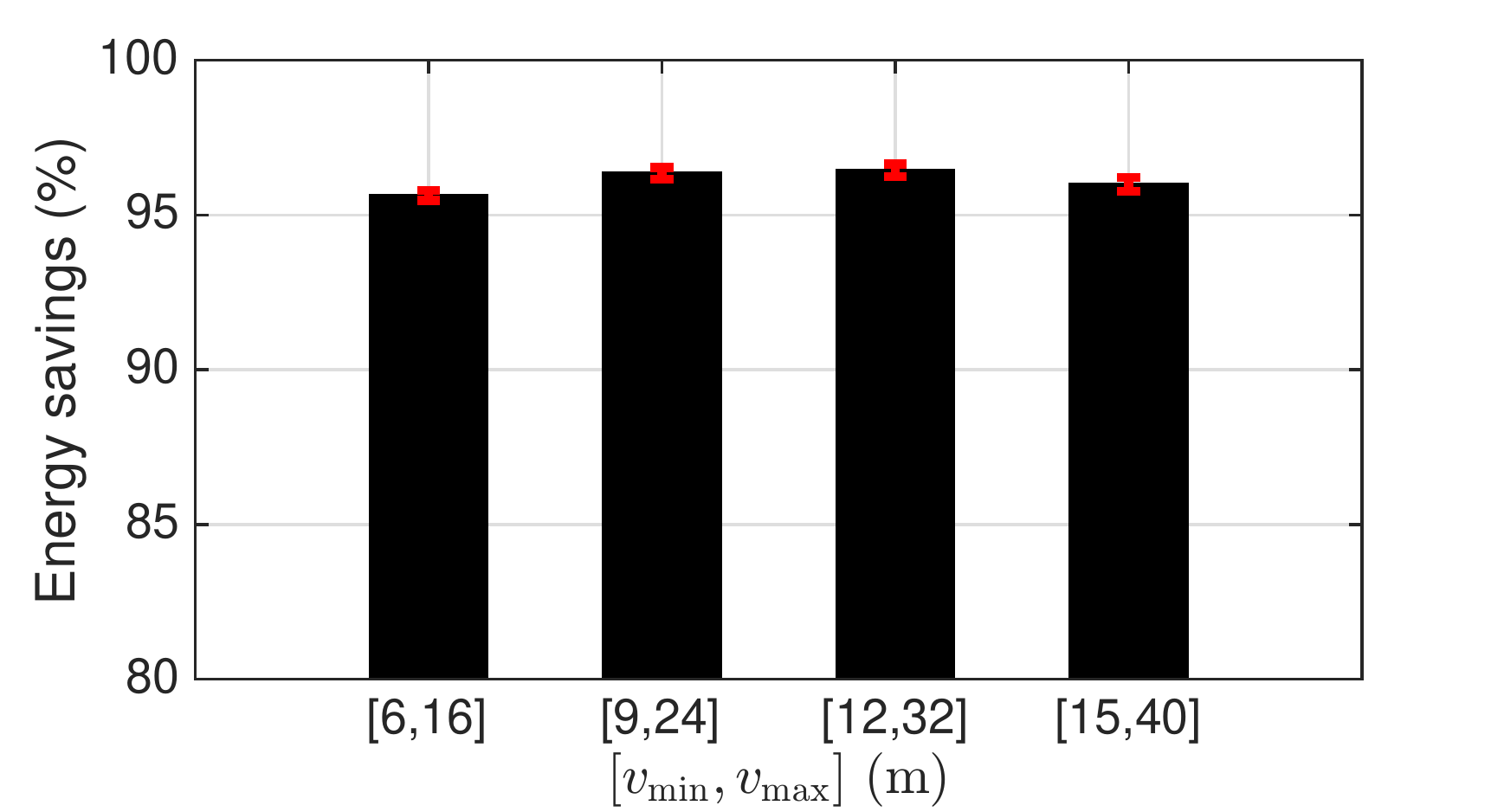} 
\par\end{centering}
}\vspace*{-2mm}
 \caption{Energy consumption reduction percentage of the proposed CMDS with
respect to a the benchmark D2D CDMS - D2D communications only}
\label{FIG_RES_6_Energy_consumption_D2D_GAIN}\vspace*{-2mm}
 
\end{figure*}

The performance gain relative to the benchmark D2D offloading system,
\emph{still taking into account both I2D transmissions and D2D ones},
is showed (with confidence intervals) in Figure~\ref{FIG_RES_4_ENERGY_GAIN_B_vs_A},
it can be seen that the gain ranges from a 2\% reduction up to 12\%
(Subfigure~\ref{FIG_RES_4_ENERGY_GAIN_B_vs_A}.a) or 17\% (Subfigure~\ref{FIG_RES_4_ENERGY_GAIN_B_vs_A}.b)\footnote{The benchmark D2D offloading scheme has itself a significant improvement
over the plain cellular system \citep{Pescosolido2018WoWMoM}, but
the CDMS proposed here further reduces the energy consumption.}. 
 Furthermore, it is worth pointing out that the overall energy consumption
is dominated by the I2D component, since the energy spent for I2D
communications is much larger than that spent for D2D ones, and the
weights associated to the two types of communications have a comparable
order of magnitude, since they are determined by the offloading efficiency,
which is 80\%, in the best case, among our selected configurations.
Therefore, the marginal impact of the proposed CMDS cannot be fully
appreciated using this performance metric. Indeed, since the mobile
devices are battery powered, and the cost associated to their energy
consumption impacts on the end user (while the cost of I2D communications
impacts on the cellular operator), \emph{it is important to single
out the gain in terms of energy consumption associated to the sole
D2D communications}. Figure~\ref{FIG_RES_5_Energy_consumption_D2D}
shows the average energy consumption of the benchmark D2D offloading
protocol and the proposed protocol. It can be seen that the proposed
protocol entails an average energy consumption, for D2D transmission,
which \emph{is a small fraction }of the energy spent by the benchmark
protocol.
 In terms of energy consumption reduction percentage, this improvement
is showed in Figure~\ref{FIG_RES_6_Energy_consumption_D2D_GAIN}.
For the D2D transmissions, the reduction in energy consumption is
always larger than 80\%, peaking at 97\% in the best case (Subfigure~\ref{FIG_RES_6_Energy_consumption_D2D_GAIN}.a)
of content timeout equal to 120 seconds.
From the analysis of the above results, it can be concluded that the
most relevant parameter is the content timeout. Intuitively, if the
type of data being transmitted is needed for a \emph{non} time-critical
application, the best thing to do, upon issuing a content request,
is to wait for a while for some devices with the content passing very
close to the requesting device, so that the transmission will be performed
at a very short distance. The statistics of the D2D transmission distance
derived in Section~\ref{sec:Analytical-model} explain why, with
an increasing content timeout, the proposed protocol outperforms the
benchmark one. In fact, for the benchmark protocol, increasing the
content timeout increases the percentage of the D2D transmission performed
with delay with respect to the request time, which (by design) are
performed as soon as an encountered PCP comes at a distance equal
to the maximum transmission distance, and hence using the maximum
transmit power for D2D transmissions. With the proposed protocol,
it is the opposite, since the PCPs have time to come very close to
the requesting node.

\vspace{-4mm}

\subsubsection{Spectrum use}

\begin{figure*}[!t]
\begin{centering}
\subfloat[{With different values of $\tau_{c}$ (and fixed parameters $r_{\max}^{\text{(D2D)}}=100$~m
and $[v_{\min},v_{\max}]=[9,24]$~m/s)}]{\begin{centering}
\includegraphics[width=0.45\columnwidth]{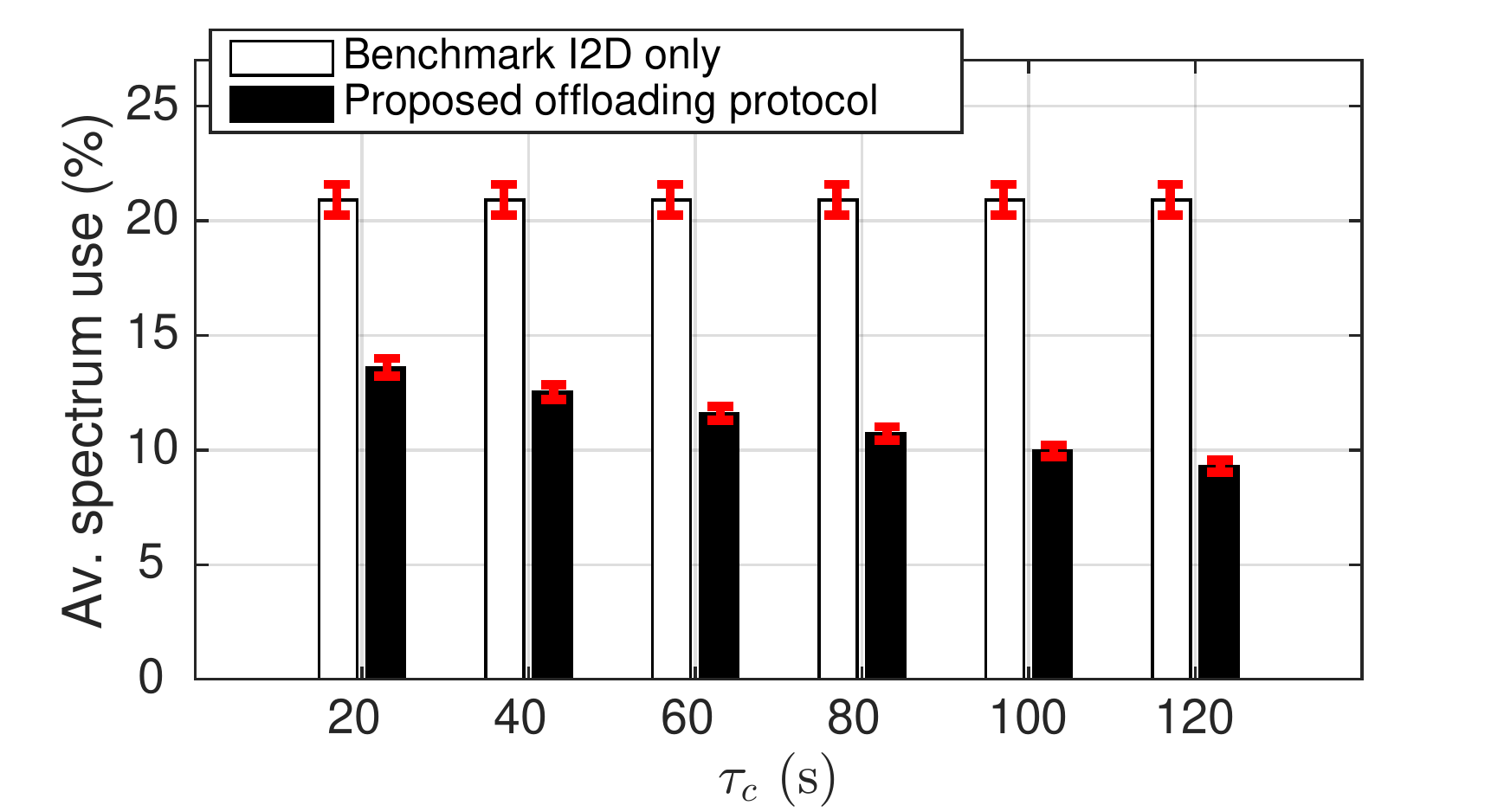}
\par\end{centering}
}\hspace{0.04\columnwidth}\subfloat[{With different values of $r_{\max}^{\text{(D2D)}}$ (and fixed parameters
$\tau_{c}=20$~s and $[v_{\min},v_{\max}]=[9,24]$~m/s)}]{\begin{centering}
\includegraphics[width=0.45\columnwidth]{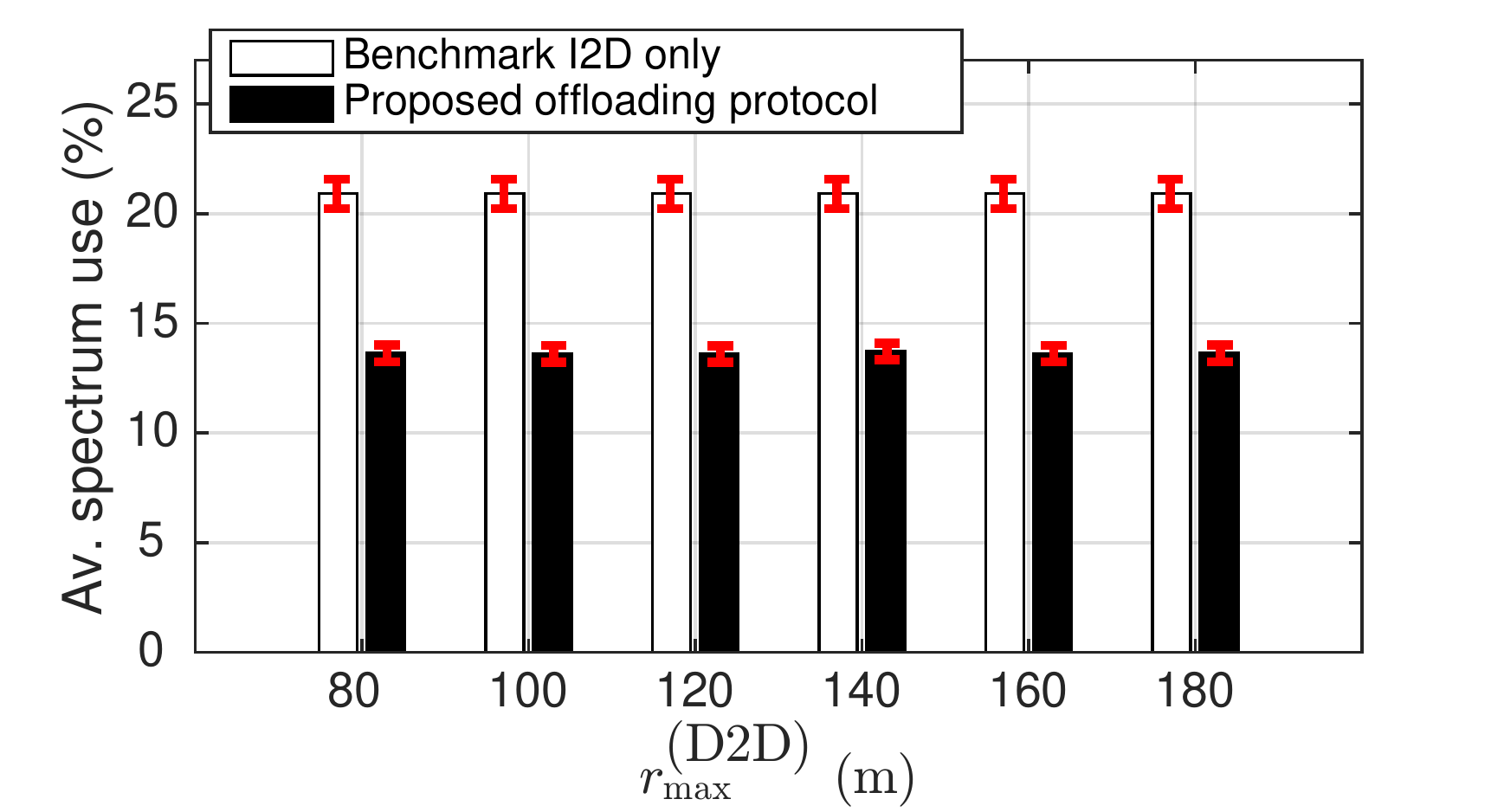} 
\par\end{centering}
}
\par\end{centering}
\centering{}\subfloat[{With different values of $[v_{\min},v_{\max}]$ (and fixed parameters
$\tau_{c}=20\text{ s}$ and $r_{\max}^{\text{(D2D)}}=100\text{ m}$)}]{\begin{centering}
\includegraphics[width=0.45\textwidth]{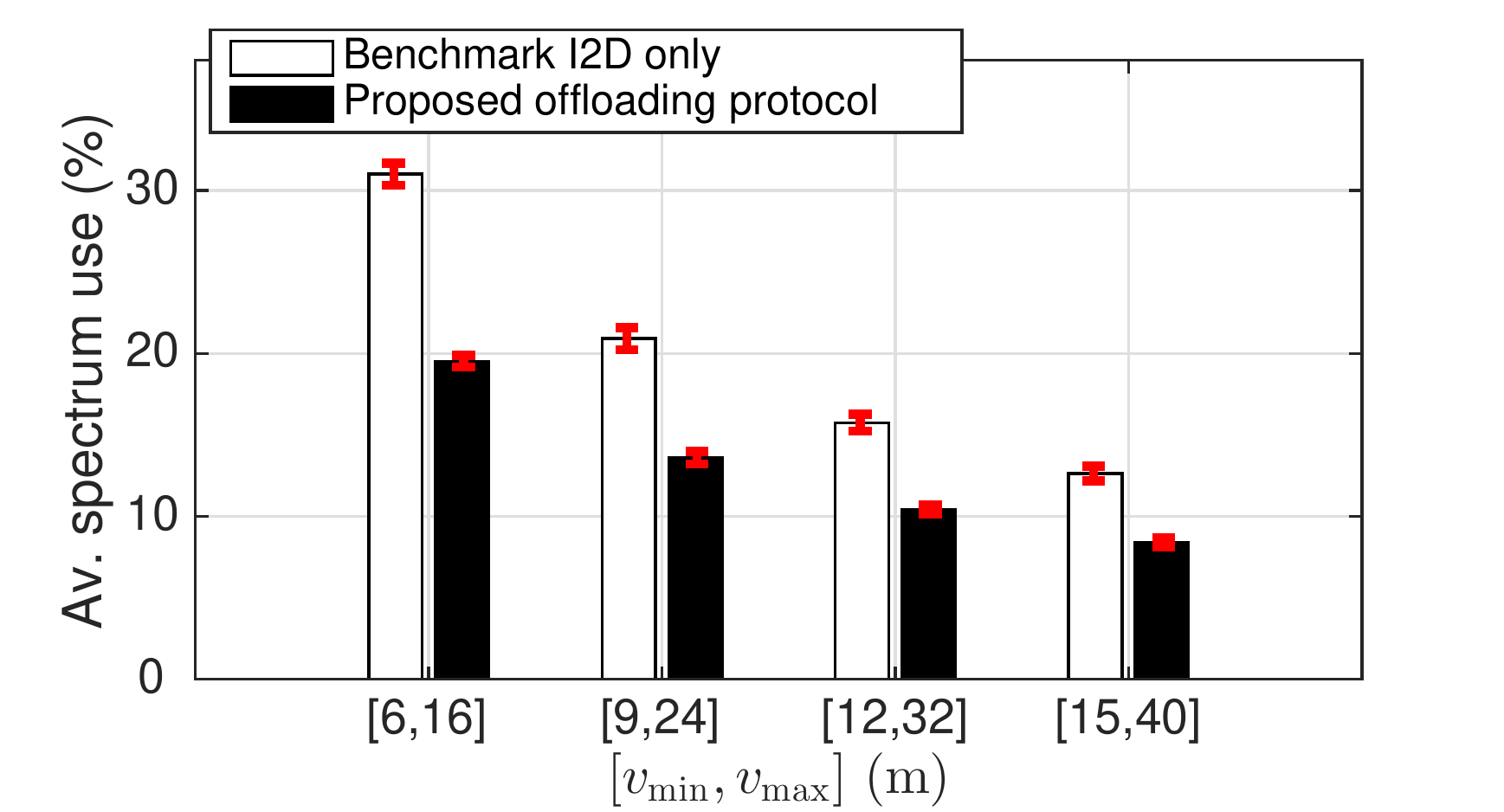} 
\par\end{centering}
}\hspace{0.04\columnwidth}\subfloat[{With different values of $[v_{\min},v_{\max}]$ (and fixed parameters
$\tau_{c}=60\text{ s}$ and $r_{\max}^{\text{(D2D)}}=100\text{ m}$)}]{\begin{centering}
\includegraphics[width=0.45\textwidth]{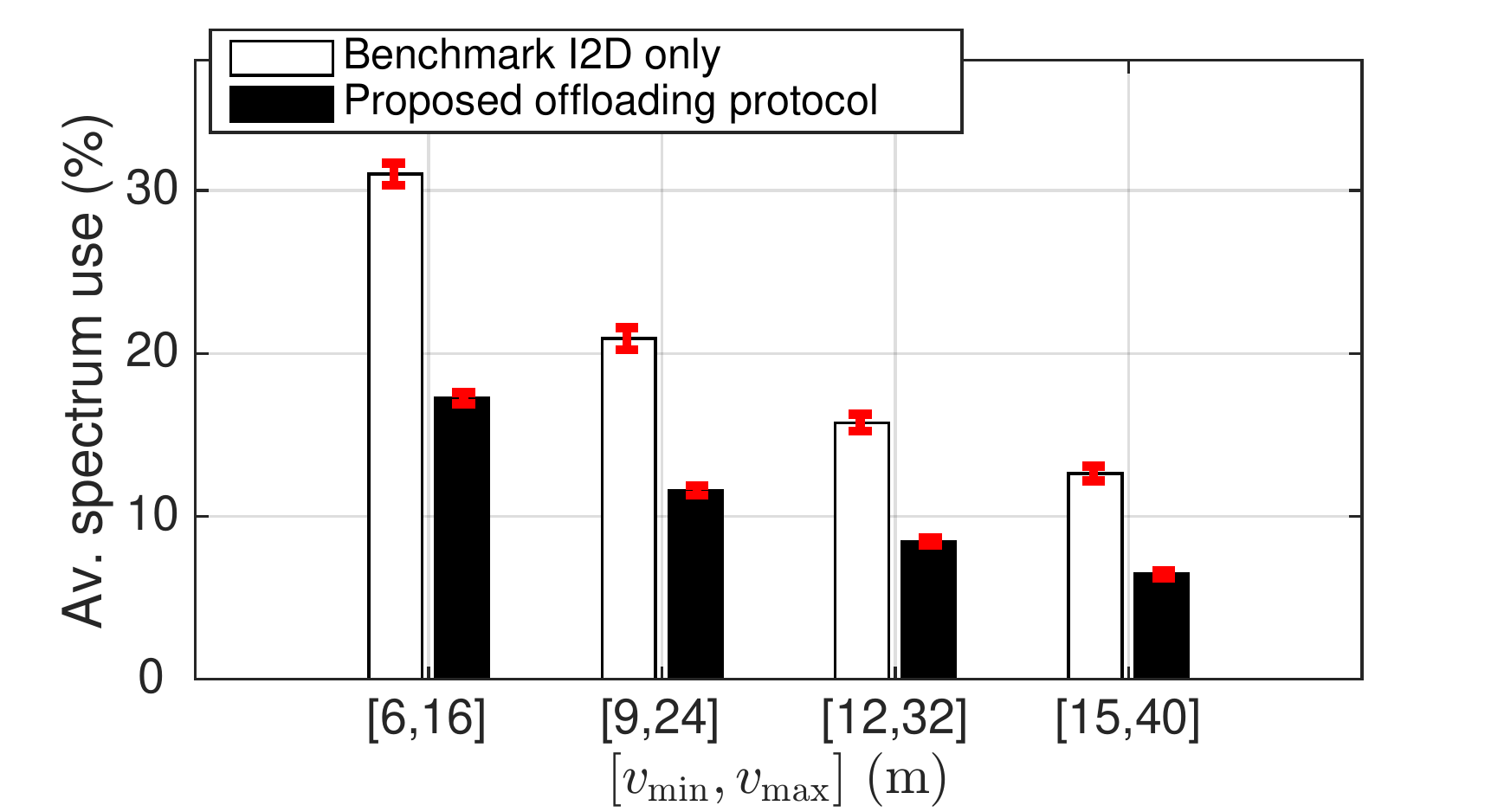} 
\par\end{centering}
}\vspace*{-2mm}
 \caption{Average spectrum use percentage of the benchmark protocols (plain
cellular and D2D offloading) and the proposed protocol}
\label{FIG_RES_7_Average_Spectrum_Use}\vspace*{-2mm}
 
\end{figure*}

Figure~\ref{FIG_RES_7_Average_Spectrum_Use} shows the average spectrum
occupation percentage (with confidence intervals) of the three considered
systems. The trends are similar to those observed for the energy consumption,
although the proportions of absolute and relative gains are different.
The offered traffic requires a spectrum occupation, for the plain
cellular system, of 26\% of the available radio resources, for the
scenario with speed range {[}6,16{]}~m/s. Increasing the speed range,
the traffic load decreases, and the spectrum occupation follows the
decrease (Subfigures~\ref{FIG_RES_7_Average_Spectrum_Use}.c and
\ref{FIG_RES_7_Average_Spectrum_Use}.d). The D2D offloading systems
succeed in using only around 20\% of the resources for the scenario
with speed range {[}6,16{]}~m/s, and the percentage decreases coherently
with increasing speed ranges. As observable in Subfigure~\ref{FIG_RES_7_Average_Spectrum_Use}.a,
and by the comparison of Subfigures~\ref{FIG_RES_7_Average_Spectrum_Use}.c
and \ref{FIG_RES_7_Average_Spectrum_Use}.d, for the spectrum use,
too, the critical parameter is the content timeout. The intuitive
reason is that shorter transmission distances allow for reusing the
same PRBs more frequently in the spatial dimension. The D2D systems
succeed in using less than 15\% of the spectrum in most of the cases,
dropping below 10\% in the most favorable conditions of $\tau_{c}=120\,\text{s}$.
\begin{figure*}[!t]
\begin{centering}
\subfloat[{With different values of $\tau_{c}$ (and fixed parameters $r_{\max}^{\text{(D2D)}}=100$~m
and $[v_{\min},v_{\max}]=[9,24]$~m/s)}]{\begin{centering}
\includegraphics[width=0.45\columnwidth]{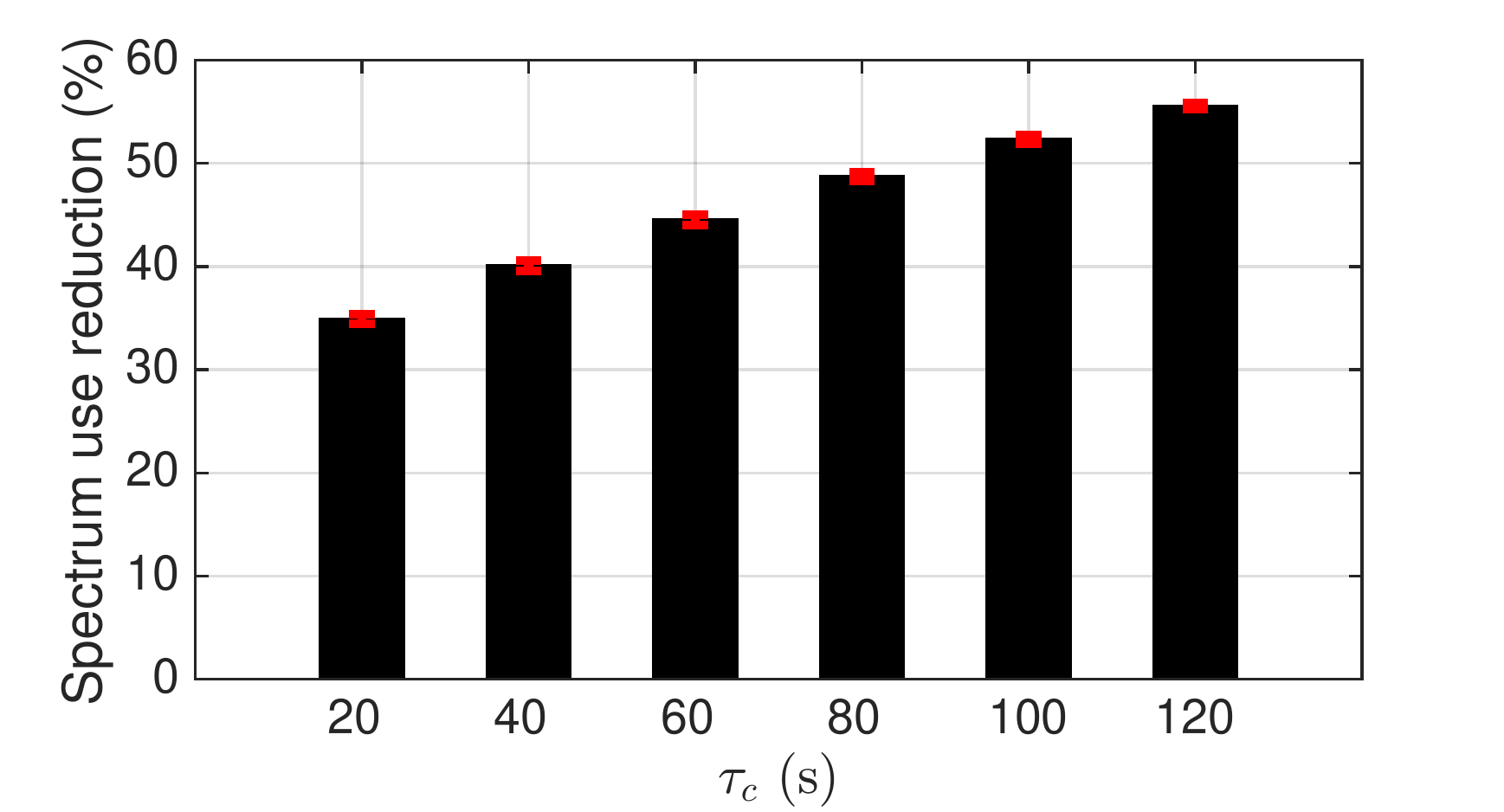}
\par\end{centering}
}\hspace{0.04\columnwidth}\subfloat[{With different values of $r_{\max}^{\text{(D2D)}}$ (and fixed parameters
$\tau_{c}=20$~s and $[v_{\min},v_{\max}]=[9,24]$~m/s)}]{\begin{centering}
\includegraphics[width=0.45\columnwidth]{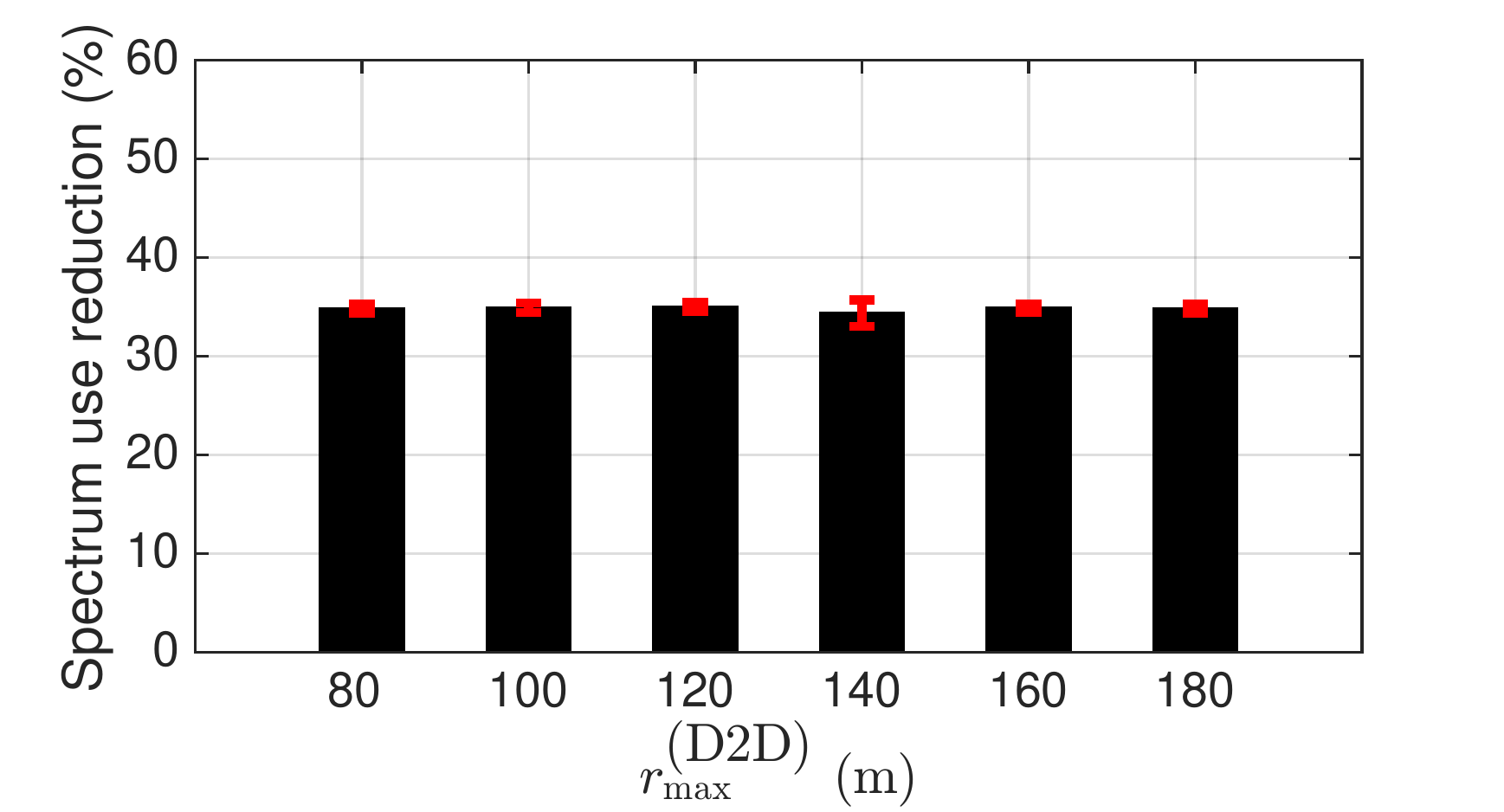} 
\par\end{centering}
}
\par\end{centering}
\centering{}\subfloat[{With different values of $[v_{\min},v_{\max}]$ (and fixed parameters
$\tau_{c}=20\text{ s}$ and $r_{\max}^{\text{(D2D)}}=100\text{ m}$)}]{\begin{centering}
\includegraphics[width=0.45\textwidth]{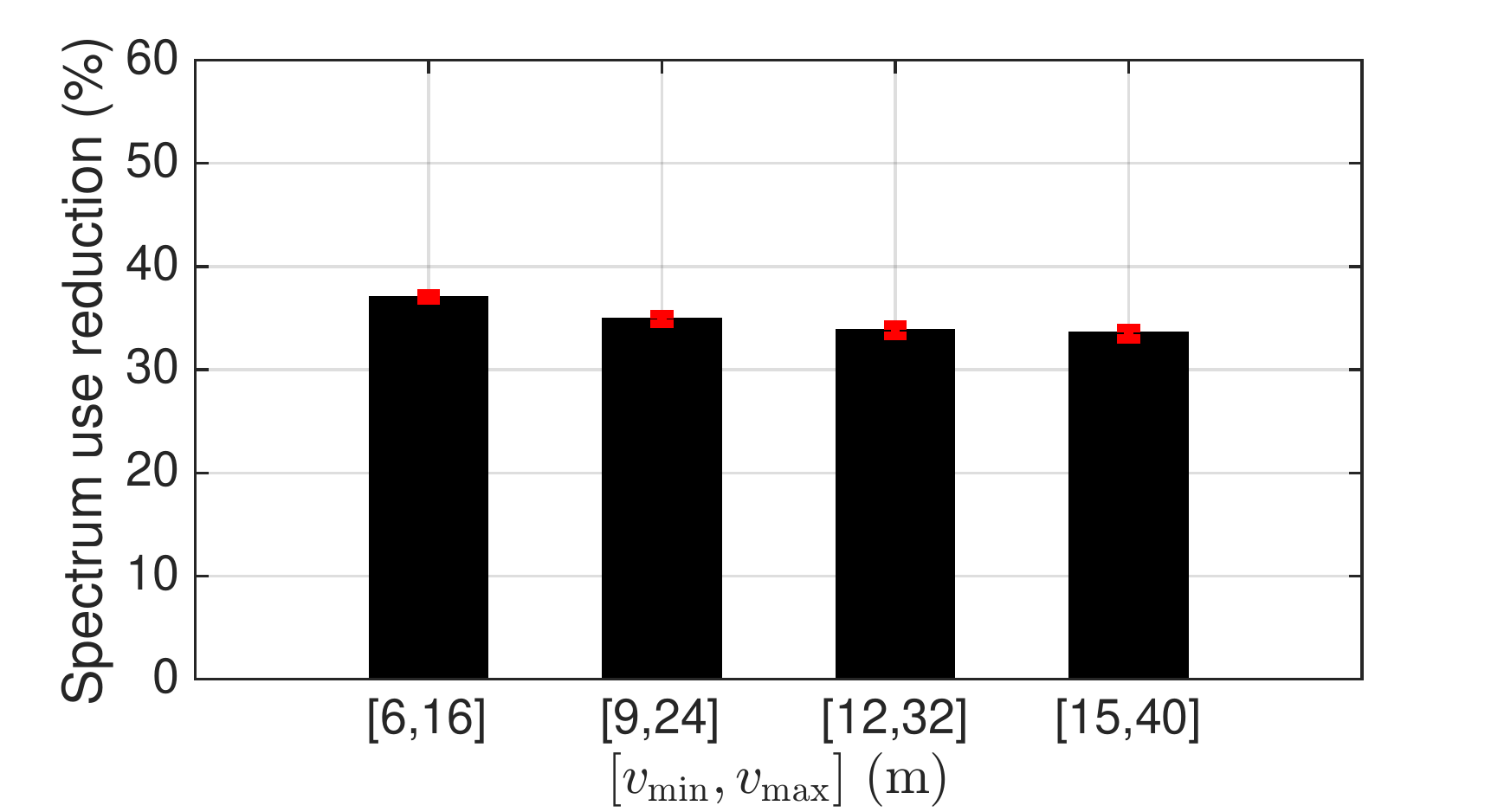} 
\par\end{centering}
}\hspace{0.04\columnwidth}\subfloat[{With different values of $[v_{\min},v_{\max}]$ (and fixed parameters
$\tau_{c}=60\text{ s}$ and $r_{\max}^{\text{(D2D)}}=100\text{ m}$)}]{\begin{centering}
\includegraphics[width=0.45\textwidth]{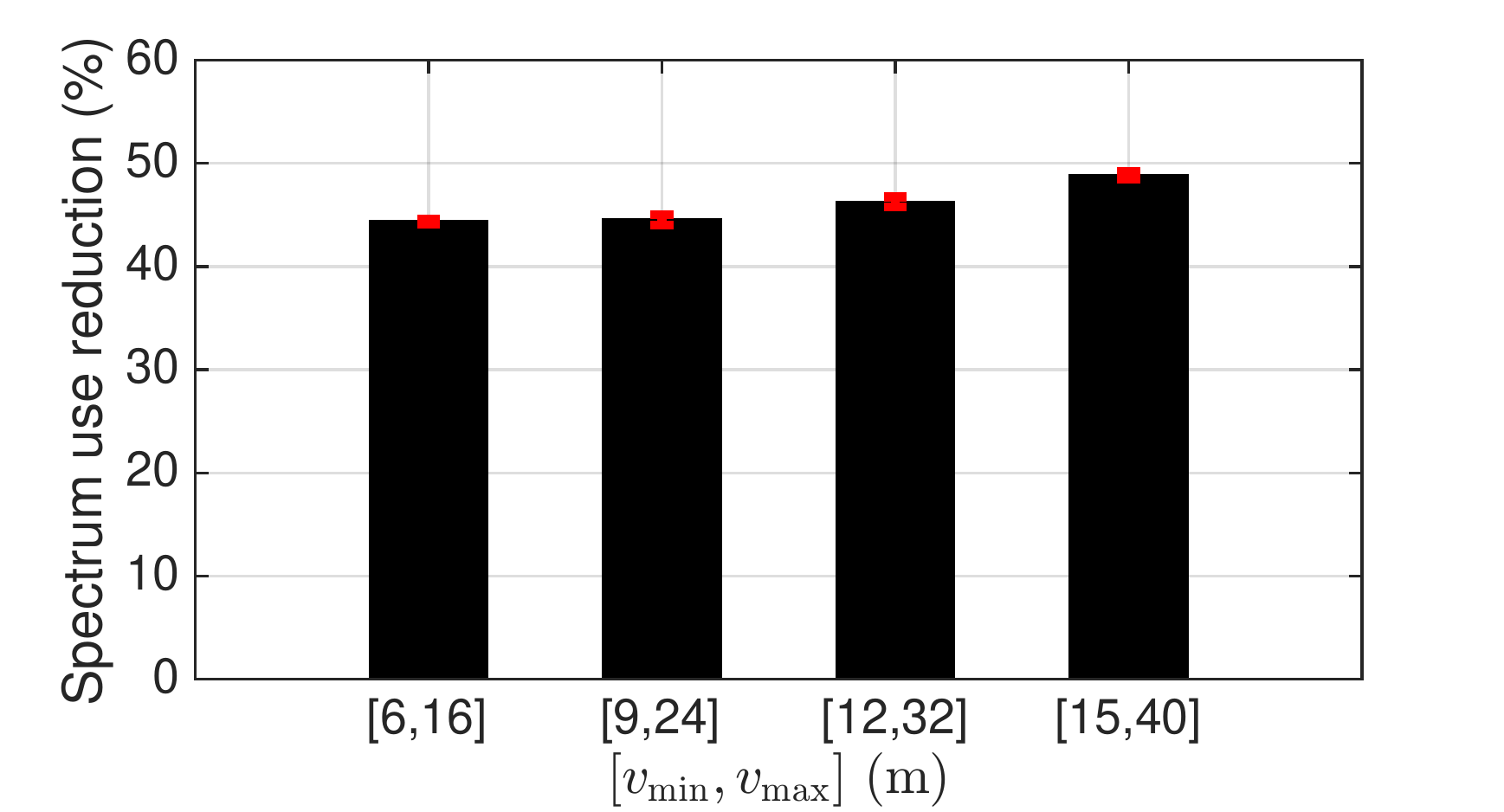} 
\par\end{centering}
}\vspace*{-2mm}
 \caption{Reduction percentage of the average spectrum use of the proposed CDMS
with respect to the cellular system.}
\label{FIG_RES_8_Average_Spectrum_Use_GAIN_AB_vs_0}\vspace*{-2mm}
 
\end{figure*}
Figure~\ref{FIG_RES_8_Average_Spectrum_Use_GAIN_AB_vs_0} shows the
percentage reduction of the spectrum occupation obtained by the D2D
system (benchmark and proposed one) against the plain cellular benchmark
system. The reduction is always above 30\%, on average 40\%, and peaking
to 50\% in the most favorable conditions. With respect to the benchmark
system, in our simulations, we observed a reduction mostly in the
range of 4 to 6\%.

Commenting our results on spectrum occupation, we remark that they
are closely related the specific implementation of the RRRM component
we used. Particularly, one aspect that the considered RRRM does not
optimize the selection of the input transmit power of the concurrent
links. In fact, the transmit power is set to satisfy a constraint
which is only function of the channel in the considered link. Furthermore,
the considered RRRM allocates the resources to I2D and D2D links in
a shared way (i.e., I2D links have no dedicated resources). It may
be the case that this coexistence prevents to fully exploit the shorter
D2D transmission distances achieved with the proposed scheme, thus
limiting the gain in terms of spectrum use. Thus, although the reduction
in spectrum use is already relevant, we believe that by using an evolved
RRRM component, which optimizes the input transmit power of the concurrent
links jointly, may result in a further performance improvement in
terms of spatial spectrum reuse. This aspect will be considered in
our future works.

\section{Conclusion\label{sec:Conclusion}}

We have proposed a content delivery management system for D2D data
offloading in cellular networks tailored to scenarios, such as vehicular
networks, where the topology varies at a fast rate, and to delay-tolerant
applications. The proposed system exploits the availability of nodes
mobility predictions at the CDMS. We have derived an analytical model
able to predict the system performance in terms of the statistics
of the D2D transmission range and the energy consumption. The analytical
model allows to rapidly evaluate the system performance in a variety
of scenarios larger than that allowed through system-level simulations
which, with the involvement of hundreds of nodes, and the MAC and
channel model implementation details, may require a very large time.

We have evaluated the system level performance using an accurate system
level simulator which includes a radio resource reuse scheme for allocating
resources over a time-frequency radio resource grid, and incorporates
a quite detailed channel model including small scale frequency selective
fading. The proposed system, in which the D2D transmission instant
is selected to minimize the transmission range, allows energy savings
at the system level (including I2D and D2D transmissions) ranging
between 30\% and 80\%, depending on the scenario parameters, with
respect to the benchmark cellular system, and mostly in the 5\%-20\%
range with respect to the D2D offloading benchmark system. However,
considering the sole energy consumed by the devices for operating
with any of the two considered D2D offloading systems (benchmark and
proposed one), the proposed system outperforms the benchmark with
a reduction of around 90\% of spent energy for transmission in most
of the considered settings, peaking at 97\% when the delay tolerance
is 2 minutes, which is a reduction of almost two orders of magnitude.
In terms of spectrum occupation, the proposed system uses an amount
of spectrum resources (for the considered configurations) 30\% to
40\% less than the plain cellular system, and up to 5\% less then
the benchmark D2D offloading system.

We emphasize that, since the energy consumption of the devices is
one of the major concerns in the evaluation of the worthiness of deploying
this kind of solutions, a performance comparison in terms of the enrgy
consumed by the devices, is the most appropriate, since this specific
metric can make a real difference in determining if a system is worth
deploying or not.\vspace{-4mm}

\section*{Acknowledgement\vspace{-2mm}
}

This work has been partially funded by the EC under the H2020 REPLICATE
(691735), SoBigData (654024) and AUTOWARE (723909) projects.

\section*{References}
\begin{center}
\bibliographystyle{elsarticle-num}

\par\end{center}

\appendixpage

\begin{appendices}

\section{Statistics of the optimal D2D transmission from a single PCP\label{subsec:APP_Theroem_1}}

\begin{proofofT}{\ref{thm:1}}

To compute of the PDF of the closest distance from a device requesting
a content achievable by a PCP within the time limit, it is convenient
to introduce an auxiliary random variable $\Delta$ defined as the
signed ``displacement'' of the optimal \emph{relative} position
of the PCP (the one at which it should transmit the content, if it
was selected for doing it) with respect to its original position $X_{0}$.
Coherently with this perspective, the optimal relative (i.e. referred
to a coordinate system integral with vehicle A motion) position for
transmission is a random variable $X^{*}=X_{0}+\Delta$. In the following,
we shall use the symbols $x_{0}$ and $x^{*}$ to indicate the realizations
of $X_{0}$ and $X^{*}$. We indicate the PDF of the random variable
$\Delta$, conditioned on the initial position of the PCP, $X_{0}$,
distinguishing between two possible cases for the conditioning realization
$x_{0}$. Specifically, we use $p_{\Delta\mid X_{0}^{+},V_{A}^{*}}\left(\delta\mid x_{0}\right)$
for the case that $x_{0}>0$, and $p_{\Delta\mid X_{0}^{-},V_{A}^{*}}\left(\delta\mid x_{0},v_{a}^{*}\right)$
for the case $x_{0}<0$. The case for $x_{0}=0$ boils down to simply
transmitting the content immediately, since it means that the PCP
is already at the closest distance, and any delay will result in an
increased distance. Accordingly, if $x_{0}=0$ the PDF of $\Delta$
is a Dirac pulse with unit mass concentrated at $r=0$.

\paragraph{Case for PCP ahead of requesting vehicle A: $x_{0}>0$}

First of all, intuition suggest that, if $x_{0}>0$, the optimal signed
displacement $\Delta$ of vehicle B (with respect to its original
position $x_{0}\triangleq x\left(t_{0}\right)$) \emph{evaluated in
a coordinate system integral with vehicle A's motion}, be either negative
or, at most, null. Specifically, we will show that, for this case,
$\Delta$ has the following properties:
\begin{lyxlist}{00.00.0000}
\item [{(i)}] With a finite positive probability, $\Delta$ is equal to
$0$, Specifically,\small{
\begin{align}
\mathbb{P}\left(\Delta=0\mid X_{0}=x_{0},V_{A}^{*}\right)\Bigr|_{x_{0}>0}=\int_{0}^{\infty}f\left(v,v_{a}^{*}\right)dv.\label{eq:Lem1_i}
\end{align}
}
\item [{(ii)}] With a finite positive probability, $\Delta$ is equal to
$-x_{0}$. Specifically,\small{
\begin{align}
\small{\mathbb{P}\left(\Delta=-x_{0}\mid X_{0}=x_{0},V_{A}^{*}\right)\Bigr|{}_{x_{0}>0}=\int_{-\infty}^{-x_{0}/\tau_{c}}\hspace{-5mm}f\left(v,v_{a}^{*}\right)dv-\frac{x_{0}}{\tau_{s}}\int_{-\infty}^{-x_{0}/\tau_{c}}\hspace{-3mm}f\left(v,v_{a}^{*}\right)\frac{1}{\left|v\right|}dv}.\label{eq:Lem1_ii}
\end{align}
}
\item [{(iii)}] For values of the realization $\delta$ in the open interval
$\left(-x_{0},0\right)$, the PDF of $\Delta$ has the following expression\small{
\begin{align}
\small{\left.p_{\Delta\mid X_{0}^{+},V_{A}^{*}}(\delta\mid x_{0},v_{a}^{*})\right|_{\delta\in(-x_{0},0)}=\frac{1}{\tau_{s}}\int_{-\tau_{c}}^{0}\hspace{-3mm}f\left(-\frac{\delta}{z},v_{a}^{*}\right)\frac{1}{\left|z\right|}dz+\left(\frac{1}{\tau_{c}}-\frac{1}{\tau_{s}}\right)f\left(\frac{\delta}{\tau_{c}},v_{a}^{*}\right)} & .\label{eq:Lem1_iii}
\end{align}
}
\end{lyxlist}
In fact, consider the three cases listed on page~\vpageref{enu:Vehicle-B-is-1}.
Under our assumptions, if the distance between B and A is constant
or increasing (case~\ref{enu:Vehicle-B-is-1} in the list), i.e.,
$V\geq0$, the optimal instant for transmitting is just $t_{0}$,
and hence $\Delta=0$. Therefore, the probability of having $\Delta=0$
is given by the probability that $V$ is zero or positive, as in Eq.~\eqref{eq:Lem1_i}.
If the distance between the two vehicles is decreasing (hence, by
our convention, $V<0$) but they will not achieve the same location
within the expiration of either the content or they sharing timeout
(case~\ref{enu:Vehicle-B-is-2} in the list), the displacement (in
the coordinate system integral with vehicle A) of vehicle B with respect
to its original position $x_{0}$ is determined by the (relative)
space travelled during an interval of duration $\Phi$. Finally, if
vehicles A and B will reach the same location before the expiration
of any of the timeouts (case~\ref{enu:Vehicle-B-is-3} in the list),
the displacement is simply given by the opposite (additive inverse)
value of the original position $x_{0}$.

Summarizing, the three cases can be easily mapped to the following
properties for the random variable $\Delta$:
\begin{equation}
\Delta=\begin{cases}
0 & \text{if }V\geq0\\
V\Phi & \text{if }V<0\text{\,and\,}V\Phi>-x_{0}\\
-x_{0} & \text{if }V<0\text{\,and\,}V\Phi\leq-x_{0}
\end{cases}\,.
\end{equation}

Eq.~\eqref{eq:Lem1_ii} can be obtained through the following steps:

\begin{align*}
\mathbb{P}\left(\Delta=-x_{0}\right)= & \mathbb{P}\left(V\Phi\leq-x_{0},V<0\right)\\
= & \mathbb{P}\left(\Phi\geq\frac{x_{0}}{\left|V\right|},V<0\right)\\
= & \int_{-\infty}^{0}f\left(v,v_{a}^{*}\right)\int_{x_{0}/\left|v\right|}^{\infty}p_{\Phi}(\phi)d\phi dv\\
= & \int_{-\infty}^{0}f\left(v,v_{a}^{*}\right)\int_{x_{0}/\left|v\right|}^{\infty}\frac{1}{\tau_{s}}u_{[0,\tau_{c}]}(\phi)+\left(1-\frac{\tau_{c}}{\tau_{s}}\right)u_{0}\left(\phi-\tau_{c}\right)d\phi dv\\
\hphantom{\mathbb{P}\left(\Delta=-x_{0}\right)}= & \int_{-\infty}^{0}f\left(v,v_{a}^{*}\right)\left(\frac{1}{\tau_{s}}\int_{x_{0}/\left|v\right|}^{\tau_{c}}d\phi\right.\\
 & \left.+\left(1-\frac{\tau_{c}}{\tau_{s}}\right)\int_{x_{0}/\left|v\right|}^{\infty}u_{0}\left(\phi-\tau_{c}\right)d\phi\right)u_{(-\infty,-x_{0}/\tau_{c}]}(v)dv\\
= & \int_{-\infty}^{-x_{0}/\tau_{c}}f\left(v,v_{a}^{*}\right)\left(\frac{1}{\tau_{s}}\left(\tau_{c}-\frac{x_{0}}{\left|v\right|}\right)+\left(1-\frac{\tau_{c}}{\tau_{s}}\right)\right)dv\\
\hphantom{\mathbb{P}\left(\Delta=-x_{0}\right)}= & \left(\frac{\tau_{c}}{\tau_{s}}+1-\frac{\tau_{c}}{\tau_{s}}\right)\int_{-\infty}^{-x_{0}/\tau_{c}}f\left(v,v_{a}^{*}\right)dv-\frac{x_{0}}{\tau_{s}}\int_{-\infty}^{-x_{0}/\tau_{c}}f\left(v,v_{a}^{*}\right)\frac{1}{\left|v\right|}dv\\
= & \int_{-\infty}^{-x_{0}/\tau_{c}}f\left(v,v_{a}^{*}\right)dv-\frac{x_{0}}{\tau_{s}}\int_{-\infty}^{-x_{0}/\tau_{c}}f\left(v,v_{a}^{*}\right)\frac{1}{\left|v\right|}dv.
\end{align*}
Eq.~\eqref{eq:Lem1_iii}, which refers to the case for negative relative
speed values (vehicles getting closer to each other), but not sufficiently
large to let vehicle B reach vehicle A, is derived as follows. Let
us indicate the CDF of $\Delta$ as $F_{\Delta}(\delta)\triangleq\mathbb{P}\left(\Delta\leq\delta\right)$,
then we have
\begin{align*}
\left.F_{\Delta}(\delta)\right|_{\delta\in(-x_{0},0)}= & \left.\mathbb{P}\left(\Delta\leq\delta\right)\right|_{\delta\in(-x_{0},0)}\\
= & \mathbb{P}\left(V\Phi\leq-x_{0},V<0\right)+\mathbb{P}\left(-x_{0}<V\Phi\leq\delta,V<0\right)\\
 & \mathbb{P}\left(\Delta=-x_{0}\right)+\mathbb{P}\left(-x_{0}<V\Phi\leq\delta,V<0\right)\\
= & \mathbb{P}\left(\Delta=-x_{0}\right)+\mathbb{P}\left(\frac{\delta}{V}<\Phi\leq\frac{-x_{0}}{V},V<0\right)\\
= & \mathbb{P}\left(\Delta=-x_{0}\right)+\int_{-\infty}^{0}f\left(v,v_{a}^{*}\right)\int_{\delta/v}^{-x_{0}/v}p_{\Phi}(\phi)d\phi dv
\end{align*}

Taking the derivative of $F_{\Delta}(\delta)$ with respect to $\delta$
we obtain

\begin{align*}
\left.p_{\Delta}(\delta)\right|_{\delta\in(-x_{0},0)}= & \frac{d}{d\delta}\left.F_{\Delta}(\delta)\right|_{\delta\in(-x_{0},0)}\\
= & \int_{-\infty}^{0}f\left(v,v_{a}^{*}\right)\frac{d}{d\delta}\int_{\delta/v}^{-x_{0}/v}p_{\Phi}(\phi)d\phi dv\\
= & \int_{-\infty}^{0}f\left(v,v_{a}^{*}\right)\frac{1}{v}\frac{d}{d(\delta/v)}\left(\int_{\delta/v}^{-x_{0}/v}p_{\Phi}\left(\phi\right)d\phi\right)dv\\
= & \int_{-\infty}^{0}f\left(v,v_{a}^{*}\right)\frac{1}{v}(-1)p_{\Phi}\left(\delta/v\right)dv\\
\hphantom{\left.p_{\Delta}(\delta)\right|_{\delta\in(-x_{0},0)}}= & \int_{-\infty}^{0}f\left(v,v_{a}^{*}\right)\frac{1}{\left|v\right|}\left(\frac{1}{\tau_{s}}u_{[0,\tau_{c}]}\left(\frac{\delta}{v}\right)+\left(1-\frac{\tau_{c}}{\tau_{s}}\right)u_{0}\left(\frac{\delta}{v}-\tau_{c}\right)\right)dv\\
 & \frac{1}{\tau_{s}}\int_{-\infty}^{\delta/\tau_{c}}f\left(v,v_{a}^{*}\right)\frac{1}{\left|v\right|}dv\\
 & +\left(1-\frac{\tau_{c}}{\tau_{s}}\right)\int_{-\infty}^{0}f\left(v,v_{a}^{*}\right)\frac{1}{\left|v\right|}u_{0}\left(\frac{\delta}{v}-\tau_{c}\right)dv\\
= & \frac{1}{\tau_{s}}\int_{-\infty}^{\delta/\tau_{c}}p_{V}(v)\frac{1}{\left|v\right|}dv\\
 & +\left(1-\frac{\tau_{c}}{\tau_{s}}\right)\int_{0}^{-\infty}f\left(-\frac{\delta}{z},v_{a}^{*}\right)\left|\frac{z}{\delta}\right|u_{0}\left(-z-\tau_{c}\right)\frac{\delta}{z^{2}}dz\\
= & \frac{1}{\tau_{s}}\int_{-\infty}^{\delta/\tau_{c}}f\left(v,v_{a}^{*}\right)\frac{1}{\left|v\right|}dv\\
 & +\left(1-\frac{\tau_{c}}{\tau_{s}}\right)(-1)\int_{-\infty}^{0}f\left(-\frac{\delta}{z},v_{a}^{*}\right)\left(-\frac{1}{\left|z\right|}\right)u_{0}\left(z+\tau_{c}\right)dz\\
\vdots
\end{align*}
\\
\begin{align*}
\vdots\\
= & \frac{1}{\tau_{s}}\int_{-\infty}^{\delta/\tau_{c}}f\left(v,v_{a}^{*}\right)\frac{1}{\left|v\right|}dv+\left(1-\frac{\tau_{c}}{\tau_{s}}\right)p_{V}\left(\frac{\delta}{\tau_{c}}\right)\frac{1}{\tau_{c}}\\
= & \frac{1}{\tau_{s}}\int_{-\infty}^{\delta/\tau_{c}}f\left(v,v_{a}^{*}\right)\frac{1}{\left|v\right|}dv+\left(\frac{1}{\tau_{c}}-\frac{1}{\tau_{s}}\right)p_{V}\left(\frac{\delta}{\tau_{c}}\right)\\
\hphantom{\left.p_{\Delta}(\delta)\right|_{\delta\in(-x_{0},0)}}= & \frac{1}{\tau_{s}}\int_{-\infty}^{0}f\left(v,v_{a}^{*}\right)\frac{1}{\left|v\right|}u_{(-\infty,\delta/\tau_{c}]}\left(v\right)dv\\
 & +\left(1-\frac{\tau_{c}}{\tau_{s}}\right)\int_{-\infty}^{0}f\left(v,v_{a}^{*}\right)\frac{1}{\left|v\right|}u_{0}\left(\frac{\delta}{v}-\tau_{c}\right)dv\\
\hphantom{\left.p_{\Delta}(\delta)\right|_{\delta\in(-x_{0},0)}}= & \frac{1}{\tau_{s}}\int_{-\infty}^{\delta/\tau_{c}}f\left(v,v_{a}^{*}\right)\frac{1}{\left|v\right|}dv\\
 & +\left(1-\frac{\tau_{c}}{\tau_{s}}\right)\int_{0}^{\infty}f\left(\frac{\delta}{z},v_{a}^{*}\right)\left|\frac{z}{\delta}\right|\left(-\frac{\delta}{z^{2}}\right)u_{0}\left(z-\tau_{c}\right)dz\\
\hphantom{\left.p_{\Delta}(\delta)\right|_{\delta\in(-x_{0},0)}}= & \frac{1}{\tau_{s}}\int_{-\infty}^{\delta/\tau_{c}}f\left(v,v_{a}^{*}\right)\frac{1}{\left|v\right|}dv\\
 & +\left(1-\frac{\tau_{c}}{\tau_{s}}\right)\int_{0}^{\infty}f\left(\frac{\delta}{z},v_{a}^{*}\right)\left|\frac{1}{z}\right|u_{0}\left(z-\tau_{c}\right)dz\\
= & \frac{1}{\tau_{s}}\int_{-\infty}^{\delta/\tau_{c}}f\left(v,v_{a}^{*}\right)\frac{1}{\left|v\right|}dv+\left(1-\frac{\tau_{c}}{\tau_{s}}\right)f\left(\frac{\delta}{\tau_{c}},v_{a}^{*}\right)\frac{1}{\tau_{c}}\\
= & \frac{1}{\tau_{s}}\int_{-\infty}^{\delta/\tau_{c}}f\left(v,v_{a}^{*}\right)\frac{1}{\left|v\right|}dv+\left(\frac{1}{\tau_{c}}-\frac{1}{\tau_{s}}\right)f\left(\frac{\delta}{\tau_{c}},v_{a}^{*}\right).
\end{align*}

Combining the Eq.~\eqref{eq:Lem1_i} through \eqref{eq:Lem1_iii},
we obtain
\begin{align}
 & \hspace{-3mm}p_{\Delta\mid X_{0}^{+},V_{A}^{*}}\left(\delta\mid x_{0},v_{a}^{*}\right)=\label{eq:APPENDIX_PDF_DELTA_X0_POS}\\
= & u_{0}\left(\delta+x_{0}\right)\left(\int_{-\infty}^{-x_{0}/\tau_{c}}f\left(v,v_{a}^{*}\right)dv-\frac{x_{0}}{\tau_{s}}\int_{-\infty}^{-x_{0}/\tau_{c}}f\left(v,v_{a}^{*}\right)\frac{1}{\left|v\right|}dv\right)\nonumber \\
 & +u_{(-x_{0},0)}\left(\delta\right)\left(\frac{1}{\tau_{s}}\int_{-\infty}^{\delta/\tau_{c}}f\left(v,v_{a}^{*}\right)\frac{1}{\left|v\right|}dv+\left(\frac{1}{\tau_{c}}-\frac{1}{\tau_{s}}\right)f\left(\frac{\delta}{\tau_{c}},v_{a}^{*}\right)\right)\nonumber \\
 & +u_{0}\left(\delta\right)\int_{0}^{\infty}f\left(v,v_{a}^{*}\right)dv\nonumber 
\end{align}
which is easy to compute, either analytically or numerically, once
a specific PDF for the relative speed $V$, $f\left(v,v_{a}^{*}\right)$,
is specified.

We observe that, in the case (considered above) that vehicle B is
in the half-line ahead of vehicle A's motion, the signed displacement
$\Delta$ can take either negative or null values, whereas the original
position of vehicle B, $x_{0}$, and its trajectory in the coordinate
system integral with vehicles A's motion, \eqref{eq:relative_trajectory},
always take positive or null values. Therefore, the expression of
the trajectory also gives the distance between the two vehicles across
time. The same holds for the optimal position $X^{*}=X_{0}+\Delta$,
which, in the case $x_{0}>0$, coincides with the optimal D2D transmission
distance.

\paragraph{Case for PCP behind the requesting vehicle A: $x_{0}<0$}

Considering, now, the case that vehicle B is in the half-line behind
vehicle A's motion, we have that $X_{0}$ and $X^{*}$ can only take
negative or null values, and the signed displacement $\Delta$ can
only take either positive or null values. Furthermore (see footnote
2), $v>0$ if the vehicles are getting closer to each other, and $v<0$
if they are getting farther. Using a line of reasoning similar to
the one used above, we obtain that, in function of the realization
$x_{0}$ of the original position of vehicle B $X_{0}$,
\begin{equation}
\Delta=\begin{cases}
0 & \text{if }V\leq0\\
V\Phi & \text{if }V>0\text{\,and\,}V\Phi<\left|x_{0}\right|\\
-x_{0} & \text{if }V>0\text{\,and\,}V\Phi\geq\left|x_{0}\right|
\end{cases}\,.
\end{equation}
In this case, however, the optimal D2D transmission distance is given
by the opposite of the (now negative) optimal relative position $X^{*}$.
With derivations similar to those presented in Appendix~\ref{subsec:APP_Theroem_1},
the following expression can be obtained for the displacement in the
case that vehicle B, at the request time, lies on the half-line behind
vehicle A:

\begin{align}
 & \hspace{-3mm}p_{\Delta\mid X_{0}^{-},V_{A}^{*}}\left(\delta\mid x_{0},v_{a}^{*}\right)=\label{eq:APPENDIX_PDF_DELTA_X0_NEG}\\
= & u_{0}\left(\delta\right)\int_{-\infty}^{0}f\left(v,v_{a}^{*}\right)dv\nonumber \\
 & +u_{(0,-x_{0})}\left(\delta\right)\left(\frac{1}{\tau_{s}}\int_{\delta/\tau_{c}}^{\infty}f\left(v,v_{a}^{*}\right)\frac{1}{\left|v\right|}dv+\left(\frac{1}{\tau_{c}}-\frac{1}{\tau_{s}}\right)f\left(\frac{\delta}{\tau_{c}},v_{a}^{*}\right)\right)\nonumber \\
 & +u_{0}\left(\delta+x_{0}\right)\left(\int_{-x_{0}/\tau_{c}}^{\infty}f\left(v,v_{a}^{*}\right)dv-\frac{\left|x_{0}\right|}{\tau_{s}}\int_{-x_{0}/\tau_{c}}^{\infty}f\left(v,v_{a}^{*}\right)\frac{1}{\left|v\right|}dv\right).\nonumber 
\end{align}

\paragraph{Minimal distance between the two vehicles within the effective time
limit}

Now, the relationship of the variable $R$ with $\Delta$ can be summarized
as follows.
\begin{align}
R & =\Bigl|X_{0}\Bigr|-\Bigl|\Delta\Bigr|=\begin{cases}
X_{0}+\Delta & \text{if\,}X_{0}>0\\
0 & \text{if\,}X_{0}=0\\
-X_{0}-\Delta & \text{if\,}X_{0}<0
\end{cases}\,.
\end{align}
Replacing $\delta$, in \eqref{eq:APPENDIX_PDF_DELTA_X0_POS} and
\eqref{eq:APPENDIX_PDF_DELTA_X0_NEG}, with the corresponding value
of the realization, $r$, of $R$, i.e., $\delta=\begin{cases}
r-x_{0} & \text{if }x_{0}>0\\
x_{0}-r & \text{if }x_{0}<0
\end{cases}$, we obtain \eqref{eq:D2D_distance_single_content_provider_POS} and
\eqref{eq:D2D_distance_single_content_provider_NEG} which, including
the unitary probability mass at $r=0$ when $x_{0}=0$, combine to
provide the desired expression \eqref{eq:Theorem1} of the conditional
PDF of the closest achievable transmission distance for the PCP.\hfill{}$\blacksquare$\end{proofofT}

\section{Physical layer model, transmit power setting, and packet error modeling\label{Appendix:Transmit-power-setting-and-error-model}}

In the majority of network level studies, the radio channel between
any two transceivers is represented as a scalar quantity, namely an
attenuation coefficient, which allows a simple mapping between the
signal to noise ratio (SNR) and the packet error probability. This
approach, however, may result in a considerable accuracy when dealing
with a multicarrier wideband system, in which channels are subject
to frequency selective fading and lognormal shadowing, with spatially
correlated statistic parameters \citep{METIS_chanmod}. Using simplistic
channels may lead to a considerable inaccuracy in the evaluation of
the transmit power required for a successful transmission. In this
work, rather than the SNR, to compute the required transmit power
we use the approach of guaranteeing a prescribed outage probability
on the Shannon capacity of the channel, referred to a nominal bandwidth.
More precisely, let $f$ be the frequency and $H(f)$ be the channel
frequency response We assume that power is allocated uniformly over
the subcarriers in use and indicate with $\mathcal{P}_{c}$ the transmit
power used on each subcarrier in use. Let $\sigma_{c}^{2}=\digamma_{\text{noise}}\mathcal{N}_{0}w_{c}$
is the noise power on a subcarrier bandwidth, where $\mathcal{N}_{0}$
is the thermal noise spectral density, $\digamma_{\text{noise}}$
is the receiver noise figure, and $w_{c}$ is the subcarrier bandwidth.
The Shannon capacity of the portion of spectrum corresponding to subcarrier
$k$ of the channel of a radio link of interest, is given by 
\[
c_{k}=w_{c}\log_{2}\left(1+\mathcal{P}_{c}\left|H(f_{k})\right|^{2}/\left(\sigma_{c}^{2}+\sum_{i=1}^{S}\mathcal{P}_{c}^{(i)}\left|H_{i}(f_{k})\right|^{2}\right)\right),
\]
where the index $i$ runs over the, say, $S$ interfering transmitters,
and $H_{i}(f_{k})$ is the transfer function, evaluated at $f_{k}$,
the $i$-th interfering transmitter and the receiver of the radio
link of interest, and $\mathcal{P}_{c}^{(i)}$ is the transmit power
of the $i$-th interfering transmitter. The achievable amount of information
that can be transferred, in an information theoretic sense, using
a PRB, is $I_{\text{PRB}}=\tau\sum_{k=1}^{K_{sc}}c_{k}$, where $K_{sc}$
is the number of subcarriers in a PRB bandwidth. Assuming that, on
each subcarrier, a fixed modulation scheme is used, in which $e$
bits are encoded in a symbol\footnote{We consider a symbol duration as the inverse of the subcarrier bandwidth.}
(e.g., a 64QAM constellation allows to encode $e=6$ bps/Hz), the
maximum achievable rate on a subcarrier is limited by $\bar{c}_{k}=\min$$\left(c_{k},ew_{c}\right)$.

We set the transmit power per subcarrier $\mathcal{P}_{c}$ for transmitting
over a range $r$, as a function of the nominal channel gain $g\left(r\right)$.
More precisely, we find the power required to support a capacity equal
the information rate $ew_{c}$ required by the constellation in use,
and, to compensate for the effect of fading and (eventual) interference,
we add a suitable link margin. More specifically, we invert the function
$c=w_{c}\log_{2}\left(1+\mathcal{P}_{c}g\left(r\right)\right)$ with
respect to $\mathcal{P}_{c}$ by imposing $c=ew_{c}$, and multiply
by $M$, obtaining\vspace*{-3mm}
\[
\mathcal{P}_{c}=M\frac{\sigma_{c}^{2}}{g\left(r\right)}\left(2^{e}-1\right).
\]

Note that both the function $g\left(r\right)$ and the required link
margin differ between D2D and I2D communications. In fact, to compute
the nominal channel gain $g\left(r\right)$ we use the formula in
\citep[Table 7-1, UMi-O2O-(BS-UE)-LOS]{METIS_chanmod} for I2D communications,
and \citep[Table 7-1, UMi-O2O-D2D/V2V]{METIS_chanmod} for D2D communications.
Furthermore we set the link margin with the same settings we used
in \citep[Table 2]{Pescosolido2018AdHocNetworks} for the ``Urban
Micro'' scenario (UMi) for I2D communications and V2V scenario for
D2D ones, which, measured in dB, are given by (considering the frequency
selective channel model in use) $M_{\text{I2D}}=10\,\text{dB}$ and
$M_{\text{D2D}}=13\,\text{dB}$.

Now, assume that $N_{\text{PRB}}$ PRBs are used for a packet transmission,
enumerated with the index $m=1,\ldots,N_{\text{PRB}}$, than the achievable
amount of information for that transmission is\small{
\begin{align}
I & =\sum_{m=1}^{N_{\text{PRB}}}I_{\text{PRB}}(m)=\tau\sum_{m=1}^{N_{\text{PRB}}}\sum_{k=1}^{K_{sc}}\bar{c}_{k,m}\label{eq:Amount of information per packet transmission}\\
 & =\tau w_{c}\sum_{m=1}^{N_{\text{PRB}}}\sum_{k=1}^{K_{sc}}\min\left(e,\log_{2}\left(1+\frac{\mathcal{P}_{c}\left|H(f_{k,m})\right|^{2}}{\sigma_{c}^{2}+\mathcal{P}_{c}^{(i)}\sum_{l=1}^{L_{i}}\left|H_{i}(f_{k,m})\right|^{2}}\right)\right).
\end{align}
}The number of bits encoded in a packet is determined by the payload
size, which we indicate with $L$, and the amount of redundancy inserted
by the forward correction code. In this work, we include forward error
correction (FEC) in the model by assuming that each payload of $L$
bits is encoded in a packet of $L/\beta$ bits, where $\beta<1$ is
the FEC coding rate of the code in use (the lower $\beta$, the stronger
the error correction code). More specifically, for a payload of $L$
bits, the number of required PRBs is\vspace*{-2mm}
\[
N_{\text{PRB}}=\frac{L}{\beta}\frac{1}{e\tau w}.
\]
We model transmission errors as the events that the achievable amount
of information $I$ in \eqref{eq:Amount of information per packet transmission}
(which is function of the actual frequency selective channel experienced
by the transmitter, and of the interfering transmissions) is less
than $L$, i.e, a successful transmission occurs if the following
inequality is satisfied
\[
w_{c}\sum_{m=1}^{N_{\text{PRB}}}\sum_{k=1}^{K_{sc}}\min\left(e,\log_{2}\left(1+\frac{\mathcal{P}_{c}\left|H(f_{k,m})\right|^{2}}{\sigma_{c}^{2}+\mathcal{P}_{c}^{(i)}\sum_{l=1}^{L_{i}}\left|H_{i}(f_{k,m})\right|^{2}}\right)\right)\geq L.
\]
\end{appendices}
\end{document}